\documentclass[sigconf]{acmart}

\usepackage{adjustbox}
\usepackage{amssymb}
\usepackage{bm}
\usepackage{caption}
\usepackage{color}
\usepackage{makecell}
\usepackage{microtype}
\usepackage{multirow}
\usepackage{paralist}
\usepackage{rotating}
\usepackage{subcaption}
\usepackage{tabularx}
\usepackage{url}
\usepackage{xcolor}

\usepackage{booktabs}
\newcommand*\cbottomrule[1]{\cmidrule[\heavyrulewidth]{#1}\addlinespace}

\newcommand{\expectation}[1]{\mathbb{E}[#1]}
\newcommand{\open}{\mathcal{O}}
\newcommand{\closed}{\mathcal{C}}
\newcommand{\given}{\;\vert\;}
\newcommand{\score}{s}
\newcommand{\scoreijk}{\score(i, j, k)}
\newcommand{\nbrhood}[1]{N(#1)}
\newcommand{\snbrhood}[1]{R(#1)}
\newcommand{\cardinality}[1]{\lvert #1 \rvert}
\newcommand{\mycheck}{\raisebox{-0.5pt}{{\sffamily X}}}

\newcommand{\validinds}{\mathcal{S}}
\newcommand{\triangleset}{\mathcal{T}}
\newcommand{\indicator}[1]{\text{Ind}[#1]}

\newcommand{\tricount}{\tau}  % all triangles
\newcommand{\opentricount}{\sigma}  % open triangles
\newcommand{\wedgecount}{\omega} % wedges
\newcommand{\edgeisocount}{\eta} % single edge + isolated node
\newcommand{\emptycount}{\phi} % 3-node empty subgraph
\newcommand{\tetracount}{\rho} % all tetrahedrons
\newcommand{\opentetracount}{\pi} % open tetrahedrons
\newcommand{\fiveedgecount}{\theta} % 4-node, 5-edge
\newcommand{\triedgecount}{\psi} % triangle + adjacent edge
\newcommand{\triisocount}{\lambda} % triangle + isolated node

\newcommand{\sppr}{S}
\newcommand{\spprgrad}{\sppr_{\text{grad}}}
\newcommand{\spprcurl}{\sppr_{\text{curl}}}
\newcommand{\spprharm}{\sppr_{\text{harm}}}

\newcommand{\edgelap}{\Delta}
\newcommand{\normedgelap}{\hat{\edgelap}}

\newcommand{\phantomsubfigure}[1]{\begin{subfigure}[b]{0.1\textwidth}\phantomcaption\label{#1}\end{subfigure}}
\newcommand{\formatsublabel}[1]{(\textbf{#1})}

\newcommand{\xhdr}[1]{\vspace{0.75mm}\noindent{{\bf #1.}}\hspace{0.5mm}}

\PassOptionsToPackage{hyphens}{url}\usepackage{hyperref} % https://tex.stackexchange.com/questions/329461/arxiv-option-clash-for-package-hyperref-when-there-are-no-options
\definecolor{mylinkcolor}{RGB}{0,0,140}
\hypersetup{colorlinks,allcolors=mylinkcolor,citecolor=mylinkcolor}
\usepackage[capitalize]{cleveref}
\crefname{figure}{Figure}{Figures}

\usepackage{tikz,nicefrac,amsmath,pifont,tkz-graph,makecell,tkz-graph}
\usetikzlibrary{arrows,backgrounds,patterns,matrix,shapes,fit,calc,shadows,plotmarks}
\usetikzlibrary{arrows.meta}

\begin{document}

%make title bold and 14 pt font (Latex default is non-bold, 16 pt)
\title{Simplicial closure and higher-order link prediction}

\author{Austin R.~Benson}
\affiliation{%
  \institution{Cornell University}
}
\email{arb@cs.cornell.edu}

\author{Rediet Abebe}
\affiliation{%
  \institution{Cornell University}
}
\email{red@cs.cornell.edu}

\author{Michael T.~Schaub}
\affiliation{%
  \institution{MIT and University of Oxford}
}
\email{mschaub@mit.edu}

\author{Ali Jadbabaie}
\affiliation{%
  \institution{MIT} 
}
\email{jadbabai@mit.edu}

\author{Jon Kleinberg}
\affiliation{%
  \institution{Cornell University}
}
\email{kleinber@cs.cornell.edu}

\maketitle

%!TEX root = higher-order-link-prediction-postprint.tex

% Please provide an abstract of no more than 250 words in a single
% paragraph. Abstracts should explain to the general reader the major
% contributions of the article. References in the abstract must be cited in full
% within the abstract itself and cited in the text.

\subsection*{Abstract}
Networks provide a powerful formalism for modeling complex systems by
using a model of pairwise interactions. But much of the
structure within these systems involves interactions that take place among more
than two nodes at once---for example, communication within a group rather than
person-to-person, collaboration among a team rather than a pair of coauthors, or
biological interaction between a set of molecules rather than just two. Such
\emph{higher-order interactions} are ubiquitous, but their empirical study has
received limited attention, and little is known about possible
organizational principles of such structures.
Here we study the temporal evolution of 19 datasets with explicit accounting
for higher-order interactions.
We show that there is a rich variety of structure in our datasets but datasets
from the same system types have consistent patterns of higher-order structure.
Furthermore, we find that tie strength and edge density are competing positive indicators
of higher-order organization, and these trends are consistent across interactions
involving differing numbers of nodes.
To systematically further the study of theories for such higher-order structures,
we propose higher-order link prediction as a benchmark problem
to assess models and algorithms that predict higher-order structure.
We find a fundamental differences from traditional pairwise link prediction,
with a greater role for local rather than long-range information
in predicting the appearance of new interactions.

%!TEX root = higher-order-link-prediction-postprint.tex

\section{Introduction}

Networks are a fundamental abstraction for complex systems and
relational data throughout the sciences~\cite{Albert-2002-survey,Easley2010,Newman-2003-survey}.
The basic premise of network models is to represent the elements of the underlying system
as nodes, and to use the links of the network to capture pairwise relationships.
In this way, a social network can represent the friendships between pairs of
people; a Web graph can encode links among Web pages or topic categories; and a
biological network can represent the interactions among pairs of biological
molecules or components~\cite{Granovetter-1973-strength,Deane-2002-PPI,Newman-2003-survey,Bullmore-2009-survey}.
But much of the structure in these systems involves {\em higher-order interactions} between more than two entities at
once~\cite{newman-affiliation,Milo-2002-motifs,Ugander-2012-diversity,Benson-2016-hoo,Grilli-2017-higher}:
people often communicate or interact in social groups, not just in pairs;
associative relations among ideas or topics often involve the intersection of
multiple concepts; and joint protein interactions in biological networks are
associated with important phenomena~\cite{Navlakha-2010-power}.

These types of higher-order, group-based interactions are apparent even in the
standard genres of datasets used for network analysis. For example, coauthorship
networks are built from data in which larger groups write papers together, and
similarly, email networks are based on messages that often have multiple
recipients. While such higher-order structure is not captured by the topology of a
graph, it may be modeled via a collection of formalisms that include set
systems~\cite{frankl-set-systems-handbook},
hypergraphs~\cite{berge-hypergraphs-book}, simplicial
complexes~\cite{Hatcher2002}, and bipartite affiliation
graphs~\cite{feld-foci,newman-affiliation}. Despite the existence of
mathematical formalisms for higher-order structure, 
there is no unifying study that analyzes the basic higher-order
structure of such datasets. This is in sharp contrast to other notions of
``higher-order models'' generalizing graph data, such as multiplex
networks~\cite{Kivela-2014-multilayer} and higher-order
Markov chain models~\cite{Xu-2016-higher-order,Rosvall-2014-memory},
which are successful but still rooted in a pairwise representation paradigm.
We study the complementary direction of group interactions, as outlined in the examples
above, and use the term ``higher-order model'' in this sense.

A key reason for the lack of large-scale studies in higher-order
models is that data is often collected directly in a network format, thus
eliminating higher-order interactions already at the data-collection stage.
Another reason is that analyzing higher-order interactions can be
computationally challenging for large datasets. Consequently, despite their
potential importance, little is known about organizational principles of
higher-order structures within real-world datasets. For instance,
one question that remains to be answered is whether higher-order interactions
enable us to differentiate different kind of datasets, or whether higher-order
properties are universal across datasets.

Here, we provide the first steps in the direction of promoting a broad, rigorous
study of higher-order topological interactions across domains. To this end, we
study the structure and temporal evolution of 19 datasets from a variety
of domains that have higher-order interactions. We find that
distinct patterns for different domains are immediately revealed with 3-way interaction
features that are not available from the graph structure of the networks alone.

%!TEX root = higher-order-link-prediction-postprint.tex

\begin{figure*}[tb]
  \phantomsubfigure{fig:example_A}
  \phantomsubfigure{fig:example_B}
  \phantomsubfigure{fig:example_C}
  \phantomsubfigure{fig:example_D}  
  \centering
  \includegraphics[width=0.7\linewidth]{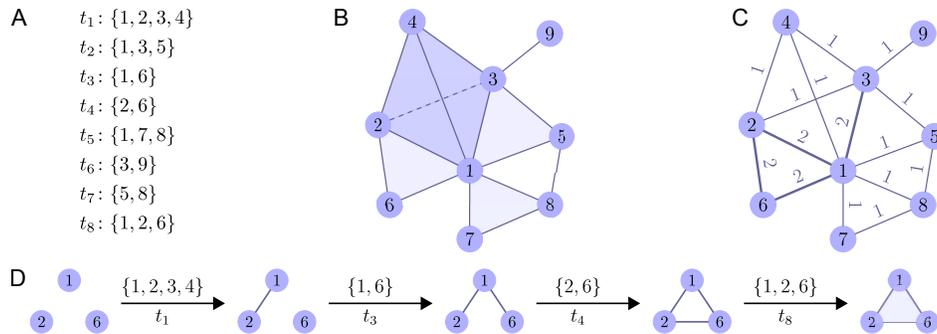}
  \caption{% 
    Higher-order network models, open and closed closed $3$-node cliques (triangles), and
    simplicial closure events.
    \formatsublabel{A} Example higher-order network dataset consisting of eight timestamped
    simplices on nine nodes. More than one simplex can appear at a given
    time, which often occurs in real-world data with coarse-grained
    temporal measurements.
    We study 19 real-world datasets of this
    type (\cref{tab:dataset_stats}).
    \formatsublabel{B} Visual representation of the dataset (ignoring timestamps). Shading
    represents the simplices (in order to highlight the difference with
    traditional graphs), and the dashed line between nodes 2 and 3 denotes
    three-dimensional perspective for the 4-node simplex $\{1,2,3,4\}$ (this 4-node
    simplex also has darker shading).
    Nodes $1$, $2$, and $3$ form a \emph{closed}
    3-node clique (i.e., closed triangle) since all three nodes appeared in the same simplex at
    time $t_1$, whereas nodes $1$, $5$, and $8$ form an \emph{open}
    triangle since all three pairs of nodes co-appeared in a simplex
    (time $t_2$ for nodes $1$ and $5$, time $t_5$ for nodes $1$ and
    $8$, and time $t_7$ for nodes $5$ and $8$) but no one simplex contains all three nodes.
    Thus, the region between nodes $1$, $5$, and $8$ is not shaded. In total, the
    dataset has seven closed triangles---$\{1, 2, 3\}$, $\{1, 2,
    4\}$, $\{1, 3, 4\}$, $\{2, 3, 4\}$, $\{1, 3, 5\}$, $\{1, 2, 6\}$,
    $\{1, 7, 8\}$---and one open triangle---$\{1, 5, 8\}$. We
    find that the fraction of triangles that are open varies widely
    depending on the dataset (\cref{fig:structure}).
    \formatsublabel{C} The ``projected graph'' of the dataset. The
    weight of an edge is the number of times its two end points have
    appeared in a simplex together. Open and closed triangles are both
    triangles in the projected graph. Traditional network science
    ideas often ignore higher-order structure and only use this graph.
    \formatsublabel{D} A simplicial closure event for nodes $1$, $2$, and
    $6$. Each transition lists the new simplex and the time it appears in the
    dataset. Before closing, the three nodes induce several subgraphs in the
    projected graph over time. For example, the nodes form an open triangle at
    time $t_4$, which persists until time $t_8$ when the simplicial closure
    event occurs. We study properties of such simplicial closure events and
    predict their future occurrence as part of a framework for evaluating
    higher-order network models.
    }
    \label{fig:example}
\end{figure*}

Motivated by the importance of triangular structures in network clustering and
the theory of triadic closure in social
networks~\cite{Granovetter-1973-strength,Newman-2001-clustering}, we study an
extension of this theory via \emph{simplicial closure}, or the way in which
groups of nodes evolve until eventually co-appearing in a higher-order
structure. In this case, we find that strong previous interactions between subsets of
a group increases the likelihood of a \emph{simplicial closure event}, where the nodes
appear in a group together. The relative importance of different types of prior
interactions depends on the dataset yet remains consistent when considering
groups of different sizes for a given dataset. To facilitate future modeling
and demonstrate that the higher-order patterns are not simple epiphenomena of
the underlying link structure, we introduce a higher-order link prediction
problem---the forecasting of future higher-order interactions---as an evaluation
framework for models and algorithms that aim to predict the emergence of
higher-order structure from existing data.

%!TEX root = higher-order-link-prediction-postprint.tex

\section{Structural analysis of higher-order networks}\label{sec:structure}
We assembled a diverse collection of 19 datasets, recording the timestamped
interactions of groups of entities. Thus, each dataset is a set of timestamped
sets of nodes. We call each set of nodes a \emph{simplex}, and the nodes in
each simplex take part in a shared interaction at a given timestamp
(\cref{fig:example_A}). For example, in a coauthorship network, a simplex
corresponds to a set of authors publishing an article at a given time.

Formally, each dataset consists of $N$ timestamped simplices, ${\{(S_i, t_i)\}}_{i=1}^{N}$, 
where $t_i \in \mathbb{R}$ is the time at which simplex $S_i$ was observed, and
$S_i$ is a set representing the nodes in the $i$th simplex.
If $\cardinality{S_i} = k$, we say that $S_i$ is a $k$-node simplex.%
\footnote{Such a structure is called a (k-1)-simplex in 
algebraic topology, and the set of all its pairs is
called a k-clique in graph theory.}
This set-based representation provides a natural format for datasets from a range of domains.
We briefly describe our datasets below (see \cref{sec:SI_data} for more complete descriptions).
\begin{itemize}
\item \emph{Coauthorship data} (coauth-DBLP; coauth-MAG-History; coauth-MAG-Geology):
  nodes are authors and a simplex is a publication; DBLP spans over 80 years and the other two datasets span about 200 years. 
\item \emph{Online tagging data} (tags-stack-overflow; tags-math-sx; tags-ask-ubuntu):
  nodes are tags (annotations) and a simplex is a set of tags for a question on online Stack Exchange forums; the data contains the complete history of the forums.
\item \emph{Online thread participation data} (threads-stack-overflow; threads-math-sx; threads-ask-ubuntu):
  nodes are users and a simplex is a set of users answering a
  question on a forum; again, the data contains
  the complete history of the forum. 
  \item \emph{Drug networks from the National Drug Code Directory}
  (NDC-classes): nodes are class labels (e.g., serotonin reuptake inhibitor)
  and a simplex is the set of class labels applied to a drug (all applied at one time).\\
  (NDC-substances): nodes are substances (e.g., testosterone) and a simplex is the set
  of substances in a drug; datasets include the complete history of the directory
\item \emph{U.S. Congress data} (congress-committees~\cite{Porter-2005-committees}; congress-bills~\cite{Fowler-2006-cosponsorship}):
  nodes are members of Congress and a simplex is the set of members in
  a committee or co-sponsoring a bill; the committees dataset spans
  1989 to 2003 and the bills dataset spans 1973 to 2016. 
\item \emph{Email networks} (email-Enron~\cite{Klimt-2004-Enron}; email-Eu~\cite{Paranjape-2017-motifs}):
  nodes are email addresses and a simplex is a set consisting of all
  recipient addresses on an email along with the sender's address;
  email-Enron spans most of the duration of a company's lifetime, and
  email-Eu spans over 2 years. 
\item \emph{Contact networks} (contact-high-school~\cite{Mastrandrea-2015-contact}; contact-primary-school~\cite{Stehle-2011-contact}):
  nodes are persons and a simplex is a set of persons in close proximity to each other
\item \emph{Drug usage in the Drug Abuse Warning Network} (DAWN):
  nodes are drugs and a simplex is the set of drugs reportedly used by a patient
  prior to an emergency department visit. 
\item \emph{Music collaboration} (music-rap-genius):
  nodes are rap artists; simplices are sets of rappers collaborating on songs.
\end{itemize}

%%%%%
% Table showing statistics of open and closed triangles.
%!TEX root = higher-order-link-prediction-postprint.tex

\begin{table}[tb]
\setlength{\tabcolsep}{3pt}
\centering
\caption{Summary statistics for our datasets. Each dataset is a collection of
  timestamped simplices (as in \cref{fig:example}).}
\scalebox{0.91}{
\begin{tabular}{l l l l l}
\toprule
Dataset                & nodes     & edges in     & timestamped & unique    \\
                       &           & proj.\ graph & simplices   & simplices \\
\midrule
coauth-DBLP            & 1,924,991 & 7,904,336    & 3,700,067   & 2,599,087 \\
coauth-MAG-Geology     & 1,256,385 & 512,0762     & 1,590,335   & 1,207,390 \\
coauth-MAG-History     & 1,014,734 & 1,156,914    & 1,812,511   & 895,668   \\
music-rap-genius       & 56,832    & 123,889      & 224,878     & 85,429    \\
tags-stack-overflow    & 49,998    & 4,147,302    & 14,458,875  & 5,675,497 \\
tags-math-sx           & 1,629     & 91,685       & 822,059     & 174,933   \\
tags-ask-ubuntu        & 3,029     & 132,703      & 271,233     & 151,441   \\
threads-stack-overflow & 2,675,955 & 20,999,838   & 11,305,343  & 9,705,709 \\
threads-math-sx        & 176,445   & 1,089,307    & 719,792     & 595,778   \\
threads-ask-ubuntu     & 125,602   & 187,157      & 192,947     & 167,001   \\
NDC-substances         & 5,311     & 88,268       & 112,405     & 10,025    \\
NDC-classes            & 1,161     & 6,222        & 49,724      & 1,222     \\
DAWN                   & 2,558     & 122,963      & 2,272,433   & 143,523   \\
congress-bills         & 1,718     & 424,932      & 260,851     & 85,082    \\
congress-committees    & 863       & 38,136       & 679         & 678       \\
email-Eu               & 998       & 29,299       & 234,760     & 25,791    \\
email-Enron            & 143       & 1,800        & 10,883      & 1,542     \\
contact-high-school    & 327       & 5,818        & 172,035     & 7,937     \\
contact-primary-school & 242       & 8,317        & 106,879     & 12,799    \\
\bottomrule
\end{tabular}
}
\label{tab:dataset_stats}
\end{table}

%%%%%

%%%%%
% Figure showing open and closed triangles.
%!TEX root = higher-order-link-prediction-postprint.tex

\begin{figure*}[tb]
  \phantomsubfigure{fig:structure_A}
  \phantomsubfigure{fig:structure_B}
  \phantomsubfigure{fig:structure_C}
  \phantomsubfigure{fig:structure_D}
  \phantomsubfigure{fig:structure_E}
  \centering
  \includegraphics[width=0.95\linewidth]{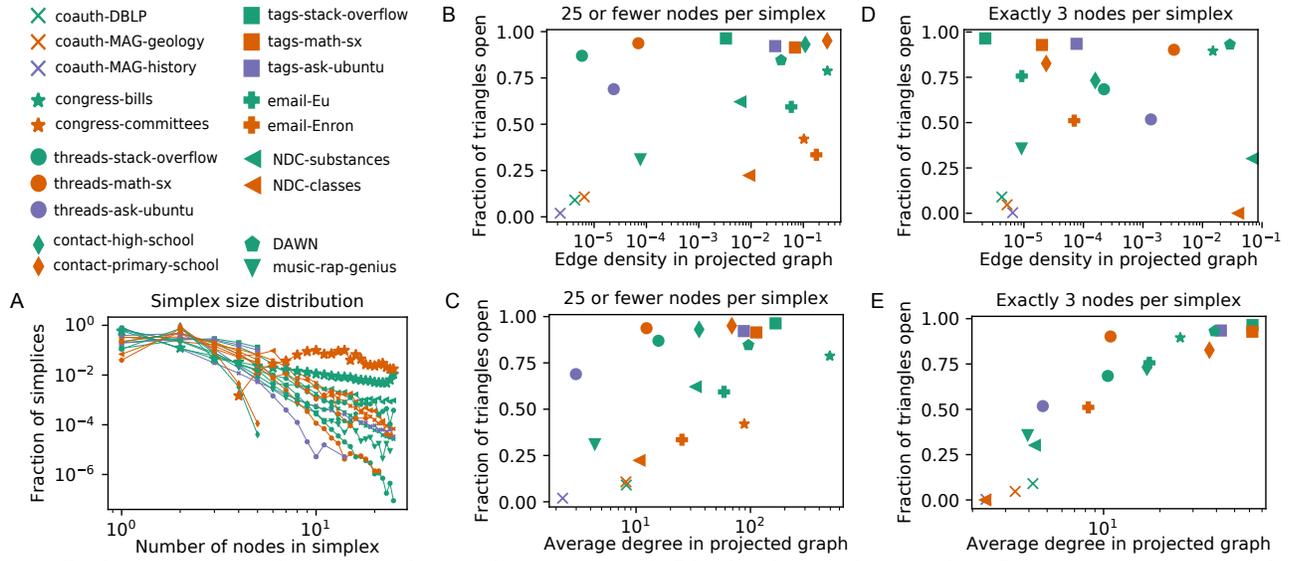}
  \caption{%
    Basic structure of higher-order interaction datasets.
    \formatsublabel{A} Distribution of simplex sizes. In most datasets, small
    simplices ($\le 4$ nodes) are the most common.
    %The congress-committees %dataset has no 3-node simplices, so it does not appear in subfigures D and E.
    %
    \formatsublabel{B--C} Dataset landscapes in terms of fraction of triangles
    that are open and either edge density (B,D) or average degree (C,E) when
    considering simplices with 25 or fewer nodes (B and C) or just 3-node
    simplices (D and E).  Datasets from the same domain tend to be similar with
    respect to these features, whether or not we include simplices with greater
    than 3 nodes. Indeed, we can predict the system domain of some datasets
    by measuring these statistics on egonets
    (\cref{tab:egonet_prediction,fig:egonet_decision_boundary}).
  }
  \label{fig:structure}
\end{figure*}

%%%%%

To provide uniformity across datasets, we restrict to simplices consisting of at most 25 nodes.
This is relevant to, e.g., the coauthorship data in which large consortia of hundreds of authors collaborate on a single paper.
However, such events are rare and not relevant for our analysis.
\Cref{tab:dataset_stats} lists some summary statistics of the datasets 
The number of unique simplices appearing in the data is minuscule compared to the total
number of possible simplices. For example, in the dataset with the smallest number of nodes (email-Enron, 143 nodes),
there are nearly 500 million possible simplices of size at most 5, whereas only 1,542 unique simplices appear in the dataset.
On the other hand, in most datasets, the number of unique simplices is within an order of magnitude of the number of pairs of nodes
that co-appear in some simplex (edges in the projected graph; to be be discussed in the next section).

\subsection{Higher-order features reveal rich structural diversity}
Our data representation distinguishes between the observation of different kinds of $k$-way interactions between a set of entities.
Stated differently, unlike in a graph representation, we do not break down each simplex into a set of (induced) pairwise interactions.
Though the specific representation is not essential provided the information of the group interaction is faithfully encoded,
it is convenient to think of our data as an abstract simplicial complex as depicted in \cref{fig:example_B}.

The simple encoding of the observed information as a graph is called the \emph{projected graph}.
Formally, in the projected graph, two nodes are joined by an edge of weight $w$ if they co-appear in $w$ simplices (\cref{fig:example_C}).
A $k$-clique in the projected graph is a set of nodes among which an edge is present between all pairs.
A $k$-cliques appear if
(i) the $k$ nodes were all part of a some simplex, or
(ii) each pair was part of some simplex, although all $k$ were never part of the same simplex. 
In the former case, we say the $k$ nodes form a {\em closed} clique, while in the latter case we say they form an {\em open} clique. 

We first study the occurrence of open and closed 3-cliques, or triangles (\cref{fig:structure}).
This is the simplest higher-order structure present in our datasets that is not captured by a graph.
Furthermore, triangles are one of the most important structural patterns in network analysis~\cite{Granovetter-1973-strength,Milo-2002-motifs,Kossinets-2006-empirical}.
As discussed above, there are two types of triangles which cannot be distinguished by the weighted projected graph alone.
In a \emph{closed triangle}, all three nodes have co-appeared in at least one simplex.
Formally, $\{u,v,w\}$ is a closed triangle if there exists some simplex $S_i$ for which $\{u,v,w\} \subset S_i$.
In an \emph{open triangle}, on the other hand, every pair of the three nodes has co-appeared in at least one simplex, but no single simplex contains all three nodes.

Every simplex with at least three nodes directly creates a closed triangle, while open triangles appear coincidental. 
Moreover, larger simplices lead to many closed triangles: for instance, a $k$-node simplex contributes ${k \choose 3}$ closed triangles.
Thus, one might intuit that closed triangles are much more common than open triangles due the presence of (potentially) large groups.
On the other hand, only a small fraction of all possible simplices are present in the network when compared to the total number of possible edges
in the projected graph, so one might expect that there are more open triangles.
Our analysis reveals that, across our datasets, there is a spectrum for the fraction of triangles that are open (\cref{fig:structure_B,fig:structure_C}).

While the distribution of simplex sizes is broadly similar in most datasets (\cref{fig:structure_A}),
jointly analyzing the edge density in the projected graph with the fraction of triangles that are open reveals a rich landscape of datasets (\cref{fig:structure_B}):
(i) low-density with a small fraction of triangles open (coauthorships and music collaboration);
(ii) low-density with a large fraction of triangles open (stack exchange threads)
(iii) high-density with a large fraction of triangles open (stack exchange tags, contact, bill co-sponsorship); and
(iv) high-density with a medium fraction of triangles open (email, Congress committee membership, NDC substances and classes).
These results are not skewed by large simplices---the landscape is broadly preserved when restricting to the $3$-node simplices (\cref{fig:structure_D}).

Measuring average unweighted degree along with fraction of open triangles also reveals substantial diversity, and 
datasets from the same domain continue to exhibit similar features (\cref{fig:structure_C}).
Restricting the data to only 3-node simplices, we find a near-linear relationship between the fraction of open triangles and the log of the average degree (\cref{fig:structure_E}). 
A linear model for the data in \cref{fig:structure_E} has $R^2 = 0.85$, compared to $R^2 = 0.38$ for a linear model of the data in \cref{fig:structure_D}.
This suggests that larger simplices bring diversity to the data.

\subsection{Higher-order egonet features discriminate system domains}

%!TEX root = higher-order-link-prediction-postprint.tex

\begin{figure}[tb]
  \centering
  \includegraphics[width=\columnwidth]{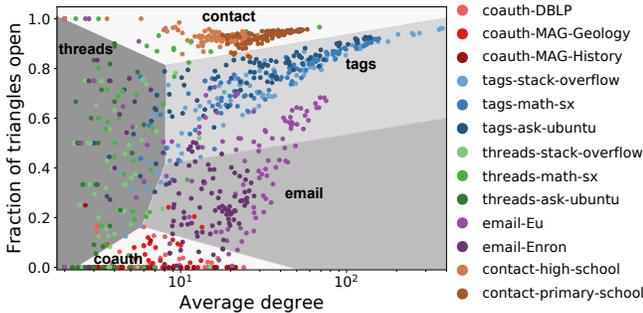}
  \vspace{-4mm}
  \caption{%
    Class decision boundaries of a learned multinomial logistic regression model
    for predicting five dataset system domains (coauthorship, threads, tags,
    email, or contact) using the log of the average degree ($\log(\bar{d})$) and
    fraction of triangles that are open ($f$) of egonets.  Markers correspond to
    sampled egonets used in model training. The two-feature linear model can
    predict the 5-class dataset domain with 75\% accuracy
    (\cref{tab:egonet_prediction}). In conjunction with the prediction
    accuracies in \cref{tab:egonet_prediction}, our analysis suggests that the
    fraction of triangles that are open (a higher-order network statistic) is an
    important covariate for analyzing and modeling the local structure of
    higher-order interaction data.
  }
  \label{fig:egonet_decision_boundary}
\end{figure}

The structural diversity of the datasets is also present at the local level of
egonets (1-hop neighborhoods of nodes), and local statistics can identify the
``system domain'' of datasets. By system domain, we simply mean the
categories identified in~\cref{fig:structure} that correspond to datasets
recorded from the same kind of system. Our collection of datasets has five
clear system domains with at least two datasets each: coauthorship, online tags,
online thread co-participation, email, and proximity contact.

We trained a multinomial logistic regression model to determine system domain as
follows. We computed (i) the fraction of open triangles, (ii) the log of the
average degree in the projected graph, and (iii) the log of edge density in the
projected graph of 100 egonets sampled uniformly at random (without replacement)
from all egonets containing at least one open or closed triangle in each of 13
datasets categorized as coauthorship, stack exchange tags, stack exchange
threads, email, or contact. Using 80 samples from each of the 13 datasets as
training data, we then trained an $\ell_2$-regularized multinomial logistic
regression classifier to predict the system domain using subsets of the three
features above and an intercept term. The model was trained using the
scikit-learn library (the regularization parameter was set to $C = 10$). Test
accuracy was computed on the remaining 20 samples for each dataset. This entire
process described was repeated 20 times, resulting in 20 different collections
of egonet samples.

The model using the fraction of triangles that are open and log of the average
degree as covariates reveals clustering structure of the system domains
(\cref{fig:egonet_decision_boundary}; the decision boundary comes from one of
the 20 samples). This simple model can predict system domain with nearly 75\%
accuracy, compared to approximately 21\% accuracy with random guessing
proportional to system domain frequency.  The prediction accuracy provides
evidence that there are different organizational mechanisms at play locally for
different systems.  In conjunction with the structure illustrated in
\cref{fig:structure}, this suggest that there is not a single ``universal''
setting of values for simplicial network statistics; the context of the
underlying the network matters, but within a given context the parameters are
quite stable.

We also trained models with the log of the edge density as a covariate, in
addition to the log of the average degree and the fraction of triangles that are open;
model accuracy mildly increased from 75\% to 78\%
(\cref{tab:egonet_prediction}; reported results are the mean and standard deviation
over the 20 trials). However, discarding the log of the average degree
as a covariate decreases model accuracy to 60\%, and only including edge density
and average degree without the fraction of triangles that are open decreases model
accuracy to 50\%. The accuracy numbers are guides in how
to model higher-order interaction data. For example, we conclude
that the fraction of triangles that are open---a network statistics that relies on
knowledge of the higher-order structure in the dataset---is a valuable covariate
for identifying system domains. Thus, simple higher-order interactions should be used when
analyzing or modeling such data. Furthermore, the average degree tends to be more valuable
than edge density when considering local organizational mechanisms.

%%%%%
% tab:egonet_predictions
%%%%%
%!TEX root = higher-order-link-prediction-postprint.tex

\begin{table}[tb]
  \setlength{\tabcolsep}{5pt}
  \centering
  \caption{Prediction of dataset type by egonet features.
    For the datasets from coauthorship, threads, tags, email, and contact system domains,
    we sampled egonets and computed the edge density ($\rho$),
    average degree ($\bar{d}$), and fraction of triangles that are open ($f$).
    Using these features, we trained a multinomial logistic regression
    model to predict the system domain of the network. We report the mean
    and standard deviation over 20 random samples of 100 egonets.
    Models incorporating the fraction of triangles that are open outperform the one
    that does not, highlighting the importance of this feature for higher-order organization.
    \Cref{fig:egonet_decision_boundary} illustrates the model that uses $\log(\rho)$ and $f$ as features.
    }
  \begin{tabular}{c c c c c c c}
    \toprule
    \multicolumn{4}{l}{model features} & & \multicolumn{2}{l}{accuracy} \\
    \cmidrule(lr){1-4} \cmidrule(lr){6-7}
    $\log(\rho)$ & $\log(\bar{d})$ & \phantom{x}$f$\phantom{x} & intercept &  & random & multinomial LR \\
    \midrule
    \mycheck & \mycheck & \mycheck & \mycheck & \phantom{x} & 0.21 & 0.78 $\pm$ 0.02 \\
             & \mycheck & \mycheck & \mycheck & \phantom{x} & 0.21 & 0.75 $\pm$ 0.02 \\
    \mycheck &          & \mycheck & \mycheck & \phantom{x} & 0.21 & 0.60 $\pm$ 0.02 \\
    \mycheck & \mycheck &          & \mycheck & \phantom{x} & 0.21 & 0.49 $\pm$ 0.03 \\    
    \bottomrule
  \end{tabular}
  \label{tab:egonet_prediction}
\end{table}

\subsection{A simple generative model for open and closed triangles}
We have now seen that there is diversity in datasets from global network
statistics and that local statistics reveal system domains of the networks.
We now provide a simple generative model of
simplices that helps describe how diversity in the datasets might arise.
The model uses the hypothesis that $3$-node simplices form independently with a fixed probability.
While extreme, this hypothesis indeed leads to diversity in the fraction of triangles that are open.
To see this, suppose that a dataset consists only of $3$-node simplices on $n$ nodes, and any
set of three nodes $\{u,v,w\}$ appears in a simplex with
probability $p = 1 / n^{b}$, where $b > 0$ is a parameter regulating the probability of this event.
Let $X_{uvw}$ be the indicator random variable that $\{u, v, w\}$ is an open
triangle.
Then, for large $n$, it follows from the independence assumption that
\begin{equation}
  \expectation{X_{uvw}} \approx {(1 - {(1 - 1/n^{b})}^n)}^3.
\end{equation}

%%%%%
% fig:simulations
%!TEX root = higher-order-link-prediction-postprint.tex

\begin{figure}[t]
  \centering
  \includegraphics[width=0.67\columnwidth]{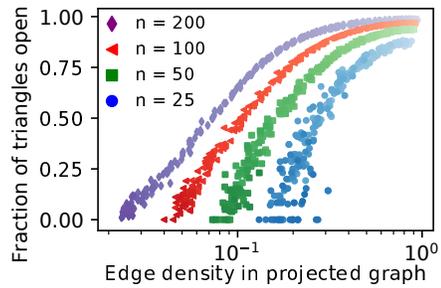}
  \caption{%
    Distribution of the fraction of triangles that are open and edge density
    in simulations from a model where each triple of $n$ total nodes forms a
    3-node simplex independently with probability $p = 1 / n^b$, $b \in [0.8, 1.8]$.
    Color scales with $b$ so that larger $p$ are lighter and smaller $p$ are
    darker. Varying $b$ creates datasets spanning all possible values of the
    fraction of triangles that are open.
  }
  \label{fig:simulations}
\end{figure}

%%%%%

%%%%%
% fig:lifecycles
%!TEX root = higher-order-link-prediction-postprint.tex

\begin{figure*}[ht]
  \phantomsubfigure{fig:lifecycles_A}
  \phantomsubfigure{fig:lifecycles_B}
  \phantomsubfigure{fig:lifecycles_C}
  \phantomsubfigure{fig:lifecycles_D}
  %\centering
  \includegraphics[width=\linewidth]{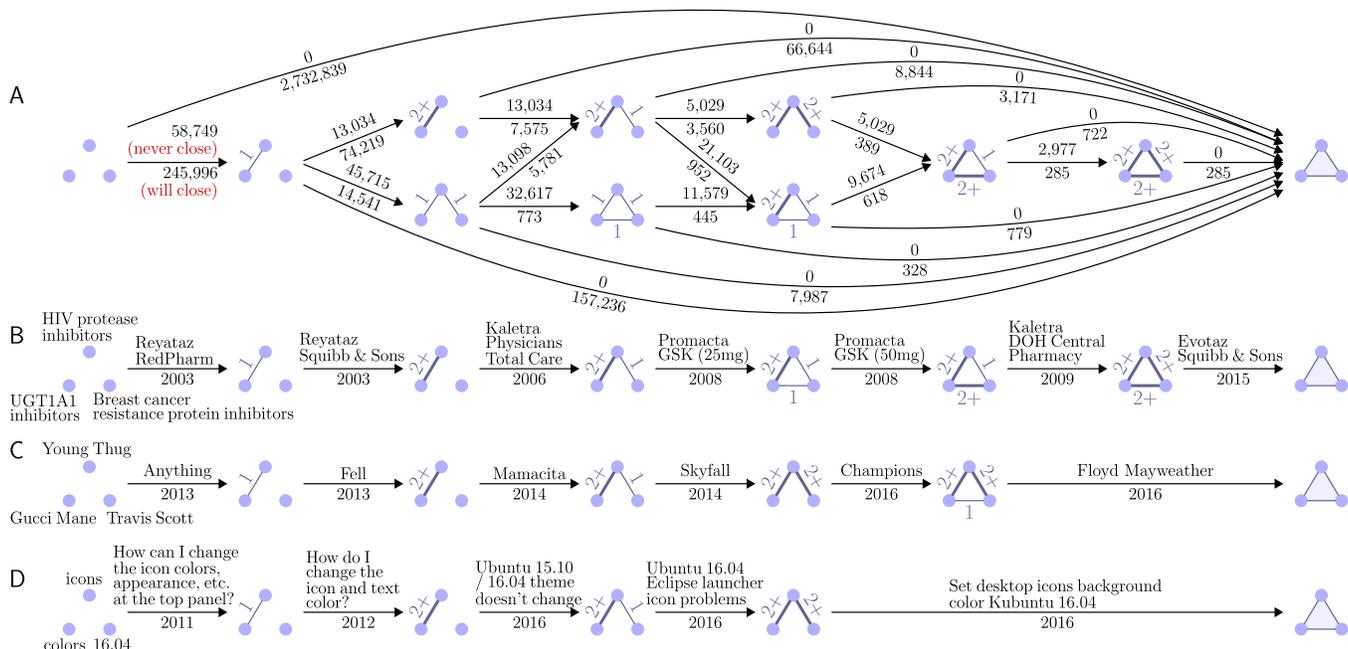}
  \caption{%
    Lifecycles of triples of nodes. Triangle edge weights
    are from the projected graph binned into weak
    ties for pairs of nodes appearing in only one simplex together (denoted ``1'')
    and strong ties for pairs of nodes appearing at least two simplices together 
    (denoted ``2+'').
    \formatsublabel{A} Lifecycles in the
    coauth-MAG-History dataset for all triples that eventually form a triangle.
    Edges represent transitions between configurations, and the numbers are counts of
    triples that follow the transition. The top number counts triples of nodes
    that never experience simplicial closure event (i.e., never reach the closed
    state on the far right), and the bottom number counts triples that do go
    through a simplicial closure event.
    \formatsublabel{B} Lifecycle of classification codes ``HIV protease inhibitors'',
    ``UGT1A1 inhibitors'', and ``Breast cancer resistance protein inhibitors''
    in the NDC-classes dataset, where simplices consist of the
    labels applied to drugs. Reyataz and Kaletra---two HIV-1 medications---produced strong ties via multiple drug
    labelers; RedPharm Drug Inc. and E.R.\ Squibb \& Sons, LLC labeled Reyataz,
    and Physicians Total Care and DOH Central Pharmacy labeled Kaletra.
    Promacta, a bone marrow stimulant classified as both a breast cancer
    resistance protein inhibitor and a UGT1A1 inhibitor, creates the open
    triangle. A strong tie is due to GlaxoSmithKline plc labeling multiple dosages
    of Promacta as products (25mg and 50mg). The introduction of Evotaz, a combination drug,
    induces a simplicial closure event for the three labels, 6 years after the open
    triangle formed.
    \formatsublabel{C} Lifecycle of rap artists Young Thug, Gucci Mane, and
    Travis Scott.  Mane and Thug first collaborated on the song ``Anything'' on
    a Mane mixtape; the two subsequently both featured on Waka Flocka Flame's
    track ``Fell''. Thug then twice featured on Travis Scott's 2014 mixtape
    ``Days Before Rodeo'', on the tracks ``Mamacita'' and ``Skyfall''. Both Mane
    and Scott featured on Kanye West's ensemble track ``Champions,'' leading to
    an open triangle.  A simplicial closure event occurred when Scott and Mane both
    featured on Thug's track ``Floyd Mayweather.''
    \formatsublabel{D} Lifecycle of tags ``icons'', ``colors'', and ``16.04''
    applied to questions on the Ask Ubuntu question-and-answer forum.  The tag
    16.04 refers to a 2016 Ubuntu release. There are questions about icons and
    colors independent of the Ubuntu version, dating back to 2011 (just one year
    after the forum was created). In 2016, users asked 16.04-specific icon
    questions related to the new release.  Finally, a 16.04-specific question on
    both icons and colors leads to a simplicial closure event.
  }
  \label{fig:lifecycles}
\end{figure*}

%%%%%

There are two asymptotic regimes here depending on the value of $b$.
If $b < 1$, then
$(1 - 1 / n^b)^n \le e^{-n^{1 - b}}$,
and $\expectation{X_{uvw}}$ approaches $1$ as $n$ gets large.
If $b > 1$, on the other hand,
\begin{align}
  \expectation{X_{uvw}} \approx (1 - (1 - 1/n^{b})^n)^3
  %\approx (1 - e^{-n^{1 - b}})^3
  %  = (1 / n^{b - 1} + O(1 / n^{2b - 2}))^3
  = O(1 / n^{3b - 3}).
\end{align}
Denote the set of open triangles by $\open$ and the set of closed triangles by $\closed$.
According to our calculations above, for large $n$, the expected number of open
triangles is
$\expectation{\lvert \open \rvert}= \sum_{\{u,v,w\}}\expectation{X_{uvw}} = O(n^3)$
if $b<1$.
For $b>1$, the expected number of open triangles for large $n$ is
$\expectation{\lvert \open \rvert} = O(n^{3(2 - b)})$.
The expected number of closed triangles is always $\expectation{\lvert \closed \rvert } = p \cdot {n \choose 3} = O(n^{3 - b})$.
Therefore, if $b< 3/2$, the number of open triangles grows faster, and if
$b>3/2$, the number of closed triangles grows faster.
To illustrate this numerically, we generated 5 random samples from this
model for $b=0.8,0.82,0.84,\ldots,1.8$ and $n = 25, 50, 100, 200$.
As suggested by the above theory, the samples have a fraction of
open triangles spanning the interval between $0$ and $1$
(\cref{fig:simulations}).

We can also use the above procedure to construct datasets with a smaller edge
density, while keeping the average degree fixed by
patching together $c$ replicates of one of these random
datasets; this creates a dataset with $c$ times as many nodes, but the same average
degree.
More formally, if a dataset with $n$ nodes has average degree $d$ and edge
density $\rho$, then the union of $c$ copies of this dataset has $cn$ nodes,
average degree $d$, and edge density
$c \rho (\binom{n}{2} - n) / (\binom{nc}{2} - nc) \approx \rho / c$
(for large $n$). Thus, our simple independent model spans the two-dimensional
feature space in \cref{fig:structure_B,fig:structure_D}, but this does not imply
that our data was generated by this model.

%!TEX root = higher-order-link-prediction-postprint.tex

\section{Temporal dynamics and simplicial closure events}\label{sec:simp_closure}
The above analysis already reveals useful information about the
organization of closed and open triangles, and studying the temporal dynamics of
the networks in detail offers additional insights. A possible hypothesis for
strong prevalence of open triangles would be temporal asynchrony in
link creation. For example, consider three Congresspersons $u$, $v$, and $w$ in the
committee membership dataset, where $u$ is in one committee
with $v$ and in another committee with $w$. If $u$ is not re-elected, there
will be no opportunity for the triple of nodes to form a closed triangle, as $u$
has effectively become inactive. An open triangle may still form if $v$ and $w$
are on the same committee in a future Congress. However, we find that temporal
asynchrony does not explain most open triangles. Depending on the dataset, the
three edges in 61.1\% to 97.4\% of open triangles have an overlapping period of
activity (including 89.5\% for Congress committees; see \cref{sec:temp_async}).

%%%%%
% fig:closure_probs
%!TEX root = higher-order-link-prediction-postprint.tex

\begin{figure*}[tb]
  \centering
  \phantomsubfigure{fig:closure_probs_A}
  \phantomsubfigure{fig:closure_probs_B}
  \phantomsubfigure{fig:closure_probs_C}
  \phantomsubfigure{fig:closure_probs_D}
  \phantomsubfigure{fig:closure_probs_E}
  \phantomsubfigure{fig:closure_probs_F}  
  \includegraphics[width=\linewidth]{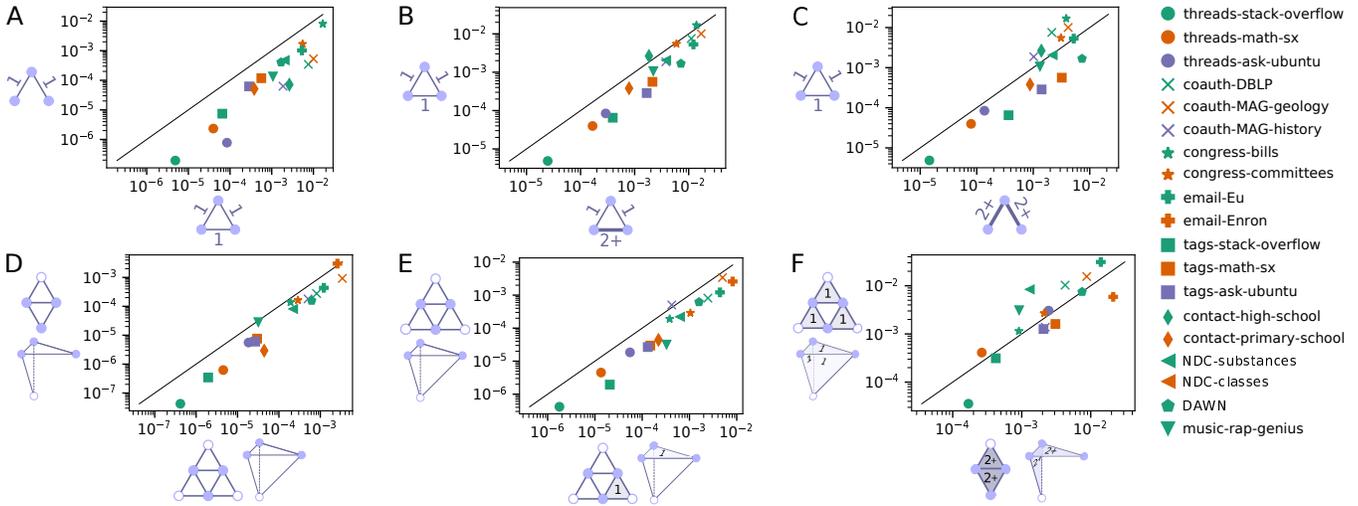}
  \caption{%
    Comparison of simplicial closure event probabilities based on configurations
    of 3-node and 4-node structures. The simplices appearing in the first 80\%
    of the time spanned by the dataset determine the configuration (appearing as
    the x-axis and y-axis labels). The scatter plots compare the probability of
    different configurations going through a simplicial closure event in the
    final 20\% of timestamped simplices.
    \formatsublabel{A--C} Comparison of simplicial closure event probabilities
    for pairs of 3-node configurations that demonstrate how increasing edge
    density (A) or tie strength (B) increases the probability of a simplicial
    closure event. However, the relative importance of edge density and tie strength
    depends on the dataset (C).
    \formatsublabel{D--F} Comparison of simplicial closure event probabilities
    for pairs of 4-node configurations.  Each axis has two labels giving two
    pictorial representations of the configuration.  The white node in the
    ``flat'' representation (left label on x-axis; top label on y-axis)
    represents the same node, so the 3-dimensional structure can be envision by
    folding the white nodes on top of each other. The other representation
    (right label on x-axis; bottom label on y-axis) shows a three-dimensional tetrahedral
    perspective of this folding.  We again see that increasing edge density (D) 
    or tie strength (E) increases the
    probability of a simplicial closure event. Here, ``tie strength'' is
    measured at the level of 3-node simplices, i.e., how often three nodes have
    appeared in a simplex (no times---not shaded; one time---shaded, denoted
    ``1''; or at least two times---shaded, denoted ``2+''). The
    relative importance of edge density and tie strength depends on the
    dataset but is consistent with the 3-node case.
    In three of the five datasets for which the configuration on the
    y-axis in (F) is significantly more likely to go through a simplicial
    closure event, the open triangle of weak ties is also significantly more
    likely to close for sets of three nodes (coauth-DBLP, coauth-MAG-Geology,
    congress-bills; c.f.\ (C); $p<10^{-5}$).  And in three of the four datasets
    for which the configuration on the x-axis in (F) is significantly more
    likely to go through a simplicial closure event the configuration with just
    two strong ties is also more likely to close than the open triangle with all
    weak ties (tags-stack-overflow, tags-math-sx, tags-ask-ubuntu; c.f.\ (C);
    $p<10^{-5}$).  Moreover, there were no datasets for which tie strength was
    significantly more indicative of simplicial closure events for one simplex
    size and density was more important for another (significance level
    $10^{-5}$).
  }
  \label{fig:closure_probs}
\end{figure*}

%%%%%

Regardless of how open triangles are created, the three associated nodes may of
course appear together in a simplex in the future as the network evolves.
Deviating from our above simple model of independent creation of closed
triangles, we find that many newly formed simplices in our data consist
of $k$ nodes that had previously constituted an open $k$-clique in the projected
graph. We say that the appearance of a new simplex containing these $k$ nodes is
an instance of a {\em simplicial closure event}, i.e., the conversion of an open
structure to a closed one, as illustrated in \cref{fig:example_D}.%
\footnote{Here we are building on terminology for datasets of static sets of
simplices~\cite{Patania-2017-collaborations}. The term ``simplicial closure''
also appears in the combinatorial topology literature but with a different
meaning~\cite{Bertrand-2011-completions}.}
In the following, we investigate the simplicial closure mechanism as an
organizational principle for higher-order interactions.

\subsection{Simplicial closure on triangles reveals competing features}
Though conceptually similar, three nodes participating in a simplicial closure
event is distinct from the well-known phenomenon of {\em triadic closure events}
in social networks \cite{Granovetter-1973-strength}. A triadic closure event
modifies the structure of the underlying pairwise interactions, whereas a
simplicial closure event adds a new higher-order interaction without necessarily changing
the pairwise structure of the projected graph.

Any induced subgraph on three nodes in the weighted projected graph can change
several times before the three nodes appear in a simplex together, i.e., go
through a simplicial closure event (\cref{fig:lifecycles}).
We call this the \emph{lifecycle} of the triple of nodes.
There are two changes that a triple of nodes can undergo during its lifecycle before a simplicial closure event. 
First, a new pairwise link can be added between two nodes $u$ and $v$. 
This corresponds to an increase in density in this induced subgraph, e.g., the introduction of the drug Promacta adds an edge in \cref{fig:lifecycles_B}.
Second, the projected graph edge weight between nodes $u$ and $v$ can increase, which we interpret as an increase in tie strength. 
For instance, in \cref{fig:lifecycles_C}, the tie strength between Gucci Mane and Young Thug increases after they collaborate on ``Fell.''
To simplify our analysis, we differentiate only between
\emph{weak ties} corresponding to a single interaction ($W_{uv}=1$ in the projected graph; denoted ``1'')
and \emph{strong ties} corresponding to multiple interactions over time ($W_{uv} \ge 2$; denoted ``2+'').
With this binning, there are 11 possible states in a lifecycle (\cref{fig:lifecycles_A}).

To get a first impression of the magnitude of these events, we examine the
lifecycle of every triple of nodes that becomes an open or closed triangle in
the coauth-MAG-History dataset (\cref{fig:lifecycles_A}). In this dataset, a
closed triangle is more likely to have come from a configuration with exactly
two strong ties edges (3,171 cases) than from an open triangle (328 + 779 + 722
+ 285 = 2,114 cases). Most closed triangles are formed by nodes that had no
previous interaction (2,732,839 cases); however, since the graph is sparse, the
\emph{fraction} of triples of nodes with no prior engagement that go through a
simplicial closure event is small (\cref{sec:closure}). Additionally, if three
nodes induce an open triangle with only weak ties at some point in time, then
the three nodes are more likely to gain a strong tie before closure (445 cases)
than to close directly from that state (328 cases).

We also analyze the probability of a simplicial closure event conditioned on the
state of the three nodes in its lifecycle. To do so, we split each dataset
based on the temporal order of appearance of the simplices into a training set,
consisting of the first 80\% of the simplices (in time) and a test set of the
remaining 20\% of the simplices. Formally, if $t_*$ is the 80th percentile of
the timestamps $t_1,\ldots,t_N$, then the training set is the set of timestamped
simplices
$\{(S_i,t_i) \given t_i \le t_*\}$
and the test set consists of
$\{(S_i,t_i) \given t_i > t_*\}$.
We then measured the probability that a triple of nodes from the training set is
a closed triangle in the test set as a function of its previous
configuration in the weighted projected graph, i.e., its lifecycle state in the
training data (\cref{sec:closure_over_time} contains all of the simplicial closure event probabilities).

We highlight four important findings. First, the simplicial closure event
probability typically increases with additional edges
(\cref{fig:closure_probs_A}). In other words, as the edge density of the
subgraph induced by the three nodes increases, the probability of a simplicial
closure event increases. We formally test this by comparing the closure
probability of a fixed weighted induced subgraph configuration and the same
configuration with an additional unit-weight edge for all suitable cases.  The
hypothesis test is conducted as follows.  Let $n_c$ and $x_c$ denote the number
of instances of an open configuration $c$ in the training set (first 80\% of
data) and the number of those instances that close in the test set (final 20\%
of data). For a pair of configurations $c$ and $c'$, we use a one-sided
hypothesis test for $x_{c} / n_{c} < x_{c'} / n_{c'}$.  We use Fisher's exact
test when $\max(x_{c}, x_{c'}) \le 5$; otherwise, we use a one-sample $z$-test.
The additional unit-weight edge configuration has a statistically significant
larger simplicial closure event probability in 102 of 113 cases over all
datasets and pairs of configurations, whereas the less dense structure is never
significantly more likely to close ($p<10^{-5}$). (Our goal here is to
illustrate general trends rather than to find a single statistically significant
result.)  This result is consistent with both
theoretical~\cite{Granovetter-1973-strength} and
empirical~\cite{Leskovec-2008-microscopic} studies of dyadic link formation in
social networks. However, several of our datasets are not social networks.

Second, the probability of a simplicial closure event typically increases with
tie strength (\cref{fig:closure_probs_B}). We test the effect of tie strength
by comparing the closure probability of a fixed weighted induced subgraph
containing at least one weak tie, and the same configuration where the weak tie
is converted to a strong tie. Increasing the tie strength significantly
increases the probability of a simplicial closure event in 82 of 113 cases over
all datasets and significantly decreases the closure probability in just 6 of
113 cases ($p < 10^{-5}$). Again, this result is consistent with both
theoretical~\cite{Granovetter-1973-strength} and
empirical~\cite{Backstrom-2006-group,Kossinets-2006-empirical} studies of social
networks, even though not all of our networks are social.

Third, neither edge density nor tie strength dominates the likelihood of
simplicial closure events (\cref{fig:closure_probs_C}). In the coauthorship and
Congress datasets, an open triangle comprised of three weak ties is more likely
to close than a 3-node subgraph with just two strong ties. The reverse is true
for the stack exchange tags and stack exchange threads datasets. Overall, the
open triangle of weak ties is significantly more likely to close than the three
nodes with two strong ties in 4 of 19 datasets, whereas the opposite is true in
6 of 19 datasets ($p<10^{-5}$).

Fourth, the results reveal varying closure dynamics over the dataset
domains. In human social interactions, simplicial closure events appear to be
driven by a topological form of triadic closure: mutual acquaintance between all
the nodes in a set increases the probability of a joint interaction. In
contrast, simplicial closure events in the discussion platform networks resemble
transitive closure: once there is a sufficiently strong co-occurrences of tags,
they become likely to be used together.

A possible concern with our analysis is that we only measured closure
probabilities at one point in time for each dataset. Furthermore, while some of
our datasets represent a complete history of the network (tags, threads, NDC)
and some span a long duration of time (coauthorship, music, congress-bills), a
few only contain a slice of the underlying network's dynamics
(email-Eu, contact). However, we find that the closure probabilities and the
results on edge density and tie strength are consistent at different points in
time (\cref{sec:closure_over_time}).

\subsection{Simplicial closure properties extend beyond triangles}
All four of the above findings hold for simplicial closure events on four nodes,
so our results are not limited to structure on three nodes
(\cref{fig:closure_probs_D,fig:closure_probs_E,fig:closure_probs_F}).
Now, a simplicial closure event is all four nodes appearing in a simplex,
and ``tie strength'' is measured on 3-node simplices, i.e., how often
the 3-node subsets of a 4-node structure have appeared together in a simplex (0, or
 ``open''; 1, or ``weak'';  at least 2 times, or ``strong'').

To measure the effect of edge density, we compare the closure probability of a
configuration consisting of a fixed number of edges to the closure probability
of the same configuration with an additional edge, keeping the tie strengths 
fixed (\cref{fig:closure_probs_D} shows one such comparison). In
180 of 228 applicable comparisons over all datasets, the closure probability
significantly increases with the edge density and significantly decreases
in only 2 cases ($p<10^{-5}$). To measure the effect of tie
strength, we compare the closure probability of a given configuration to the
closure probability of the same configuration where the tie strength
increases from an open tie to a weak tie or from a weak tie to a
strong tie (\cref{fig:closure_probs_E} shows a case where the tie strength increases
from open to weak). The closure probability significantly
increases with simplicial tie strength in 26 of 38 cases for 3-edge
configurations, 31 of 38 cases for 4-edge configurations, 77 of 114 cases for
5-edge configurations, and 177 of 359 cases for 6-edge configurations; compared
to a significant decrease in closure probability in just 2 of 38, 1 of 38, 1 of
114, and 4 of 359 cases ($p<10^{-5}$). Therefore, tie strength
is also a positive indicator of simplicial closure in 4-node configurations.

There is also tension between the influence
of sparser configurations with strong ties and denser configurations with weak
ties. \Cref{fig:closure_probs_F} shows one such comparison. In this case, three out of five
datasets for which edge density is significantly more indicative than tie
strength in the 3-node comparison of \cref{fig:closure_probs_C}, edge density is also
significantly more important in the 4-node case ($p<10^{-5}$). And in three of
the four datasets for which tie strength is significantly more indicative than
edge density in the same 3-node case, the same is true in the 4-node case. 
Finally, there is no dataset for which tie strength was significantly more
influential for one simplex size and density was significantly more
influential for another.

%!TEX root = higher-order-link-prediction.tex

\section{Higher-order link prediction}\label{sec:prediction}
Thus far, we have showed that higher-order interactions provide a rich source of
additional information beyond traditional network modeling. Our analysis leaves open many
questions, such as the development of better mechanistic models
for the emergence of these interactions. To facilitate this process, we
propose an analog of link prediction for higher-order structure. 

%%%%%
% tab:prediction_performance
%!TEX root = higher-order-link-prediction-postprint.tex

\begin{table*}[tb]
\setlength{\tabcolsep}{3pt}
\centering
\caption{
  Open triangle closure prediction performance based on eight models:
  harmonic, geometric, and arithmetic means of the 3 edge weights;
  3-way Adamic-Adar coefficient;
  preferential attachment;
  Katz similarity;
  personalized PageRank similarity (PPR); and
  a feature-based supervised logistic regression model.
  Performance is AUC-PR relative to the random baseline, i.e., relative to the
  faction of open triangles that close.
  The top performance number for each dataset is bolded.
}
\begin{tabular}{l c @{\quad} c c c c c c c c c c c c c c c c}
\toprule
Dataset                & Harm. mean & Geom. mean & Arith. mean & Adamic-Adar   & Pref.\ Attach  & Katz similarity  & PPR  & Logistic Regression   \\
\midrule
coauth-DBLP            & 1.49  & 1.59  & 1.50   & 1.60  & 0.74 & 1.51  & 1.83 & \textbf{3.37}   \\
coauth-MAG-History     & 1.69  & 2.72  & 3.20   & 5.82  & 2.49 & 3.40  & 1.88 & \textbf{6.75}   \\
coauth-MAG-Geology     & 2.01  & 1.97  & 1.69   & 2.71  & 0.97 & 1.74  & 1.26 & \textbf{4.74}   \\
music-rap-genius       & 5.44  & \textbf{6.92}  & 1.98   & 2.10  & 2.15 & 2.00  & 2.09 & 2.67   \\
tags-stack-overflow    & \textbf{13.08} & 10.42 & 3.97   & 6.63  & 2.74 & 3.60  & 1.85 & 3.37   \\
tags-math-sx           & 9.08  & 8.67  & 2.88   & 6.34  & 2.81 & 2.71  & 1.55 & \textbf{13.99}  \\
tags-ask-ubuntu        & 12.29 & \textbf{12.64} & 4.24   & 7.51  & 5.63 & 4.15  & 2.54 & 7.48   \\
threads-stack-overflow & 23.85 & \textbf{31.12} & 12.97  & 3.19  & 3.89 & 11.54 & 4.06 & 1.53   \\
threads-math-sx        & 20.86 & 16.01 & 5.03   & 23.32 & 7.46 & 4.86  & 1.18 & \textbf{47.18}  \\
threads-ask-ubuntu     & 78.12 & \textbf{80.94} & 29.00  & 30.82 & 6.62 & 32.31 & 1.51 & 9.82   \\
NDC-substances         & 4.90  & 5.27  & 2.90   & 5.97  & 4.46 & 2.93  & 1.83 & \textbf{8.17}   \\
NDC-classes            & \textbf{4.43}  & 3.38  & 1.82   & 0.99  & 2.14 & 1.34  & 0.91 & 0.62   \\
DAWN                   & 4.43  & 3.86  & 2.13   & \textbf{4.77}  & 1.45 & 2.04  & 1.37 & 2.86   \\
congress-committees    & 3.59  & 3.28  & 2.48   & 5.04  & 1.31 & 2.59  & 3.89 & \textbf{7.67}   \\
congress-bills         & 0.93  & 0.90  & 0.88   & 0.66  & 0.55 & 0.78  & 1.07 & \textbf{107.19} \\
email-Enron            & 1.78  & 1.62  & 1.33   & 0.87  & 0.83 & 1.28  & \textbf{3.16} & 0.72   \\ 
email-Eu               & 1.98  & 2.15  & 1.78   & 1.37  & 1.55 & 1.79  & 1.75 & \textbf{3.47}   \\
contact-high-school    & 3.86  & \textbf{4.16}  & 2.54   & 2.00  & 1.13 & 2.53  & 2.41 & 2.86   \\
contact-primary-school & 5.63  & 6.40  & 3.96   & 3.21  & 0.94 & 4.02  & 4.31 & \textbf{6.91}   \\
\bottomrule
\end{tabular}
\label{tab:prediction_performance}
\end{table*}

%%%%%

\subsection{Model evaluation framework}
The basic premise in link prediction---whether pairwise or higher-order---is to
use structural network properties up to some time $t$ to predict the appearance of
new interactions after $t$. In traditional network analysis, link prediction is
a cornerstone problem and a highly successful evaluation framework for comparing
different models via a well-calibrated prediction
task~\cite{LibenNowell-2007-link-prediction,Lu-2011-linkpred-survey}.
Specifically, link prediction examines data that evolves over time and sees how
well a given model predicts the appearance of new links---for example, new
coauthorships appearing in a coauthor network, or new messages between pairs of
people in an email network.

In this context, a \emph{model} is interpreted broadly and may be
mechanistic (e.g., preferential attachment~\cite{Barabasi-2002-evolution}),
statistical (e.g., probabilistic hierarchical
models~\cite{Clauset-2008-hierarchical}), or implicitly encapsulated by a
principled heuristic algorithm. For instance, personalized
PageRank is a model capturing the fact
that a large number of walks between two nodes drives up the connection probability between
them~\cite{LibenNowell-2007-link-prediction}. A key advantage of link prediction 
as an evaluation framework is precisely that it can handle these various kinds of models.
This holds even in the absence of a likelihood expression,
which would be required for a more standard statistical evaluation of goodness
of fit. While ultimately we may want to arrive at a generative, causal
description of the emergence of higher-order patterns, the flexibility of
link prediction enables us to probe the importance of features of the network
data in a simple manner without having to create a formal statistical model.

Link prediction has proved valuable both for methodological reasons and also in
concrete applications. Methodologically, asking whether one model is
better than another at predicting new links provides a data-driven
way of assessing the effectiveness of the
models~\cite{LibenNowell-2007-link-prediction,Grover-2016-node2vec,Santolini-2018-predicting}. 
Link prediction also has a number of direct applications that cut across disciplines,
including predicting friendships in social
networks~\cite{Backstrom-2011-supervised}, inferring new relationships between
genes and diseases~\cite{Wang-2011-disease}, and suggesting novel connections in
the scientific community~\cite{Tang-2012-collaboration}.

Link prediction is also used within model selection tools for evaluating community detection
algorithms~\cite{ghasemian2018evaluating,kawamoto2017cross}. In these cases,
link prediction may be interpreted as the smallest possible test for the fit of
a model as we need to predict only one edge at a time. However, if one were to
consider all edges in a cross-validation assessment, good link prediction
performance indicates a good model fit for other structure in the data. Our
higher-order link prediction task probes a larger set of features, in that it
requires us to be able to predict more aspects of the data (any higher-order
interaction, in principle).

For simplicity of presentation and scalability reasons, we predict simplicial
closure events on triples of nodes. Thus, the higher-order link prediction
problem examined here is predicting which triples of nodes that have not yet
appeared in a simplex together will be a subset of some simplex in the
future. Our above analysis suggests that open triangles or triples of nodes with
strong ties are the most likely to close in the future. For our experiments, we
predict which open triangles will go through a simplicial closure event in the
future. Thus, this is a problem completely ignored by traditional link
prediction, which would just view the triangle as already part of the
graph. From a computational view, this restriction also makes it feasible to
enumerate all open structures upon which the algorithms will make a prediction,
using only modest computational resources. Thus, we avoid a common problem in
link prediction of how to pare down an enormous candidate set of potential
links, which itself is an active research
topic~\cite{Ballard-2015-diamond,Sharma-2017-hashes}.

\subsection{Simple local features predict well}\label{sec:pred_perf}
We first split the data into training (first 80\% of simplices in time) and test
(final 20\%) sets. Then, we evaluated the prediction performance of several new
models (several inspired from classical link-prediction) on each dataset by the
area under the precision-recall curve (AUC-PR) metric
(\cref{tab:prediction_performance}).
AUC-PR is appropriate for prediction problems with class imbalance~\cite{Davis-2006-pr-roc}, which is the case for our datasets. 
We use random scores as a baseline, which, with respect to AUC-PR, corresponds
to the proportion of open triangles in the training set that go through a
simplicial closure event in the test set.

We compare eight models here and provide additional comparisons in \cref{sec:all_score_funcs}.
Three are heuristics based on our finding that tie strength is indicative of
closure; these are the harmonic, geometric, and arithmetic means of the three
edge weights in the open triangle.  Two more are based on the Adamic-Adar
model~\cite{Adamic-2003-friends} and the preferential attachment model.
The latter has been suggested as a growth mechanism of coauthorship
networks~\cite{Newman-2001-clustering,Barabasi-2002-evolution}.  Two are based
on longer path counts (Katz and personalized PageRank), which are models known
for providing good prediction in dyadic link
prediction~\cite{LibenNowell-2007-link-prediction}. Lastly, we use a supervised
logistic regression model based on features from the other models.

No single model performs the best over all
datasets, but our proposed baseline algorithms can achieve much better performance than randomly guessing
which open triangles go through a simplicial closure event.
In the threads datasets, we achieve between one and two orders of magnitude
performance improvements with the harmonic and geometric means, which indicates that local
tie strength is relatively more important for these datasets than others.
The absolute performance of the algorithms
is far from perfect (see \cref{sec:all_score_funcs}), as the higher-order link
prediction is challenging. This finding is consistent with recent
research on subgraph prediction in pairwise networks~\cite{meng2018subgraph}.
However, our goal here is to identify some of the important structural features
of the problem, rather than to predict with perfect accuracy.

The harmonic and geometric means of edge weights perform well across many
datasets, which further highlights the importance of tie strength in predicting
simplicial closure events. This finding is fundamentally different from
traditional link prediction with pairwise interactions (i.e., for the edges in a
graph). In traditional link prediction, a key principle is that it
is valuable to use information contained in paths of non-trivial length between
two nodes $u$ and $v$ for predicting a link between them---for example,
PageRank and Katz measures are effective~\cite{LibenNowell-2007-link-prediction,Lu-2011-linkpred-survey}.
In this sense, higher-order link prediction is fundamentally more local in its
overall structure. This arises from the ability of a $k$-tuple of nodes, for $k \geq 3$,
to contain rich local information in its interactions among subsets of
size $k-1$, a phenomenon that has no natural analogue when $k = 2$.

%%%%%
% fig:genmeans
%!TEX root = higher-order-link-prediction-postprint.tex

\begin{figure}[tb]
  \centering
  \includegraphics[width=\columnwidth]{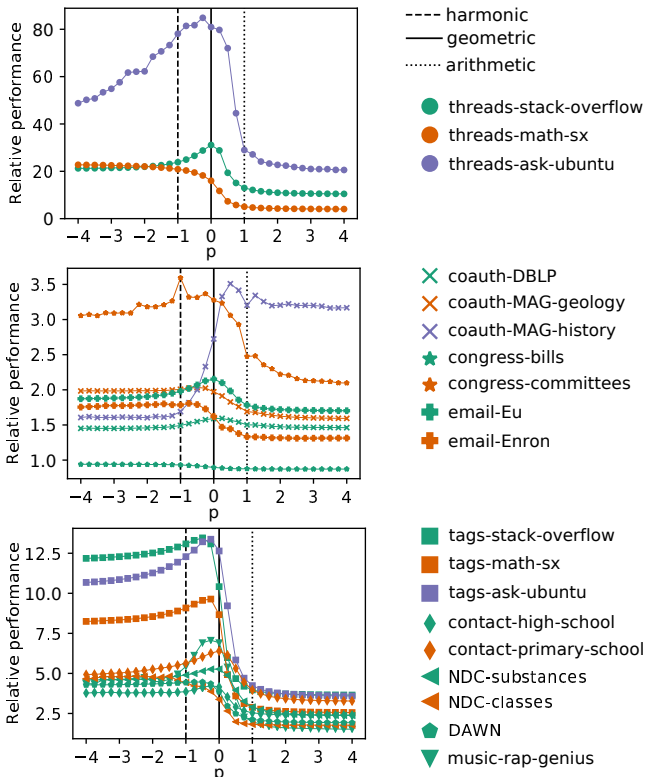}
  \caption{%
    AUC-PR relative to random predictions as a function of the parameter $p$ in
    the generalized mean heuristic model for higher-order link prediction.
  }
  \label{fig:genmeans}
\end{figure}

%%%%%

The arithmetic mean performs the worst of the three means in all but one dataset.
We further analyze the performance of edge weight means
using the generalized mean with parameter $p$ as score functions:
$s_p(u, v, w) = [(W_{uv}^p + W_{uw}^p + W_{vw}^p) / 3]^{1/p}$,
where $W_{ab}$ is the weight between nodes $a$ and $b$ in the projected graph.
The harmonic, arithmetic, and geometric means are the special cases where $p = -1$, $p = 1$, and the limit $p \to 0$.
Generally, prediction performance is
(i) unimodal in $p$,
(ii) maximized for $p \in [-1, 0]$, and
(iii) better for $p < -1$ than for $p > 1$ (\cref{fig:genmeans}).
Two exceptions are NDC-classes and coauth-MAG-History.
The former is the only dataset without an open triangle
with exactly one strong tie to close.
Thus, smaller $p$ should perform better, as this accounts more for the minimum edge weight value.
The latter is the dataset with the smallest average degree in the projected graph (\cref{fig:structure_C}). 
Therefore, a single strong edge could provide the signal for closure, in which
case a larger $p$ is a better score function.

The supervised learning approach also performs well broadly, especially in the
larger datasets such as the coauthorship datasets, which have sufficient
training data to learn a good model. However, even when including the
features of the other models, the method does not always
perform the best. This is likely a case of overfitting.
In the case of the Congress bills data, the supervised method captures a
unique feature of this dataset---nodes appearing in fewer simplices are
\emph{more} likely to go through a simplicial closure event. This is possibly
due to the ambition of junior Congresspersons. The fact that combinations of
features prove effective in many domains highlights the richness of the
underlying problem, and the array of methods and findings presented here can
guide progress on better models.

%!TEX root = higher-order-link-prediction-postprint.tex

\section{Discussion}

The dyadic network modeling paradigm has been successful but fails to capture
natural higher-order interactions. Here, we established the foundation for
analyzing the basic structure of temporal networks with higher-order structure.
We found rich structural variety in our datasets in terms of the fraction of
triangles that are open, the average degree, and the edge density. Local
statistics at the level of egonets can identify system domain, which suggests
that these features are key to the organizing principles of the systems.  Recent
research shows the small fraction of triangles that are open in coauthorship
networks~\cite{Patania-2017-collaborations}; our results are consistent but
reveal that open triangles are extremely common in other domains. Prior research
has also identified the distinction between open and closed triangles when
projecting bipartite networks but have not studied the idea
of simplicial closure events~\cite{newman-affiliation,Opsahl-2013-closure-two-mode}.

We found that common principles from dyadic network evolution also hold for
higher-order structure, namely, tie strength and edge density are positive
indicators of simplicial closure events amongst sets of three and four
nodes. However, there is tension between these features---the more influential
feature depends on the dataset, suggesting different mechanisms for simplicial
closure events. For example, edge density matters more in human interaction, but
tie strength matters more for tagging on online discussion platforms.

Higher-order link prediction provides a general methodology
for evaluating models in any data where higher-order structure
evolves over time, such as predicting which sets of authors will write a
paper together or which sets of people will appear as joint recipients on an
email. We anticipate that higher-order link prediction will validate emerging
higher-order network modeling techniques, such as multipartite
networks~\cite{Lind-2007-new}, meta paths~\cite{Sun-2012-when},
embeddings~\cite{Goyal-2017-embedding}, and connect to ideas in computational
topology, such as random walks on simplicial
complexes~\cite{Mukherjee-2016-walks,Parzanchevski-2016-simplicial}.
Related higher-order models for different
data~\cite{Rosvall-2014-memory,Xu-2016-higher-order} can also use higher-order link
prediction for model evaluation. For example, in the absence of temporal
information, higher-order link prediction could be used to find missing data,
similar to how dyadic link prediction can find missing data in static
networks~\cite{Clauset-2008-hierarchical}.
Our higher-order link prediction framework also provides a way to study more
sophisticated models where the underlying network is also dynamic, e.g., with
arrival and departure of nodes. Specifically, such models should be able to
predict higher-order links.

Our prediction problem examined a structure that is not even
considered in traditional network analysis, where no distinction is made between
open and closed triangles. From this setup, we found that simple local measures
(generalized means of edge weights) are effective predictors. This finding differs
from traditional link prediction, where long paths are
important~\cite{LibenNowell-2007-link-prediction} and suggests that the temporal
evolution of higher-order network data is fundamentally different than dyadic
network evolution.

\xhdr{Data and software availability}
All datasets except the Congress and music ones are available at \url{http://www.cs.cornell.edu/~arb/data/}.
Software is available at \url{https://github.com/arbenson/ScHoLP-Tutorial}.

\subsection*{Acknowledgements}
We thank Mason Porter and Peter Mucha for providing the Congress committees
dataset. We thank Paul Horn, Gabor Lippner, and Jaros\l{}aw B\l{}asiok for
helpful discussion. This research was supported in part by a Simons
Investigator Award. 
ARB was supported in part by NSF Award DMS-1830274.
RA was supported in part by a Google scholarship and a
Facebook scholarship. AJ received funding from the Vannevar Bush Fellowship from
the office of the Secretary of Defense. MTS received funding from the European
Union's Horizon 2020 research and innovation programme under the Marie
Sk\l{}odowska-Curie grant agreement No 702410.

\bibliographystyle{ACM-Reference-Format}
\bibliography{refs-full}

\appendix

\colorlet{TufteRed}{red!80!black}
\colorlet{MyEdgeColor}{gray!80!blue}
\colorlet{MyNodeColor}{blue!30!white}
\renewcommand*{\VertexInterMinSize}{16pt}
\renewcommand*{\VertexSmallMinSize}{16pt}
\renewcommand*{\VertexLineColor}{MyNodeColor}
\renewcommand*{\VertexLightFillColor}{MyNodeColor}
\newcommand{\edgewidth}{0.75}

\newcommand{\ax}{10}
\newcommand{\ay}{10}
\newcommand{\bx}{8}
\newcommand{\by}{11}
\newcommand{\cx}{10.5}
\newcommand{\cy}{12}
\newcommand{\dx}{8.75}
\newcommand{\dy}{13.25}
\newcommand{\ex}{12}
\newcommand{\ey}{10.75}
\newcommand{\fx}{8.25}
\newcommand{\fy}{9.25}
\newcommand{\gx}{10}
\newcommand{\gy}{8.5}
\newcommand{\hx}{11.75}
\newcommand{\hy}{9.25}
\newcommand{\ix}{11.75}
\newcommand{\iy}{13}

\clearpage
%!TEX root = higher-order-link-prediction-postprint.tex

\section{Dataset collection and construction}\label{sec:SI_data}
Here we provide a more complete description of the datasets.

\xhdr{Coauthorship} In these datasets, the nodes correspond to authors,
and each simplex represents the authors on a scientific publication. The
timestamp is the year of publication. We analyze three coauthorship
networks---one derived from DBLP, an online bibliography for computer science,
and two derived from the Microsoft Academic Graph (MAG). We used the September 3, 2017
release of DBLP\footnote{\url{http://dblp.org/xml/release/}}) and the
MAG version released with the Open Academic Graph\footnote{\url{https://www.openacademic.ai/oag/}}~\cite{Sinha-2015-overview}.
We constructed two field-specific datasets by filtering the MAG data according to
keywords in the ``field of study'' information. One dataset consisted of all
papers with ``History'' as a field of study and the other all papers
with ``Geology'' as a field of study.

\xhdr{Stack Exchange tags} Stack Exchange is a collection of
question-and-answer web sites.\footnote{\url{https://stackexchange.com}} Users
post questions and may annotate each question with up to 5 tags that specify
topic areas spanned by the question. We derive tag networks where nodes
correspond to tags and each simplex represents the tags on a question. The
timestamp for a simplex is the time that the question was posted on the web
site. We derived three datasets corresponding to three stack exchange web sites:
Stack Overflow,\footnote{\url{https://stackoverflow.com}}
Mathematics Stack Exchange,\footnote{\url{https://math.stackexchange.com}} and
Ask Ubuntu.\footnote{\url{https://askubuntu.com}}
The raw data was downloaded from the Stack Exchange
data dump,\footnote{\url{https://archive.org/details/stackexchange}; downloaded September 20, 2017.}
which contains the complete history of the content on the stack exchange web sites.

\xhdr{Stack Exchange threads} We also formed user interaction datasets
from the stack exchange web sites. Users post answers to questions, creating a
question-and-answer ``thread.'' The nodes are users and simplices correspond to
the users asking a question or posting an answer on a single thread. We only
considered threads where the question and all answers were posted within 24
hours. The timestamps of the simplices are the times that the question was
posted.

\xhdr{National Drug Code Directory (NDC)} Under the Drug Listing Act of
1972, the U.S. Food and Drug Administration releases information on all
commercial drugs going through the regulation of the agency. We constructed two
datasets from this data where simplices correspond to drugs. In one, the nodes
are classification labels (e.g., serotonin reuptake inhibitor), and simplices
are comprised of all labels applied to a drug; in the other, the nodes are
substances (e.g., testosterone) and simplices are constructed from all
substances in a drug. In both derived datasets, the timestamps are the days when
the drugs were first marketed.

\xhdr{United States Congress} We derived two datasets from political
networks, where the nodes are congresspersons in the U.S. congress. In the
first, simplices represent all members of committees and sub-committees in the
House of Representatives (Congresses 101 to 107, from 1989 to 2003), and the
timestamp of the simplex is the year that the committee
formed~\cite{Porter-2005-committees,Porter-2007-committees}. In the second
dataset, simplices are comprised of the sponsor and co-sponsors of legislative
bills put forth in the House of Representatives and the
Senate~\cite{Fowler-2006-cosponsorship,Fowler-2006-connecting}, and the
timestamps are the days that the bills were introduced.

\xhdr{Email} In email communication, messages can be sent to multiple
recipients. We analyze two email datasets---one from communication between
Enron employees~\cite{Klimt-2004-Enron} and the other from a European research
institution~\cite{Paranjape-2017-motifs}. In both datasets, nodes are email
addresses. In the Enron dataset, a simplex consists of the sender and all
recipients of an email. The data source for the European research institution
only contains (sender, receiver, timestamp) tuples, where timestamps are
recorded at 1-second resolution~\cite{Paranjape-2017-motifs}. Simplices consist
of a sender and all receivers such that the email between the two has the same
timestamp.

\xhdr{Human contact} The human contact networks are constructed from
interactions recorded by wearable sensors in a high
school~\cite{Mastrandrea-2015-contact} and a primary
school~\cite{Stehle-2011-contact}. The sensors record proximity-based contacts
every 20 seconds. We construct a graph for each interval, where nodes $i$ and
$j$ are connected if they are in contact during the interval. Simplices are all
maximal cliques in the graph at each time interval.

\xhdr{DAWN} The Drug Abuse Warning Network (DAWN) is a national health
surveillance system that records drug use contributing to hospital emergency
department visits throughout the United States. Simplices in our dataset are the
drugs used by a patient (as reported by the patient) in an emergency department
visit. The drugs include illicit substances, prescription and over-the-counter
medication, and dietary supplements. Timestamps of visits are recorded at the
resolution of quarter-years, spanning a total duration of 8 years. For a period
of time, the recording system only recorded the first 16 drugs reported by a
patient, so we only use (at most) the first 16 drugs reported by a patient for
the entire dataset.

\xhdr{Music collaboration} Musical artists often collaborate on
individual songs. We derive a dataset where nodes are artists and a simplex
consists of all artists collaborating on a song. The songs were obtained from a
web crawl of the music lyrics web site Genius.\footnote{\url{https://genius.com/}}
We consider the collaborating artists to be the lead artist along with any ``featured''
artists (this excludes some cases where lyrics from an artist are included but
that artist is not listed as a featured artist). The timestamps are the release
dates of the songs. We only collected data from songs that contain the ``rap''
tag on the web site and discarded songs without a specified release date. The
crawler ran for several days and collected over 500,000 songs.

\clearpage
%!TEX root = higher-order-link-prediction-postprint.tex

\section{Temporal asynchrony and open triangles}\label{sec:temp_async}
Our datasets contain temporal dynamics, so edges may only be ``active'' for
certain periods in the total time spanned by the dataset. This provides one
plausible explanation for the existence of open triangles. For example, in
coauthorship networks, an open triangle may arise when three separate
collaborations occurred in disjoint time periods. To investigate the importance
of such effects, we analyze the temporal asynchrony in open triangles in our
datasets. Let the ``active interval'' of an edge in the projected graph
be the interval bounded by the earliest and latest timestamps of simplices
containing the two nodes in the edge. Recall that our datasets are defined by a
collection of timestamped simplices $\{(S_i, t_i)\}$, where each $S_i \subset V$
is the simplex and each $t_i \in \mathbb{R}$ is a timestamp. The active
interval of an edge $(u, v)$ is then
\begin{equation}\label{eqn:active_interval}
  I_{u, v} = [\min \{ t_i \given u, v \in S_i \},\; \max \{ t_i \given u, v \in S_i \}].
\end{equation}

For each open triangle in each dataset, we compute the number of pairwise
overlapping active intervals amongst the three edges in the triangle
(\cref{tab:temp_async}). In the majority of cases, all three pairs of intervals
overlap. By Helly's theorem, there is an interval of time for which all three
edges in the open triangle are simultaneously active. Stated differently, in the
coauthorship example, the collaborators could have theoretically formed a closed
triangle during this time period, but they did not. We conclude that temporal
asynchrony is not a major reason for the presence of open triangles in our
datasets.

%!TEX root = higher-order-link-prediction-postprint.tex

\begin{table}[tb]
  \setlength{\tabcolsep}{2pt}
  \centering
  \caption{Temporal asynchrony and open triangles. For each open triangle in
    each dataset, we find the number of overlaps between the active intervals of
    the three edges, where an active interval of an edge has end points given by
    the earliest and latest timestamps of simplices containing the two nodes in
    the edge (\cref{eqn:active_interval}). The edges in most open triangles
    have three pairwise overlapping intervals. In these cases, there is a time
    period where all three edges were simultaneously active by Helly's theorem.}
  \begin{tabular}{l c c c c c}
    \toprule
    & &  \multicolumn{4}{c}{\# overlaps} \\
    \cmidrule(lr){3-6}
    Dataset                & \# open triangles & 0     & 1     & 2     & 3     \\
    \midrule
    coauth-DBLP            & 1,295,214         & 0.012 & 0.143 & 0.123 & 0.722 \\
    coauth-MAG-history     & 96,420            & 0.002 & 0.055 & 0.059 & 0.884 \\
    coauth-MAG-geology     & 2,494,960         & 0.010 & 0.128 & 0.109 & 0.753 \\
    tags-stack-overflow    & 300,646,440       & 0.002 & 0.067 & 0.071 & 0.860 \\
    tags-math-sx           & 2,666,353         & 0.001 & 0.040 & 0.049 & 0.910 \\
    tags-ask-ubuntu        & 3,288,058         & 0.002 & 0.088 & 0.085 & 0.825 \\
    threads-stack-overflow & 99,027,304        & 0.001 & 0.034 & 0.037 & 0.929 \\
    threads-math-sx        & 11,294,665        & 0.001 & 0.038 & 0.039 & 0.922 \\
    threads-ask-ubuntu     & 136,374           & 0.000 & 0.020 & 0.023 & 0.957 \\
    NDC-substances         & 1,136,357         & 0.020 & 0.196 & 0.151 & 0.633 \\
    NDC-classes            & 9,064             & 0.022 & 0.191 & 0.136 & 0.652 \\
    DAWN                   & 5,682,552         & 0.027 & 0.216 & 0.155 & 0.602 \\
    congress-committees    & 190,054           & 0.001 & 0.046 & 0.058 & 0.895 \\
    congress-bills         & 44,857,465        & 0.003 & 0.063 & 0.113 & 0.821 \\
    email-Enron            & 3,317             & 0.008 & 0.130 & 0.151 & 0.711 \\
    email-Eu               & 234,600           & 0.010 & 0.131 & 0.132 & 0.727 \\
    contact-high-school    & 31,850            & 0.000 & 0.015 & 0.019 & 0.966 \\
    contact-primary-school & 98,621            & 0.000 & 0.012 & 0.014 & 0.974 \\
    music-rap-genius       & 70,057            & 0.028 & 0.221 & 0.141 & 0.611 \\
    \bottomrule
  \end{tabular}
  \label{tab:temp_async}
  \end{table}

\clearpage
%!TEX root = higher-order-link-prediction-postprint.tex

\section{Simplicial closure probabilities}\label{sec:closure}

We now present simplicial closure event probabilities on three and four
nodes. The setup is the same as in \cref{sec:simp_closure}. We split the data
into training and test sets, corresponding to the first 80\% and final 20\% of
time-ordered simplices, respectively. Given the configuration of a set of three
or four nodes in the training data, we measure the probability that the nodes go
through a simplicial closure event in the final 20\% of the data.

\begin{figure*}[tb]
  \centering
  \phantomsubfigure{fig:closure_probs_higher_A}
  \phantomsubfigure{fig:closure_probs_higher_B}
  \phantomsubfigure{fig:closure_probs_higher_C}
  \phantomsubfigure{fig:closure_probs_higher_D}
  \includegraphics[width=\linewidth]{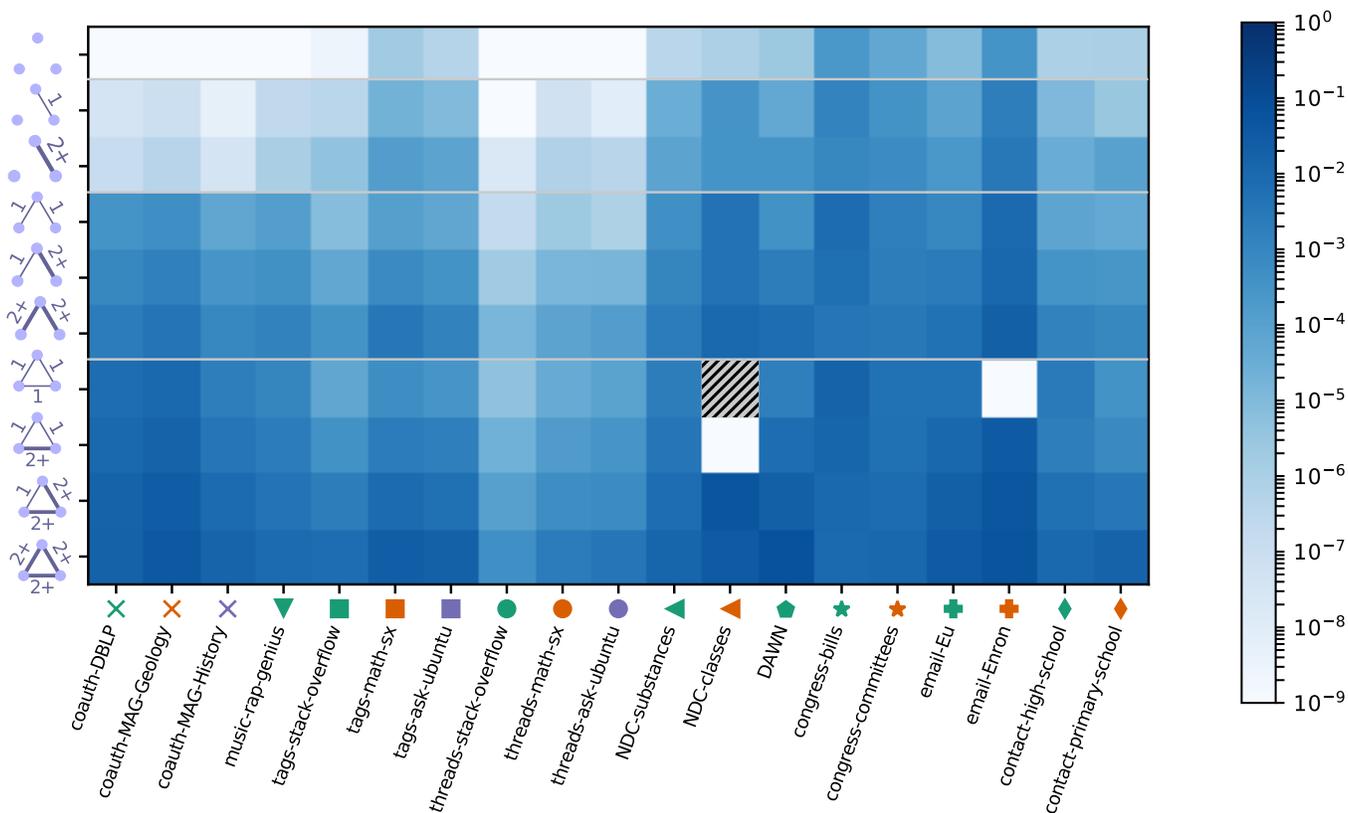}
  \caption{%
    Heat map of the probabilities of simplicial closure events as a function of
    the 3-node open configuration. We use the first 80\% of the timestamped data
    to determine the configuration of every 3-node set and compute the
    probability that the set appears in a simplex in the final 20\% of the data,
    conditioned on the open configuration. Shaded boxes are configurations that
    appear 20 or fewer times in the first 80\% of the data. The four sections
    of the heat map correspond to 0, 1, 2, or 3 edges in the induced subgraph.
  }
  \label{fig:closure_probs3}
\end{figure*}

In the case of $3$-node simplicial closure events, we determine the open
configuration in the training set by examining the three nodes in the 
projected graph. Effectively, we examined how many times each of the three
\emph{$2$-node subsets} co-appeared in a simplex. 
\Cref{fig:closure_probs3} shows a heat map of the closure probabilities
as a function of the open 3-node configuration.

For $4$-node open configurations we proceed analogously, using the corresponding 
\emph{$3$-node subsets}. Specifically, for a given set of 4 nodes, every triangle in the
projected graph is classified as either (i) an open simplicial tie, i.e., the
triangle is open; (ii) a weak simplicial tie, meaning that the three nodes have
appeared in just one simplex together; or (iii) a strong simplicial tie, meaning
that the three nodes have appeared in at least two simplices together. In
contrast to the $3$-node case, these $4$-node configurations are not completely
determined by the weighted projected graph, since the projected graph (as
defined) does not contain information on whether or not 3 nodes induce a closed
triangle. Thus, with $4$-node simplices, we make use of additional topological
information provided in our set-valued datasets.

\begin{figure*}[tb]
  \centering
  \phantomsubfigure{fig:closure_probs_higher_A}
  \phantomsubfigure{fig:closure_probs_higher_B}
  \phantomsubfigure{fig:closure_probs_higher_C}
  \phantomsubfigure{fig:closure_probs_higher_D}
  \includegraphics[width=\linewidth]{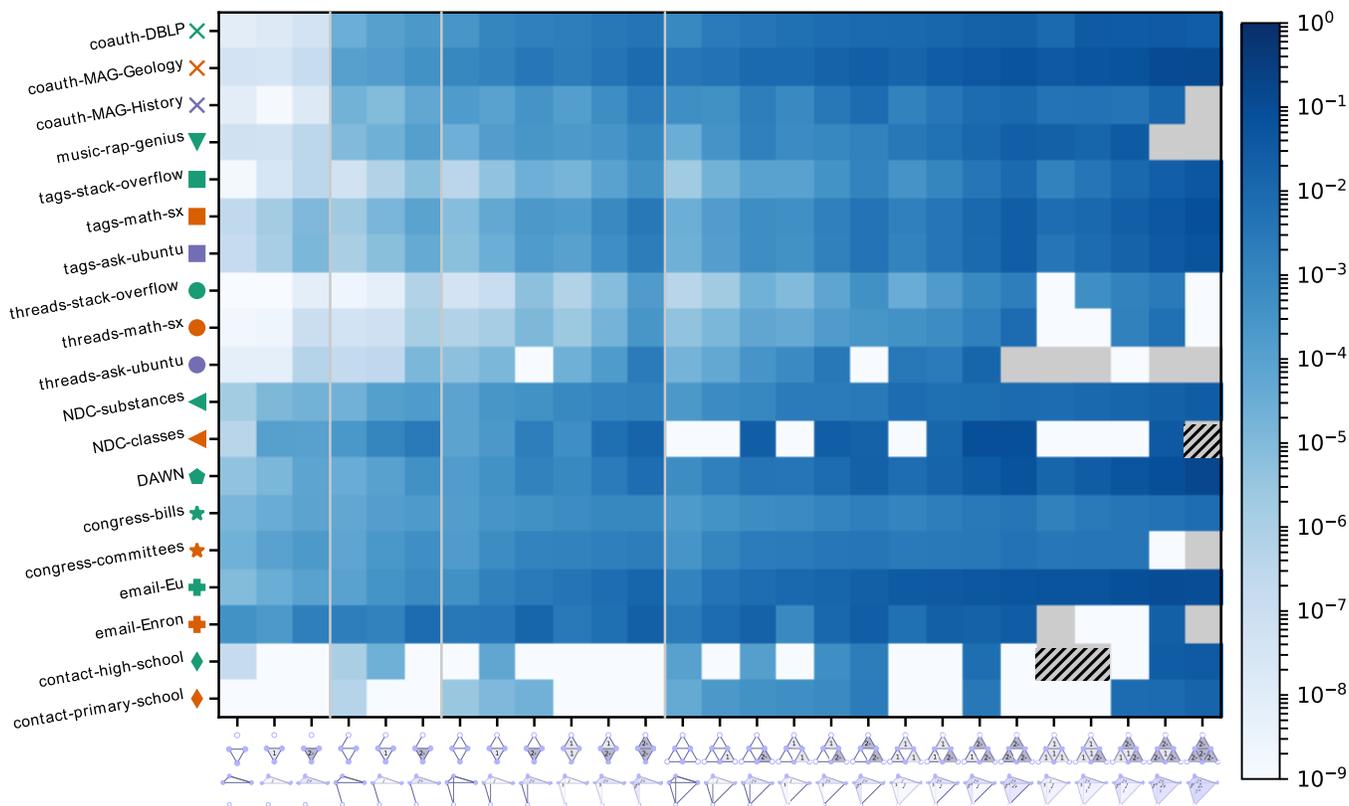}
  \caption{%
    Heat map of the probabilities of simplicial closure events as a function of
    the 4-node open configuration. We use the first 80\% of the timestamped data
    to determine the configuration of every 4-node set that contains at least 1
    triangle and does not appear in a simplex. We then compute the probability
    that a 4-node set appears in a simplex in the final 20\% of the data,
    conditioned on the open configuration. In an open configuration, there are
    three types of simplicial tie strengths for a triangle---open, weak, and
    strong---given by the number of times the three nodes in the triangle have
    co-appeared in a simplex (zero, one, or at least two times). Shaded boxes
    are configurations that appear 20 or fewer times in the first 80\% of the
    data. We illustrate each subgraph configuration on the x-axis with a
    projection of the simplex onto two dimensions (top line---the unfilled
    circle represents the same node) as well as a tetrahedral three-dimensional
    perspective figure (bottom line). The four sections of the heat map
    correspond to 3, 4, 5, or 6 edges in the configuration.
  }
  \label{fig:closure_probs4}
\end{figure*}

One could also account for the tie strengths of the $2$-node subsets (edges) in
a $4$-node configuration---a complete characterization of the possible induced
configurations leading to a simplicial closure event is much more complex for
the $4$-node case than the $3$-node case. For an accessible study on the
closure patterns of $4$-node simplices, we measure the probability of closure
with respect to the 27 open configurations where the induced $4$-node subgraph
in the projected graph contains at least one triangle. Our classification
distinguishes triangles by tie strength (open, weak, or closed), but not by edge
tie strengths, other than by what is implied by the triangles.
\Cref{fig:closure_probs4} shows a heat map of the closure probabilities
as a function of the open 4-node configuration.

%%%%%
% fig:closure_probs3
%!TEX root = higher-order-link-prediction-postprint.tex

%%%%%

%%%%%
% fig:closure_probs4
%!TEX root = higher-order-link-prediction-postprint.tex

%%%%%

\clearpage
%!TEX root = higher-order-link-prediction-postprint.tex

\section{Simplicial closure events at different points in time}\label{sec:closure_over_time}

In \cref{sec:simp_closure}, we studied simplicial closure events by counting the 3-node and
4-node configuration patterns in the first 80\% of timestamped simplices and
then measuring the fraction of instances that experience a simplicial close event
in the final 20\% of timestamped simplices. Here, we show that our
results are consistent when examining different time slices of the data. We
first filtered each dataset to contain only the first $X$\% of timestamped
simplices, for $X=40,60,80$ (the original dataset is the case of $X = 100$.)
We then split the filtered dataset into the first 80\% and last 20\%
of timestamped simplices (within the time frame of the filtered dataset)
and computed the probabilities of simplicial closure events.

\begin{sidewaysfigure*}[tb]
\centering
\includegraphics[width=\linewidth]{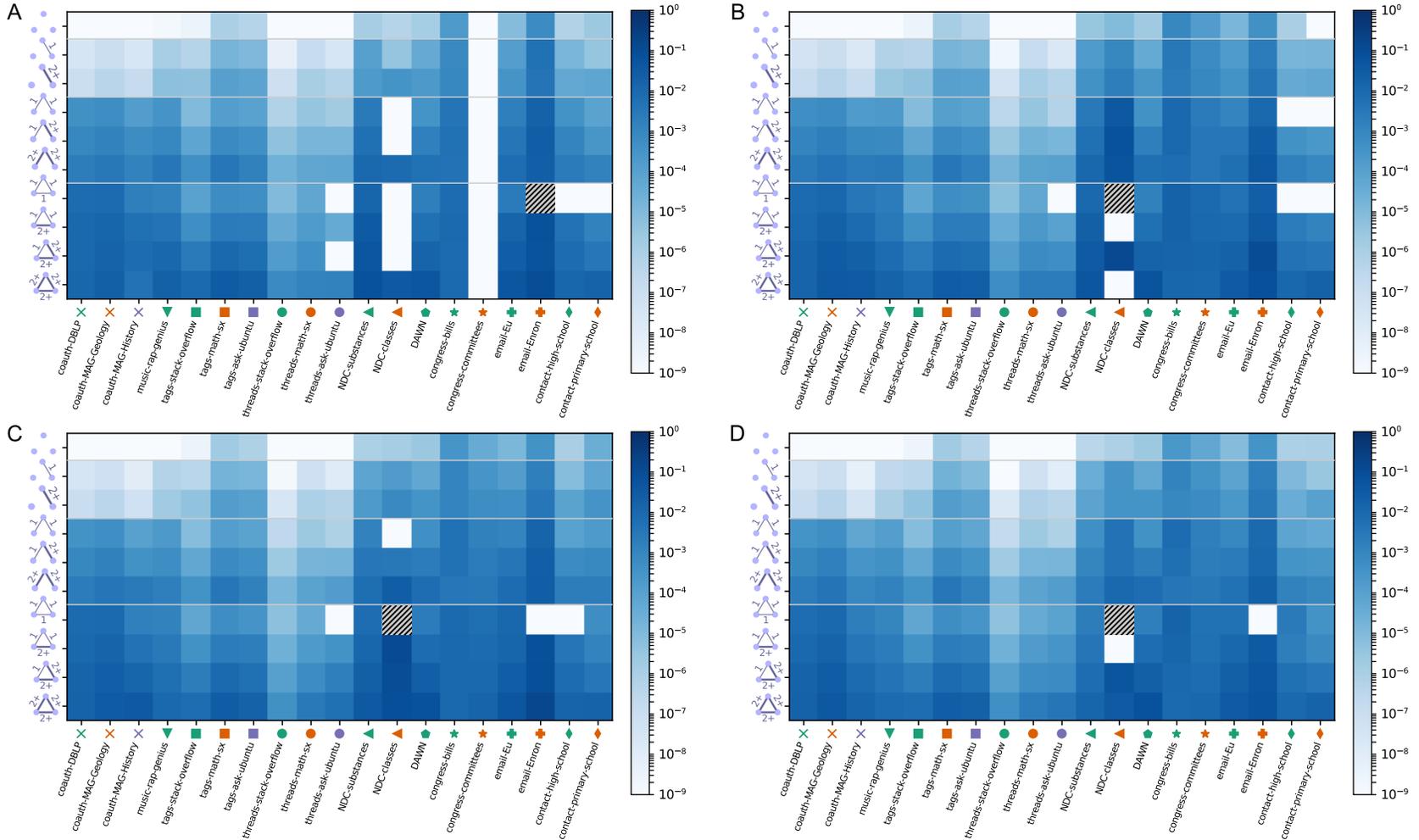}
\caption{%
  Heat maps of simplicial closure event probabilities of different configurations at
  different points in time. We first filtered each dataset to contain only the
  first $X$\% of timestamped simplices, for
  $X=40$ \formatsublabel{A},
  $X=60$ \formatsublabel{B},
  $X=80$ \formatsublabel{C}, and
  $X=100$ \formatsublabel{D}.
  We then split the filtered dataset into the first 80\% and last 20\% of
  timestamped simplices (within the time frame of the filtered dataset) and
  compute the probability of closure in last 20\% conditioned on the open
  configuration in the first 80\%. Overall, the four heat maps of simplicial
  closure event probabilities exhibit similar trends (actual probabilities are
  listed in \cref{tab:closure_over_time}). Shaded boxes are cases with fewer
  than 25 samples. The four sections of the heat map correspond to 0, 1, 2, or
  3 edges in the induced subgraph.
}
\label{fig:closure_probs_3_nodes_time}
\end{sidewaysfigure*}

\Cref{tab:closure_over_time} lists the 3-node simplicial closure event
probabilities as a function of the configuration of the 3 nodes in the first
80\% of the data for $X = 40,60,80,100$, and
\cref{fig:closure_probs_3_nodes_time} provides the heat maps for each value of
$X$ (analogous to the heat map in \cref{fig:closure_probs3}). Broadly, the closure
probabilities remain similar for different values of $X$. We also find that edge
density and tie strength are always positive indicators of simplicial closure
events, regardless of $X$ (\cref{tab:closure_indicators}). Thus, these features
are important for simplicial closure events throughout the history of the
network dynamics.

The tension between these features is also consistent over time. The weak open
triangle (where all three edges are weak ties) is more likely to
close than the strong wedge (the 3-node configuration with exactly two strong
ties) in the coauth-DBLP, coauth-MAG-Geology, and congress-bills datasets for
all values of $X$ as well as in the congress-committees dataset for
$X=60,80,100$. On the other hand, the strong wedge is more likely to close in
the three stack exchange tags networks, DAWN, and threads-stack-overflow for all
values of $X$ as well as in threads-math-sx for $X=80,100$.

%%%%%
% tab:closure_indicators
%!TEX root = SI-higher-order-link-prediction.tex

\begin{table}[tb]
\setlength{\tabcolsep}{5pt}
\centering
\caption{Consistency in the effects of tie strength and edge density in 3-node
  configurations on simplicial closure events at different points in time.
  For edge density, we tested whether or not the closure event probability of a fixed
  weighted induced subgraph configuration and the same configuration with an
  additional unit-weight edge significantly increases or decreases the closure
  probability (at significance level $10^{-5}$).
  For tie strength, we tested whether the closure event probability significantly
  increases or decreases when comparing a fixed weighted induced subgraph
  containing at least one weak tie, and the same configuration where the weak
  tie is converted to a strong tie (edge weight at least 2 in the projected graph).
   The ``total" column is the number of tested hypotheses.
  We apply the tests to filtered datasets that only contain the first X\% of
  timestamped simplices (in time order). We only consider cases where the
  configuration has at least 25 samples in the first 80\% of timestamped
  simplices of a filtered dataset.
  Increasing either edge density or tie strength significantly increases the
  closure probability for all values of $X$, suggesting that these features are
  positive indicators of simplicial closure over time.
}
\begin{tabular}{l r c c c c c c}
\toprule
& & \multicolumn{3}{c}{edge density increases} & \multicolumn{3}{c}{tie strength increases}  \\
\cmidrule(lr){3-5}\cmidrule(lr){6-8}
&  & sig.\ incr.\ & sig.\ decr.\ & total
& sig.\ incr.\ & sig.\ decr.\ & total  \\
X \\
\midrule
40  &  & 89  & 0 & 113 & 71 & 2 & 113 \\
60  &  & 101 & 0 & 113 & 80 & 7 & 113 \\
80  &  & 102 & 0 & 113 & 86 & 2 & 113 \\
100 &  & 96  & 0 & 107 & 76 & 6 & 107 \\
\bottomrule
\end{tabular}
\label{tab:closure_indicators}
\end{table}

%%%%%

%%%%%
% tab:closure_over_time
%!TEX root = higher-order-link-prediction-postprint.tex

\newcommand{\trifig}[1]{
  \begin{adjustbox}{valign=t}
  \scalebox{0.7}{
    \renewcommand*{\VertexInterMinSize}{8pt}
    \renewcommand*{\VertexSmallMinSize}{8pt}
    \begin{tikzpicture}\input{#1}\end{tikzpicture}
  }
  \end{adjustbox}
}

\begin{sidewaystable*}[tb]
\setlength{\tabcolsep}{2pt}
\centering
\caption{%
  Simplicial closure event probabilities of different configurations at
  different points in time. We first filtered each dataset to contain only the
  first $X$\% of timestamped simplices, for $X=40,60,80,100$. We then split the
  filtered dataset into the first 80\% and last 20\% of timestamped simplices
  (within the time frame of the filtered dataset). We record the probability of
  closure in last 20\% conditioned on the open configuration in the first 80\%.
}
\scalebox{0.75}{
\begin{tabular}{l   c c c c @{\quad} c c c c @{\quad} c c c c @{\quad} c c c c @{\quad} c c c c @{\quad} c c c c}
\toprule
 & \multicolumn{4}{c}{
  \begin{adjustbox}{valign=t}
  \scalebox{0.7}{
    \renewcommand*{\VertexInterMinSize}{8pt}
    \renewcommand*{\VertexSmallMinSize}{8pt}
    \begin{tikzpicture}\node[VertexStyle](i) at (0, 0) {};
\node[VertexStyle](j) at (0.5, 0.865) {};
\node[VertexStyle](k) at (1, 0) {};
\end{tikzpicture}
  }
  \end{adjustbox}
} &
   \multicolumn{4}{c}{
  \begin{adjustbox}{valign=t}
  \scalebox{0.7}{
    \renewcommand*{\VertexInterMinSize}{8pt}
    \renewcommand*{\VertexSmallMinSize}{8pt}
    \begin{tikzpicture}\node[VertexStyle](i) at (0, 0) {};
\node[VertexStyle](j) at (0.5, 0.865) {};
\node[VertexStyle](k) at (1, 0) {};
\path[-, MyEdgeColor, line width=1.25pt] (j) edge node[pos=0.5, sloped, above] {\Large \textbf 1}  (k);
\end{tikzpicture}
  }
  \end{adjustbox}
} &
   \multicolumn{4}{c}{
  \begin{adjustbox}{valign=t}
  \scalebox{0.7}{
    \renewcommand*{\VertexInterMinSize}{8pt}
    \renewcommand*{\VertexSmallMinSize}{8pt}
    \begin{tikzpicture}\node[VertexStyle](i) at (0, 0) {};
\node[VertexStyle](j) at (0.5, 0.865) {};
\node[VertexStyle](k) at (1, 0) {};
\path[-, MyEdgeColor, line width=3pt] (j) edge node[pos=0.5, sloped, above] {\Large \textbf 2+}  (k);
\end{tikzpicture}
  }
  \end{adjustbox}
} &
   \multicolumn{4}{c}{
  \begin{adjustbox}{valign=t}
  \scalebox{0.7}{
    \renewcommand*{\VertexInterMinSize}{8pt}
    \renewcommand*{\VertexSmallMinSize}{8pt}
    \begin{tikzpicture}\node[VertexStyle](i) at (0, 0) {};
\node[VertexStyle](j) at (0.5, 0.865) {};
\node[VertexStyle](k) at (1, 0) {};
\path[-, MyEdgeColor, line width=1.25pt] (i) edge node[pos=0.5, sloped, above] {\Large \textbf 1}  (j);
\path[-, MyEdgeColor, line width=1.25pt] (j) edge node[pos=0.5, sloped, above] {\Large \textbf 1}  (k);
\end{tikzpicture}
  }
  \end{adjustbox}
} &
   \multicolumn{4}{c}{
  \begin{adjustbox}{valign=t}
  \scalebox{0.7}{
    \renewcommand*{\VertexInterMinSize}{8pt}
    \renewcommand*{\VertexSmallMinSize}{8pt}
    \begin{tikzpicture}\node[VertexStyle](i) at (0, 0) {};
\node[VertexStyle](j) at (0.5, 0.865) {};
\node[VertexStyle](k) at (1, 0) {};
\path[-, MyEdgeColor, line width=1.25pt] (i) edge node[pos=0.5, sloped, above] {\Large \textbf 1}  (j);
\path[-, MyEdgeColor, line width=3pt] (j) edge node[pos=0.5, sloped, above] {\Large \textbf 2+}  (k);
\end{tikzpicture}
  }
  \end{adjustbox}
} &
   \multicolumn{4}{c}{
  \begin{adjustbox}{valign=t}
  \scalebox{0.7}{
    \renewcommand*{\VertexInterMinSize}{8pt}
    \renewcommand*{\VertexSmallMinSize}{8pt}
    \begin{tikzpicture}\node[VertexStyle](i) at (0, 0) {};
\node[VertexStyle](j) at (0.5, 0.865) {};
\node[VertexStyle](k) at (1, 0) {};
\path[-, MyEdgeColor, line width=3pt] (i) edge node[pos=0.5, sloped, above] {\Large \textbf 2+}  (j);
\path[-, MyEdgeColor, line width=3pt] (j) edge node[pos=0.5, sloped, above] {\Large \textbf 2+}  (k);
\end{tikzpicture}
  }
  \end{adjustbox}
} \\
   \cmidrule(lr){2-5}\cmidrule(lr){6-9}\cmidrule(lr){10-13}\cmidrule(lr){14-17}\cmidrule(lr){18-21}\cmidrule(lr){22-25}
   & 40 & 60 & 80 & 100 & 40 & 60 & 80 & 100 & 40 & 60 & 80 & 100 & 40 & 60 & 80 & 100 & 40 & 60 & 80 & 100 & 40 & 60 & 80 & 100 \\
   \cmidrule(lr){1-25}   
   coauth-DBLP & 8.2e-13 & 1.2e-12 & 8.3e-13 & 9.3e-13 & 3.1e-08 & 4.2e-08 & 3.4e-08 & 3.6e-08 & 1.1e-07 & 1.5e-07 & 1.2e-07 & 1.3e-07 & 4.4e-04 & 5.2e-04 & 3.8e-04 & 3.5e-04 & 9.7e-04 & 1.2e-03 & 9.4e-04 & 8.8e-04 & 2.0e-03 & 2.5e-03 & 2.2e-03 & 2.1e-03 \\
   coauth-MAG-Geology & 7.9e-12 & 5.0e-12 & 3.2e-12 & 4.2e-12 & 1.1e-07 & 9.9e-08 & 7.6e-08 & 8.9e-08 & 4.1e-07 & 4.6e-07 & 3.6e-07 & 4.5e-07 & 5.8e-04 & 6.8e-04 & 5.1e-04 & 5.3e-04 & 1.4e-03 & 1.8e-03 & 1.5e-03 & 1.6e-03 & 3.0e-03 & 4.4e-03 & 3.7e-03 & 4.1e-03 \\
   coauth-MAG-History & 1.2e-12 & 8.8e-13 & 3.3e-13 & 1.8e-13 & 2.7e-08 & 2.0e-08 & 1.0e-08 & 5.6e-09 & 1.6e-07 & 1.4e-07 & 6.3e-08 & 3.9e-08 & 1.4e-04 & 1.8e-04 & 1.0e-04 & 6.3e-05 & 6.2e-04 & 8.5e-04 & 4.1e-04 & 2.9e-04 & 1.3e-03 & 2.3e-03 & 2.5e-03 & 1.0e-03 \\
   music-rap-genius & 1.6e-09 & 8.6e-10 & 4.3e-10 & 1.2e-10 & 1.3e-06 & 6.2e-07 & 5.7e-07 & 2.3e-07 & 5.5e-06 & 2.9e-06 & 1.9e-06 & 1.1e-06 & 3.3e-04 & 2.2e-04 & 2.2e-04 & 1.3e-04 & 1.0e-03 & 8.0e-04 & 5.6e-04 & 4.4e-04 & 3.3e-03 & 2.9e-03 & 1.7e-03 & 1.3e-03 \\
   tags-stack-overflow & 3.6e-09 & 3.5e-09 & 2.9e-09 & 2.8e-09 & 4.8e-07 & 4.5e-07 & 3.8e-07 & 3.8e-07 & 5.0e-06 & 4.6e-06 & 4.2e-06 & 4.2e-06 & 9.6e-06 & 8.5e-06 & 7.4e-06 & 7.4e-06 & 6.9e-05 & 6.3e-05 & 5.8e-05 & 5.7e-05 & 4.6e-04 & 4.1e-04 & 3.8e-04 & 3.7e-04 \\
   tags-math-sx & 1.0e-06 & 1.4e-06 & 1.9e-06 & 1.9e-06 & 1.7e-05 & 1.7e-05 & 1.9e-05 & 2.1e-05 & 1.1e-04 & 1.1e-04 & 1.5e-04 & 1.4e-04 & 1.3e-04 & 1.2e-04 & 1.3e-04 & 1.2e-04 & 7.0e-04 & 7.0e-04 & 7.2e-04 & 6.9e-04 & 3.4e-03 & 3.3e-03 & 3.4e-03 & 3.2e-03 \\
   tags-ask-ubuntu & 4.9e-07 & 5.4e-07 & 7.9e-07 & 4.8e-07 & 1.1e-05 & 1.3e-05 & 1.4e-05 & 9.8e-06 & 8.1e-05 & 9.7e-05 & 1.1e-04 & 7.9e-05 & 9.4e-05 & 8.8e-05 & 8.6e-05 & 6.2e-05 & 4.8e-04 & 4.6e-04 & 4.7e-04 & 3.8e-04 & 1.5e-03 & 1.6e-03 & 1.5e-03 & 1.4e-03 \\
   threads-stack-overflow & 3.2e-12 & 8.4e-13 & 4.3e-13 & 2.2e-13 & 5.5e-09 & 2.2e-09 & 1.4e-09 & 9.0e-10 & 8.4e-08 & 4.2e-08 & 2.7e-08 & 2.1e-08 & 6.4e-07 & 3.3e-07 & 2.5e-07 & 1.9e-07 & 4.8e-06 & 3.0e-06 & 2.3e-06 & 2.0e-06 & 2.5e-05 & 1.7e-05 & 1.5e-05 & 1.5e-05 \\
   threads-math-sx & 2.3e-10 & 1.4e-10 & 6.4e-11 & 4.2e-11 & 1.5e-07 & 1.3e-07 & 7.8e-08 & 6.0e-08 & 1.2e-06 & 1.3e-06 & 9.4e-07 & 7.4e-07 & 3.6e-06 & 3.8e-06 & 2.5e-06 & 2.3e-06 & 1.6e-05 & 2.0e-05 & 1.7e-05 & 1.5e-05 & 6.4e-05 & 8.9e-05 & 9.0e-05 & 7.9e-05 \\
   threads-ask-ubuntu & 5.3e-11 & 1.4e-11 & 5.3e-12 & 3.2e-12 & 2.8e-08 & 2.1e-08 & 1.2e-08 & 9.0e-09 & 5.7e-07 & 5.2e-07 & 5.7e-07 & 4.1e-07 & 1.7e-06 & 5.6e-07 & 7.7e-07 & 7.8e-07 & 1.6e-05 & 1.4e-05 & 1.8e-05 & 1.6e-05 & 6.8e-05 & 1.0e-04 & 1.6e-04 & 1.4e-04 \\
   NDC-substances & 1.4e-06 & 3.4e-06 & 9.6e-07 & 3.8e-07 & 1.1e-04 & 1.9e-04 & 7.5e-05 & 3.3e-05 & 1.9e-04 & 4.2e-04 & 1.9e-04 & 7.8e-05 & 2.5e-03 & 2.9e-03 & 1.4e-03 & 4.8e-04 & 6.0e-03 & 6.0e-03 & 3.2e-03 & 1.1e-03 & 1.0e-02 & 1.2e-02 & 6.7e-03 & 2.2e-03 \\
   NDC-classes & 5.3e-07 & 9.0e-06 & 1.5e-06 & 8.4e-07 & 3.7e-06 & 8.8e-04 & 1.7e-04 & 3.1e-04 & 3.6e-04 & 1.8e-03 & 7.7e-04 & 3.4e-04 & 0.0e+00 & 4.0e-02 & 0.0e+00 & 4.8e-03 & 0.0e+00 & 7.1e-02 & 3.0e-03 & 4.8e-03 & 9.1e-03 & 5.4e-02 & 2.4e-02 & 1.1e-02 \\
   DAWN & 1.5e-06 & 2.1e-06 & 2.7e-06 & 2.5e-06 & 4.2e-05 & 4.6e-05 & 6.0e-05 & 5.4e-05 & 2.1e-04 & 2.7e-04 & 3.5e-04 & 3.4e-04 & 3.4e-04 & 3.9e-04 & 4.7e-04 & 4.1e-04 & 1.6e-03 & 1.8e-03 & 2.2e-03 & 2.1e-03 & 5.2e-03 & 6.3e-03 & 7.6e-03 & 7.3e-03 \\
   congress-bills & 1.9e-04 & 9.2e-04 & 3.0e-04 & 2.5e-04 & 7.6e-04 & 3.0e-03 & 1.2e-03 & 1.3e-03 & 9.9e-04 & 2.4e-03 & 1.2e-03 & 9.1e-04 & 4.2e-03 & 1.2e-02 & 5.4e-03 & 8.1e-03 & 5.4e-03 & 1.0e-02 & 5.2e-03 & 6.1e-03 & 5.0e-03 & 7.2e-03 & 3.4e-03 & 3.8e-03 \\
   congress-committees & 0.0e+00 & 1.5e-04 & 3.7e-05 & 5.8e-05 & 0.0e+00 & 6.7e-04 & 2.9e-04 & 3.6e-04 & 0.0e+00 & 9.9e-04 & 5.7e-04 & 6.3e-04 & 0.0e+00 & 2.4e-03 & 1.5e-03 & 1.7e-03 & 0.0e+00 & 3.2e-03 & 2.2e-03 & 2.1e-03 & 0.0e+00 & 3.4e-03 & 3.0e-03 & 3.1e-03 \\
   email-Eu & 8.2e-06 & 1.5e-05 & 1.4e-05 & 8.4e-06 & 1.3e-04 & 1.8e-04 & 1.4e-04 & 8.3e-05 & 3.3e-04 & 3.6e-04 & 5.0e-04 & 2.4e-04 & 1.7e-03 & 1.1e-03 & 1.2e-03 & 1.0e-03 & 3.6e-03 & 3.3e-03 & 3.9e-03 & 2.4e-03 & 7.8e-03 & 6.4e-03 & 8.1e-03 & 5.2e-03 \\
   email-Enron & 6.3e-04 & 4.3e-04 & 3.8e-04 & 3.4e-04 & 5.4e-03 & 2.2e-03 & 1.8e-03 & 1.9e-03 & 4.1e-03 & 4.3e-03 & 3.3e-03 & 3.1e-03 & 1.9e-02 & 1.1e-02 & 1.5e-02 & 9.4e-03 & 2.4e-02 & 2.6e-02 & 2.3e-02 & 1.2e-02 & 2.4e-02 & 3.9e-02 & 2.5e-02 & 2.1e-02 \\
   contact-high-school & 6.4e-07 & 1.1e-06 & 1.1e-06 & 9.4e-07 & 1.5e-05 & 1.6e-05 & 7.5e-06 & 1.2e-05 & 5.3e-05 & 4.3e-05 & 8.6e-05 & 3.7e-05 & 3.8e-04 & 0.0e+00 & 9.1e-05 & 7.2e-05 & 1.1e-03 & 4.1e-04 & 6.7e-04 & 3.5e-04 & 2.4e-03 & 1.6e-03 & 2.1e-03 & 1.4e-03 \\
   contact-primary-school & 2.6e-06 & 0.0e+00 & 3.2e-05 & 1.0e-06 & 3.7e-06 & 1.7e-05 & 6.9e-05 & 3.2e-06 & 6.7e-05 & 5.6e-05 & 3.2e-04 & 1.0e-04 & 9.0e-05 & 0.0e+00 & 1.9e-04 & 5.1e-05 & 2.7e-04 & 2.6e-04 & 8.6e-04 & 2.6e-04 & 1.4e-03 & 9.9e-04 & 2.1e-03 & 8.8e-04 \\
   \midrule \\
  & \multicolumn{4}{c}{
  \begin{adjustbox}{valign=t}
  \scalebox{0.7}{
    \renewcommand*{\VertexInterMinSize}{8pt}
    \renewcommand*{\VertexSmallMinSize}{8pt}
    \begin{tikzpicture}\node[VertexStyle](i) at (0, 0) {};
\node[VertexStyle](j) at (0.5, 0.865) {};
\node[VertexStyle](k) at (1, 0) {};
\path[-, MyEdgeColor, line width=1.25pt] (i) edge node[pos=0.5, sloped, above] {\Large \textbf 1}  (j);
\path[-, MyEdgeColor, line width=1.25pt] (j) edge node[pos=0.5, sloped, above] {\Large \textbf 1}  (k);
\path[-, MyEdgeColor, line width=1.25pt] (k) edge node[pos=0.5, sloped, below] {\Large \textbf 1}  (i);\end{tikzpicture}
  }
  \end{adjustbox}
} &
   \multicolumn{4}{c}{
  \begin{adjustbox}{valign=t}
  \scalebox{0.7}{
    \renewcommand*{\VertexInterMinSize}{8pt}
    \renewcommand*{\VertexSmallMinSize}{8pt}
    \begin{tikzpicture}\node[VertexStyle](i) at (0, 0) {};
\node[VertexStyle](j) at (0.5, 0.865) {};
\node[VertexStyle](k) at (1, 0) {};
\path[-, MyEdgeColor, line width=1.25pt] (i) edge node[pos=0.5, sloped, above] {\Large \textbf 1}  (j);
\path[-, MyEdgeColor, line width=1.25pt] (j) edge node[pos=0.5, sloped, above] {\Large \textbf 1}  (k);
\path[-, MyEdgeColor, line width=3pt] (k) edge node[pos=0.5, sloped, below] {\Large \textbf 2+}  (i);\end{tikzpicture}
  }
  \end{adjustbox}
} &
   \multicolumn{4}{c}{
  \begin{adjustbox}{valign=t}
  \scalebox{0.7}{
    \renewcommand*{\VertexInterMinSize}{8pt}
    \renewcommand*{\VertexSmallMinSize}{8pt}
    \begin{tikzpicture}\node[VertexStyle](i) at (0, 0) {};
\node[VertexStyle](j) at (0.5, 0.865) {};
\node[VertexStyle](k) at (1, 0) {};
\path[-, MyEdgeColor, line width=1.25pt] (i) edge node[pos=0.5, sloped, above] {\Large \textbf 1}  (j);
\path[-, MyEdgeColor, line width=3pt] (j) edge node[pos=0.5, sloped, above] {\Large \textbf 2+}  (k);
\path[-, MyEdgeColor, line width=3pt] (k) edge node[pos=0.5, sloped, below] {\Large \textbf 2+}  (i);\end{tikzpicture}
  }
  \end{adjustbox}
} &
   \multicolumn{4}{c}{
  \begin{adjustbox}{valign=t}
  \scalebox{0.7}{
    \renewcommand*{\VertexInterMinSize}{8pt}
    \renewcommand*{\VertexSmallMinSize}{8pt}
    \begin{tikzpicture}\node[VertexStyle](i) at (0, 0) {};
\node[VertexStyle](j) at (0.5, 0.865) {};
\node[VertexStyle](k) at (1, 0) {};
\path[-, MyEdgeColor, line width=3pt] (i) edge node[pos=0.5, sloped, above] {\Large \textbf 2+}  (j);
\path[-, MyEdgeColor, line width=3pt] (j) edge node[pos=0.5, sloped, above] {\Large \textbf 2+}  (k);
\path[-, MyEdgeColor, line width=3pt] (k) edge node[pos=0.5, sloped, below] {\Large \textbf 2+}  (i);\end{tikzpicture}
  }
  \end{adjustbox}
} \\
   \cmidrule(lr){2-5}\cmidrule(lr){6-9}\cmidrule(lr){10-13}\cmidrule(lr){14-17} 
   & 40 & 60 & 80 & 100 & 40 & 60 & 80 & 100 & 40 & 60 & 80 & 100 & 40 & 60 & 80 & 100 \\
   \cmidrule(lr){1-17}
   coauth-DBLP & 8.3e-03 & 8.6e-03 & 8.5e-03 & 7.6e-03 & 1.0e-02 & 1.1e-02 & 1.1e-02 & 1.1e-02 & 1.2e-02 & 1.5e-02 & 1.5e-02 & 1.7e-02 & 1.3e-02 & 1.6e-02 & 1.7e-02 & 1.9e-02 \\
   coauth-MAG-Geology & 8.6e-03 & 1.2e-02 & 9.4e-03 & 1.0e-02 & 1.4e-02 & 2.0e-02 & 1.5e-02 & 1.7e-02 & 1.7e-02 & 2.9e-02 & 2.3e-02 & 2.7e-02 & 2.2e-02 & 3.7e-02 & 3.0e-02 & 4.0e-02 \\
   coauth-MAG-History & 1.7e-03 & 3.4e-03 & 1.6e-03 & 1.9e-03 & 6.0e-03 & 9.1e-03 & 4.9e-03 & 3.8e-03 & 8.3e-03 & 2.1e-02 & 1.4e-02 & 9.1e-03 & 5.1e-03 & 3.6e-02 & 3.8e-02 & 1.5e-02 \\
   music-rap-genius & 1.4e-03 & 2.3e-03 & 1.3e-03 & 1.1e-03 & 4.8e-03 & 2.9e-03 & 2.8e-03 & 2.2e-03 & 1.2e-02 & 7.9e-03 & 6.5e-03 & 4.6e-03 & 2.0e-02 & 1.7e-02 & 1.2e-02 & 8.6e-03 \\
   tags-stack-overflow & 9.2e-05 & 7.0e-05 & 6.5e-05 & 6.5e-05 & 5.2e-04 & 4.4e-04 & 4.0e-04 & 3.9e-04 & 2.7e-03 & 2.3e-03 & 2.1e-03 & 2.1e-03 & 9.6e-03 & 8.4e-03 & 7.7e-03 & 7.6e-03 \\
   tags-math-sx & 6.3e-04 & 5.6e-04 & 5.4e-04 & 5.6e-04 & 2.4e-03 & 2.3e-03 & 2.3e-03 & 2.1e-03 & 1.0e-02 & 9.1e-03 & 9.2e-03 & 8.6e-03 & 2.9e-02 & 2.7e-02 & 2.7e-02 & 2.6e-02 \\
   tags-ask-ubuntu & 5.3e-04 & 4.5e-04 & 3.3e-04 & 2.9e-04 & 2.2e-03 & 2.2e-03 & 1.8e-03 & 1.7e-03 & 6.9e-03 & 7.1e-03 & 5.9e-03 & 5.9e-03 & 2.1e-02 & 2.3e-02 & 1.8e-02 & 1.9e-02 \\
   threads-stack-overflow & 9.6e-06 & 6.1e-06 & 5.7e-06 & 4.9e-06 & 4.2e-05 & 3.5e-05 & 2.6e-05 & 2.5e-05 & 1.5e-04 & 1.3e-04 & 1.1e-04 & 1.1e-04 & 6.9e-04 & 5.8e-04 & 4.3e-04 & 4.7e-04 \\
   threads-math-sx & 6.3e-05 & 5.9e-05 & 4.3e-05 & 4.0e-05 & 1.8e-04 & 2.3e-04 & 2.0e-04 & 1.7e-04 & 4.6e-04 & 6.4e-04 & 6.1e-04 & 5.4e-04 & 1.3e-03 & 2.3e-03 & 2.7e-03 & 2.2e-03 \\
   threads-ask-ubuntu & 0.0e+00 & 0.0e+00 & 0.0e+00 & 8.4e-05 & 4.5e-05 & 9.2e-05 & 3.8e-04 & 2.9e-04 & 0.0e+00 & 5.2e-04 & 1.7e-03 & 6.5e-04 & 1.4e-03 & 2.2e-03 & 5.7e-03 & 3.6e-03 \\
   NDC-substances & 1.5e-02 & 1.1e-02 & 7.6e-03 & 2.1e-03 & 3.1e-02 & 2.2e-02 & 1.3e-02 & 3.8e-03 & 4.8e-02 & 3.5e-02 & 2.2e-02 & 7.2e-03 & 7.3e-02 & 4.8e-02 & 3.9e-02 & 1.5e-02 \\
   NDC-classes & 0.0e+00 & 0.0e+00 & 0.0e+00 & 0.0e+00 & 0.0e+00 & 0.0e+00 & 1.1e-01 & 0.0e+00 & 0.0e+00 & 1.3e-01 & 9.1e-02 & 5.2e-02 & 3.6e-02 & 0.0e+00 & 8.7e-02 & 3.4e-02 \\
   DAWN & 1.7e-03 & 1.5e-03 & 2.3e-03 & 1.7e-03 & 5.7e-03 & 6.0e-03 & 7.9e-03 & 7.2e-03 & 1.5e-02 & 1.7e-02 & 2.2e-02 & 2.1e-02 & 4.8e-02 & 5.5e-02 & 7.3e-02 & 6.9e-02 \\
   congress-bills & 7.9e-03 & 1.9e-02 & 9.5e-03 & 1.7e-02 & 9.5e-03 & 1.8e-02 & 1.1e-02 & 1.4e-02 & 8.4e-03 & 1.6e-02 & 1.0e-02 & 1.1e-02 & 9.6e-03 & 1.3e-02 & 1.1e-02 & 9.3e-03 \\
   congress-committees & 0.0e+00 & 9.8e-03 & 6.1e-03 & 5.5e-03 & 0.0e+00 & 1.2e-02 & 8.5e-03 & 5.9e-03 & 0.0e+00 & 1.4e-02 & 1.1e-02 & 8.4e-03 & 0.0e+00 & 1.2e-02 & 1.6e-02 & 1.2e-02 \\
   email-Eu & 2.5e-03 & 5.2e-03 & 1.2e-02 & 5.3e-03 & 1.6e-02 & 9.9e-03 & 1.5e-02 & 1.2e-02 & 2.6e-02 & 1.8e-02 & 2.4e-02 & 2.1e-02 & 4.7e-02 & 3.4e-02 & 4.8e-02 & 3.3e-02 \\
   email-Enron & 0.0e+00 & 2.3e-02 & 0.0e+00 & 0.0e+00 & 6.6e-02 & 3.9e-02 & 7.6e-02 & 3.1e-02 & 5.9e-02 & 9.9e-02 & 8.2e-02 & 4.8e-02 & 9.2e-02 & 8.0e-02 & 1.4e-01 & 5.5e-02 \\
   contact-high-school & 0.0e+00 & 0.0e+00 & 0.0e+00 & 2.6e-03 & 5.2e-03 & 2.2e-03 & 4.0e-03 & 1.8e-03 & 7.9e-03 & 4.9e-03 & 7.9e-03 & 5.7e-03 & 1.3e-02 & 1.2e-02 & 1.5e-02 & 1.2e-02 \\
   contact-primary-school & 0.0e+00 & 0.0e+00 & 5.6e-04 & 3.8e-04 & 1.4e-03 & 1.3e-03 & 1.3e-03 & 7.9e-04 & 3.4e-03 & 3.6e-03 & 3.2e-03 & 3.4e-03 & 1.7e-02 & 1.8e-02 & 1.7e-02 & 1.7e-02 \\
   \cbottomrule{1-17}
\end{tabular}
}
\label{tab:closure_over_time}
\end{sidewaystable*}

%%%%%

%%%%%
% fig:closure_probs_3_nodes_time
%!TEX root = higher-order-link-prediction-postprint.tex

%%%%%

\clearpage
%!TEX root = higher-order-link-prediction-postprint.tex

\section{Efficient counting of simplicial closure event probabilities}

%%%%%
% Table showing statistics of open and closed triangles.
%!TEX root = higher-order-link-prediction-postprint.tex

\newcommand{\tritp}[1]{
  \begin{adjustbox}{valign=t}
  \scalebox{0.6}{
    \renewcommand*{\VertexInterMinSize}{8pt}
    \renewcommand*{\VertexSmallMinSize}{8pt}
    \begin{tikzpicture}\input{#1}\end{tikzpicture}
  }
  \end{adjustbox}
}

\newcommand{\tetratp}[1]{
  \scalebox{0.5}{
    \renewcommand*{\VertexInterMinSize}{9pt}
    \renewcommand*{\VertexSmallMinSize}{9pt}
    \begin{tikzpicture}\input{#1}\end{tikzpicture}
  }
}

\newcommand{\alttetratp}[1]{
  \scalebox{0.25}{
    \renewcommand*{\VertexInterMinSize}{16pt}
    \renewcommand*{\VertexSmallMinSize}{16pt}
    \begin{tikzpicture}[line join = round, line cap = round]\input{#1}\end{tikzpicture}
  }
}

\colorlet{MyEdgeColor}{gray!80!blue}
\colorlet{MyNodeColor}{blue!30!white}
\colorlet{MySimplexColor}{blue!20!gray}
\colorlet{FaceColor}{blue!20}
\renewcommand*{\VertexLineColor}{MyNodeColor}
\renewcommand*{\VertexLightFillColor}{MyNodeColor}

\newcommand{\plus}{\raisebox{.3\height}{\scalebox{.7}{+}}}

\begin{table*}[tb]
\setlength{\tabcolsep}{5pt}
\centering
\caption{%
  The 10 open 3-node configurations analyzed and the 27 open
  4-node configurations analyzed in \cref{fig:closure_probs3,fig:closure_probs4}. We
  illustrate each 4-node configuration with a projection onto two dimensions
  (top---the unfilled circle represents the same node) as well as a
  tetrahedral three-dimensional perspective figure (bottom). For each
  configuration, we list (i) a reference number of the configuration and
  (ii) our notation for the number of instances of the configuration.
  For $3$-node configurations, the subscripts $1$ and $2$ denote weak and strong simplicial ties, and for
  $4$-node configurations, the subscripts $0$, $1$, and $2$ denote open, weak,
  and strong simplicial ties. We also use $\tricount_{i,j,k}$ to denote the sum
  of counts of open and closed $3$-node, triangles ($1 \le i \le j \le k \le 2$)
  and $\tetracount_{i,j,k,l}$ to denote the sum of counts of open and closed
  $4$-node, $6$-edge tetrahedral wireframe configurations
  ($0 \le i \le j \le k \le l \le 2$).
}
\begin{tabular}{c c c c c c c c c c c}
  \toprule
  
  \begin{adjustbox}{valign=t}
  \scalebox{0.6}{
    \renewcommand*{\VertexInterMinSize}{8pt}
    \renewcommand*{\VertexSmallMinSize}{8pt}
    \begin{tikzpicture}\node[VertexStyle](i) at (0, 0) {};
\node[VertexStyle](j) at (0.5, 0.865) {};
\node[VertexStyle](k) at (1, 0) {};
\end{tikzpicture}
  }
  \end{adjustbox}
 &
  
  \begin{adjustbox}{valign=t}
  \scalebox{0.6}{
    \renewcommand*{\VertexInterMinSize}{8pt}
    \renewcommand*{\VertexSmallMinSize}{8pt}
    \begin{tikzpicture}\node[VertexStyle](i) at (0, 0) {};
\node[VertexStyle](j) at (0.5, 0.865) {};
\node[VertexStyle](k) at (1, 0) {};
\path[-, MyEdgeColor, line width=1.25pt] (j) edge node[pos=0.5, sloped, above] {\Large \textbf 1}  (k);
\end{tikzpicture}
  }
  \end{adjustbox}
 &
  
  \begin{adjustbox}{valign=t}
  \scalebox{0.6}{
    \renewcommand*{\VertexInterMinSize}{8pt}
    \renewcommand*{\VertexSmallMinSize}{8pt}
    \begin{tikzpicture}\node[VertexStyle](i) at (0, 0) {};
\node[VertexStyle](j) at (0.5, 0.865) {};
\node[VertexStyle](k) at (1, 0) {};
\path[-, MyEdgeColor, line width=3pt] (j) edge node[pos=0.5, sloped, above] {\Large \textbf 2+}  (k);
\end{tikzpicture}
  }
  \end{adjustbox}
 &
  
  \begin{adjustbox}{valign=t}
  \scalebox{0.6}{
    \renewcommand*{\VertexInterMinSize}{8pt}
    \renewcommand*{\VertexSmallMinSize}{8pt}
    \begin{tikzpicture}\input{triangle-0-1-1-tikz.tex}\end{tikzpicture}
  }
  \end{adjustbox}
 &
  
  \begin{adjustbox}{valign=t}
  \scalebox{0.6}{
    \renewcommand*{\VertexInterMinSize}{8pt}
    \renewcommand*{\VertexSmallMinSize}{8pt}
    \begin{tikzpicture}\input{triangle-0-1-2-tikz.tex}\end{tikzpicture}
  }
  \end{adjustbox}
 &
  
  \begin{adjustbox}{valign=t}
  \scalebox{0.6}{
    \renewcommand*{\VertexInterMinSize}{8pt}
    \renewcommand*{\VertexSmallMinSize}{8pt}
    \begin{tikzpicture}\input{triangle-0-2-2-tikz.tex}\end{tikzpicture}
  }
  \end{adjustbox}
 &
  
  \begin{adjustbox}{valign=t}
  \scalebox{0.6}{
    \renewcommand*{\VertexInterMinSize}{8pt}
    \renewcommand*{\VertexSmallMinSize}{8pt}
    \begin{tikzpicture}\node[VertexStyle](i) at (0, 0) {};
\node[VertexStyle](j) at (0.5, 0.865) {};
\node[VertexStyle](k) at (1, 0) {};
\path[-, MyEdgeColor, line width=1.25pt] (i) edge node[pos=0.5, sloped, above] {\Large \textbf 1}  (j);
\path[-, MyEdgeColor, line width=1.25pt] (j) edge node[pos=0.5, sloped, above] {\Large \textbf 1}  (k);
\path[-, MyEdgeColor, line width=1.25pt] (k) edge node[pos=0.5, sloped, below] {\Large \textbf 1}  (i);\end{tikzpicture}
  }
  \end{adjustbox}
 &
  
  \begin{adjustbox}{valign=t}
  \scalebox{0.6}{
    \renewcommand*{\VertexInterMinSize}{8pt}
    \renewcommand*{\VertexSmallMinSize}{8pt}
    \begin{tikzpicture}\node[VertexStyle](i) at (0, 0) {};
\node[VertexStyle](j) at (0.5, 0.865) {};
\node[VertexStyle](k) at (1, 0) {};
\path[-, MyEdgeColor, line width=1.25pt] (i) edge node[pos=0.5, sloped, above] {\Large \textbf 1}  (j);
\path[-, MyEdgeColor, line width=1.25pt] (j) edge node[pos=0.5, sloped, above] {\Large \textbf 1}  (k);
\path[-, MyEdgeColor, line width=3pt] (k) edge node[pos=0.5, sloped, below] {\Large \textbf 2+}  (i);\end{tikzpicture}
  }
  \end{adjustbox}
 &  
  
  \begin{adjustbox}{valign=t}
  \scalebox{0.6}{
    \renewcommand*{\VertexInterMinSize}{8pt}
    \renewcommand*{\VertexSmallMinSize}{8pt}
    \begin{tikzpicture}\node[VertexStyle](i) at (0, 0) {};
\node[VertexStyle](j) at (0.5, 0.865) {};
\node[VertexStyle](k) at (1, 0) {};
\path[-, MyEdgeColor, line width=1.25pt] (i) edge node[pos=0.5, sloped, above] {\Large \textbf 1}  (j);
\path[-, MyEdgeColor, line width=3pt] (j) edge node[pos=0.5, sloped, above] {\Large \textbf 2+}  (k);
\path[-, MyEdgeColor, line width=3pt] (k) edge node[pos=0.5, sloped, below] {\Large \textbf 2+}  (i);\end{tikzpicture}
  }
  \end{adjustbox}
 &  
  
  \begin{adjustbox}{valign=t}
  \scalebox{0.6}{
    \renewcommand*{\VertexInterMinSize}{8pt}
    \renewcommand*{\VertexSmallMinSize}{8pt}
    \begin{tikzpicture}\node[VertexStyle](i) at (0, 0) {};
\node[VertexStyle](j) at (0.5, 0.865) {};
\node[VertexStyle](k) at (1, 0) {};
\path[-, MyEdgeColor, line width=3pt] (i) edge node[pos=0.5, sloped, above] {\Large \textbf 2+}  (j);
\path[-, MyEdgeColor, line width=3pt] (j) edge node[pos=0.5, sloped, above] {\Large \textbf 2+}  (k);
\path[-, MyEdgeColor, line width=3pt] (k) edge node[pos=0.5, sloped, below] {\Large \textbf 2+}  (i);\end{tikzpicture}
  }
  \end{adjustbox}
  \\  
  1; $\emptycount$ & 2; $\eta_1$ & 3; $\eta_2$ & 4; $\wedgecount_{1,1}$ & 5; $\wedgecount_{1,2}$ & 6; $\wedgecount_{2,2}$ &
  7; $\opentricount_{1,1,1}$ & 8; $\opentricount_{1,1,2}$ & 9; $\opentricount_{1,2,2}$ & 10; $\opentricount_{2,2,2}$ \\ \midrule
  
  \scalebox{0.5}{
    \renewcommand*{\VertexInterMinSize}{9pt}
    \renewcommand*{\VertexSmallMinSize}{9pt}
    \begin{tikzpicture}\node[VertexStyle](a) at (0.5, 0.866) {};
\node[VertexStyle](b) at (1.5, 0.866) {};
\node[VertexStyle](c) at (1, 0) {};
\renewcommand*{\VertexLightFillColor}{white}
\node[VertexStyle](y) at (1, 1.732) {};
\renewcommand*{\VertexLineColor}{white}
\node[VertexStyle](x) at (0, 0) {};
\node[VertexStyle](z) at (2, 0) {};
\path[-, MyEdgeColor, line width=1.5pt] (a) edge (b);
\path[-, MyEdgeColor, line width=1.5pt] (b) edge (c);
\path[-, MyEdgeColor, line width=1.5pt] (c) edge (a);
\end{tikzpicture}
  }
 &
  
  \scalebox{0.5}{
    \renewcommand*{\VertexInterMinSize}{9pt}
    \renewcommand*{\VertexSmallMinSize}{9pt}
    \begin{tikzpicture}\fill[opacity=0.15pt,MySimplexColor](0.5,0.866)--(1.5,0.866)--(1,0)--cycle;
\node[](arg) at (1, 0.866/2+0/2+0.12) {\large \textbf 1};
\node[VertexStyle](a) at (0.5, 0.866) {};
\node[VertexStyle](b) at (1.5, 0.866) {};
\node[VertexStyle](c) at (1, 0) {};
\renewcommand*{\VertexLightFillColor}{white}
\node[VertexStyle](y) at (1, 1.732) {};
\renewcommand*{\VertexLineColor}{white}
\node[VertexStyle](x) at (0, 0) {};
\node[VertexStyle](z) at (2, 0) {};
\path[-, MyEdgeColor, line width=1.5pt] (a) edge (b);
\path[-, MyEdgeColor, line width=1.5pt] (b) edge (c);
\path[-, MyEdgeColor, line width=1.5pt] (c) edge (a);
\end{tikzpicture}
  }
 &
  
  \scalebox{0.5}{
    \renewcommand*{\VertexInterMinSize}{9pt}
    \renewcommand*{\VertexSmallMinSize}{9pt}
    \begin{tikzpicture}\fill[opacity=0.45pt,MySimplexColor](0.5,0.866)--(1.5,0.866)--(1,0)--cycle;
\node[](arg) at (1, 0.866/2+0/2+0.12) {\large \textbf 2\plus};
\node[VertexStyle](a) at (0.5, 0.866) {};
\node[VertexStyle](b) at (1.5, 0.866) {};
\node[VertexStyle](c) at (1, 0) {};
\renewcommand*{\VertexLightFillColor}{white}
\node[VertexStyle](y) at (1, 1.732) {};
\renewcommand*{\VertexLineColor}{white}
\node[VertexStyle](x) at (0, 0) {};
\node[VertexStyle](z) at (2, 0) {};
\path[-, MyEdgeColor, line width=1.5pt] (a) edge (b);
\path[-, MyEdgeColor, line width=1.5pt] (b) edge (c);
\path[-, MyEdgeColor, line width=1.5pt] (c) edge (a);
\end{tikzpicture}
  }
 &
  
  \scalebox{0.5}{
    \renewcommand*{\VertexInterMinSize}{9pt}
    \renewcommand*{\VertexSmallMinSize}{9pt}
    \begin{tikzpicture}\node[VertexStyle](a) at (0.5, 0.866) {};
\node[VertexStyle](b) at (1.5, 0.866) {};
\node[VertexStyle](c) at (1, 0) {};
\renewcommand*{\VertexLightFillColor}{white}
\node[VertexStyle](y) at (1, 1.732) {};
\renewcommand*{\VertexLineColor}{white}
\node[VertexStyle](x) at (0, 0) {};
\node[VertexStyle](z) at (2, 0) {};
\path[-, MyEdgeColor, line width=1.5pt] (a) edge (y);
\path[-, MyEdgeColor, line width=1.5pt] (a) edge (b);
\path[-, MyEdgeColor, line width=1.5pt] (b) edge (c);
\path[-, MyEdgeColor, line width=1.5pt] (c) edge (a);
\end{tikzpicture}
  }
 &
  
  \scalebox{0.5}{
    \renewcommand*{\VertexInterMinSize}{9pt}
    \renewcommand*{\VertexSmallMinSize}{9pt}
    \begin{tikzpicture}\fill[opacity=0.15pt,MySimplexColor](0.5,0.866)--(1.5,0.866)--(1,0)--cycle;
\node[](arg) at (1, 0.866/2+0/2+0.12) {\large \textbf 1};
\node[VertexStyle](a) at (0.5, 0.866) {};
\node[VertexStyle](b) at (1.5, 0.866) {};
\node[VertexStyle](c) at (1, 0) {};
\renewcommand*{\VertexLightFillColor}{white}
\node[VertexStyle](y) at (1, 1.732) {};
\renewcommand*{\VertexLineColor}{white}
\node[VertexStyle](x) at (0, 0) {};
\node[VertexStyle](z) at (2, 0) {};
\path[-, MyEdgeColor, line width=1.5pt] (a) edge (y);
\path[-, MyEdgeColor, line width=1.5pt] (a) edge (b);
\path[-, MyEdgeColor, line width=1.5pt] (b) edge (c);
\path[-, MyEdgeColor, line width=1.5pt] (c) edge (a);
\end{tikzpicture}
  }
 &
  
  \scalebox{0.5}{
    \renewcommand*{\VertexInterMinSize}{9pt}
    \renewcommand*{\VertexSmallMinSize}{9pt}
    \begin{tikzpicture}\fill[opacity=0.45pt,MySimplexColor](0.5,0.866)--(1.5,0.866)--(1,0)--cycle;
\node[](arg) at (1, 0.866/2+0/2+0.12) {\large \textbf 2\plus};
\node[VertexStyle](a) at (0.5, 0.866) {};
\node[VertexStyle](b) at (1.5, 0.866) {};
\node[VertexStyle](c) at (1, 0) {};
\renewcommand*{\VertexLightFillColor}{white}
\node[VertexStyle](y) at (1, 1.732) {};
\renewcommand*{\VertexLineColor}{white}
\node[VertexStyle](x) at (0, 0) {};
\node[VertexStyle](z) at (2, 0) {};
\path[-, MyEdgeColor, line width=1.5pt] (a) edge (y);
\path[-, MyEdgeColor, line width=1.5pt] (a) edge (b);
\path[-, MyEdgeColor, line width=1.5pt] (b) edge (c);
\path[-, MyEdgeColor, line width=1.5pt] (c) edge (a);
\end{tikzpicture}
  }
 \\
  
  \scalebox{0.25}{
    \renewcommand*{\VertexInterMinSize}{16pt}
    \renewcommand*{\VertexSmallMinSize}{16pt}
    \begin{tikzpicture}[line join = round, line cap = round]\coordinate [] (AA) at (2,0,-1.414);
\coordinate [] (BB) at (-2,0,-1.414);
\coordinate [] (CC) at (0,2,1.414);
\coordinate [] (DD) at (0,-2,1.414);
\node[VertexStyle](A) at (2,0,-1.414) {};
\node[VertexStyle](B) at (-2,0,-1.414) {};
\node[VertexStyle](C) at (0,2,1.414) {};
\renewcommand*{\VertexLightFillColor}{white};
\node[VertexStyle](D) at (0,-2,1.414) {};
\path[-, MyEdgeColor, line width=1.25pt] (A) edge (B);
\path[-, MyEdgeColor, line width=1.25pt] (B) edge (C);
\path[-, MyEdgeColor, line width=1.25pt] (C) edge (A);
\path[-, MyEdgeColor, line width=1.25pt] (C) edge (A);
\end{tikzpicture}
  }
 &
  
  \scalebox{0.25}{
    \renewcommand*{\VertexInterMinSize}{16pt}
    \renewcommand*{\VertexSmallMinSize}{16pt}
    \begin{tikzpicture}[line join = round, line cap = round]\coordinate [] (AA) at (2,0,-1.414);
\coordinate [] (BB) at (-2,0,-1.414);
\coordinate [] (CC) at (0,2,1.414);
\coordinate [] (DD) at (0,-2,1.414);
\fill[opacity=0.15,FaceColor] (AA)--(BB)--(CC)--cycle;
\node[sloped,xslant=0.6,yslant=0] at (0.4,1.2,0.8) {\LARGE \textbf 1};
\node[VertexStyle](A) at (2,0,-1.414) {};
\node[VertexStyle](B) at (-2,0,-1.414) {};
\node[VertexStyle](C) at (0,2,1.414) {};
\renewcommand*{\VertexLightFillColor}{white};
\node[VertexStyle](D) at (0,-2,1.414) {};
\path[-, MyEdgeColor, line width=1.25pt] (A) edge (B);
\path[-, MyEdgeColor, line width=1.25pt] (B) edge (C);
\path[-, MyEdgeColor, line width=1.25pt] (C) edge (A);
\path[-, MyEdgeColor, line width=1.25pt] (C) edge (A);
\end{tikzpicture}
  }
 &
  
  \scalebox{0.25}{
    \renewcommand*{\VertexInterMinSize}{16pt}
    \renewcommand*{\VertexSmallMinSize}{16pt}
    \begin{tikzpicture}[line join = round, line cap = round]\coordinate [] (AA) at (2,0,-1.414);
\coordinate [] (BB) at (-2,0,-1.414);
\coordinate [] (CC) at (0,2,1.414);
\coordinate [] (DD) at (0,-2,1.414);
\fill[opacity=0.45,FaceColor] (AA)--(BB)--(CC)--cycle;
\node[sloped,xslant=0.6,yslant=0] at (0.4,1.2,0.8) {\LARGE \textbf 2+};
\node[VertexStyle](A) at (2,0,-1.414) {};
\node[VertexStyle](B) at (-2,0,-1.414) {};
\node[VertexStyle](C) at (0,2,1.414) {};
\renewcommand*{\VertexLightFillColor}{white};
\node[VertexStyle](D) at (0,-2,1.414) {};
\path[-, MyEdgeColor, line width=1.25pt] (A) edge (B);
\path[-, MyEdgeColor, line width=1.25pt] (B) edge (C);
\path[-, MyEdgeColor, line width=1.25pt] (C) edge (A);
\path[-, MyEdgeColor, line width=1.25pt] (C) edge (A);
\end{tikzpicture}
  }
 &
  
  \scalebox{0.25}{
    \renewcommand*{\VertexInterMinSize}{16pt}
    \renewcommand*{\VertexSmallMinSize}{16pt}
    \begin{tikzpicture}[line join = round, line cap = round]\coordinate [] (AA) at (2,0,-1.414);
\coordinate [] (BB) at (-2,0,-1.414);
\coordinate [] (CC) at (0,2,1.414);
\coordinate [] (DD) at (0,-2,1.414);
\node[VertexStyle](A) at (2,0,-1.414) {};
\node[VertexStyle](B) at (-2,0,-1.414) {};
\node[VertexStyle](C) at (0,2,1.414) {};
\renewcommand*{\VertexLightFillColor}{white};
\node[VertexStyle](D) at (0,-2,1.414) {};
\path[-, MyEdgeColor, line width=1.25pt] (A) edge (B);
\path[-, MyEdgeColor, line width=1.25pt] (B) edge (C);
\path[-, MyEdgeColor, line width=1.25pt] (C) edge (A);
\path[-, MyEdgeColor, line width=1.25pt] (C) edge (A);
\path[-, MyEdgeColor, line width=1.25pt] (B) edge (D);
\end{tikzpicture}
  }
 &
  
  \scalebox{0.25}{
    \renewcommand*{\VertexInterMinSize}{16pt}
    \renewcommand*{\VertexSmallMinSize}{16pt}
    \begin{tikzpicture}[line join = round, line cap = round]\coordinate [] (AA) at (2,0,-1.414);
\coordinate [] (BB) at (-2,0,-1.414);
\coordinate [] (CC) at (0,2,1.414);
\coordinate [] (DD) at (0,-2,1.414);
\fill[opacity=0.15,FaceColor] (AA)--(BB)--(CC)--cycle;
\node[sloped,xslant=0.6,yslant=0] at (0.4,1.2,0.8) {\LARGE \textbf 1};
\node[VertexStyle](A) at (2,0,-1.414) {};
\node[VertexStyle](B) at (-2,0,-1.414) {};
\node[VertexStyle](C) at (0,2,1.414) {};
\renewcommand*{\VertexLightFillColor}{white};
\node[VertexStyle](D) at (0,-2,1.414) {};
\path[-, MyEdgeColor, line width=1.25pt] (A) edge (B);
\path[-, MyEdgeColor, line width=1.25pt] (B) edge (C);
\path[-, MyEdgeColor, line width=1.25pt] (C) edge (A);
\path[-, MyEdgeColor, line width=1.25pt] (C) edge (A);
\path[-, MyEdgeColor, line width=1.25pt] (B) edge (D);
\end{tikzpicture}
  }
 &
  
  \scalebox{0.25}{
    \renewcommand*{\VertexInterMinSize}{16pt}
    \renewcommand*{\VertexSmallMinSize}{16pt}
    \begin{tikzpicture}[line join = round, line cap = round]\coordinate [] (AA) at (2,0,-1.414);
\coordinate [] (BB) at (-2,0,-1.414);
\coordinate [] (CC) at (0,2,1.414);
\coordinate [] (DD) at (0,-2,1.414);
\fill[opacity=0.45,FaceColor] (AA)--(BB)--(CC)--cycle;
\node[sloped,xslant=0.6,yslant=0] at (0.4,1.2,0.8) {\LARGE \textbf 2+};
\node[VertexStyle](A) at (2,0,-1.414) {};
\node[VertexStyle](B) at (-2,0,-1.414) {};
\node[VertexStyle](C) at (0,2,1.414) {};
\renewcommand*{\VertexLightFillColor}{white};
\node[VertexStyle](D) at (0,-2,1.414) {};
\path[-, MyEdgeColor, line width=1.25pt] (A) edge (B);
\path[-, MyEdgeColor, line width=1.25pt] (B) edge (C);
\path[-, MyEdgeColor, line width=1.25pt] (C) edge (A);
\path[-, MyEdgeColor, line width=1.25pt] (C) edge (A);
\path[-, MyEdgeColor, line width=1.25pt] (B) edge (D);
\end{tikzpicture}
  }
 \\
  1; $\triisocount_{0}$ & 2; $\triisocount_{1}$ & 3; $\triisocount_{2}$ & 4; $\triedgecount_{0}$ & 5; $\triedgecount_{1}$ & 6; $\triedgecount_{2}$ \\ \midrule

  \scalebox{0.5}{
    \renewcommand*{\VertexInterMinSize}{9pt}
    \renewcommand*{\VertexSmallMinSize}{9pt}
    \begin{tikzpicture}\node[VertexStyle](a) at (0.5, 0.866) {};
\node[VertexStyle](b) at (1.5, 0.866) {};
\node[VertexStyle](c) at (1, 0) {};
\renewcommand*{\VertexLightFillColor}{white}
\node[VertexStyle](y) at (1, 1.732) {};
\renewcommand*{\VertexLineColor}{white}
\node[VertexStyle](x) at (0, 0) {};
\node[VertexStyle](z) at (2, 0) {};
\path[-, MyEdgeColor, line width=1.5pt] (a) edge (y);
\path[-, MyEdgeColor, line width=1.5pt] (y) edge (b);
\path[-, MyEdgeColor, line width=1.5pt] (a) edge (b);
\path[-, MyEdgeColor, line width=1.5pt] (b) edge (c);
\path[-, MyEdgeColor, line width=1.5pt] (c) edge (a);
\end{tikzpicture}
  }
 &

  \scalebox{0.5}{
    \renewcommand*{\VertexInterMinSize}{9pt}
    \renewcommand*{\VertexSmallMinSize}{9pt}
    \begin{tikzpicture}\fill[opacity=0.15pt,MySimplexColor](0.5,0.866)--(1.5,0.866)--(1,0)--cycle;
\node[](arg) at (1, 0.866/2+0/2+0.12) {\large \textbf 1};
\node[VertexStyle](a) at (0.5, 0.866) {};
\node[VertexStyle](b) at (1.5, 0.866) {};
\node[VertexStyle](c) at (1, 0) {};
\renewcommand*{\VertexLightFillColor}{white}
\node[VertexStyle](y) at (1, 1.732) {};
\renewcommand*{\VertexLineColor}{white}
\node[VertexStyle](x) at (0, 0) {};
\node[VertexStyle](z) at (2, 0) {};
\path[-, MyEdgeColor, line width=1.5pt] (a) edge (y);
\path[-, MyEdgeColor, line width=1.5pt] (y) edge (b);
\path[-, MyEdgeColor, line width=1.5pt] (a) edge (b);
\path[-, MyEdgeColor, line width=1.5pt] (b) edge (c);
\path[-, MyEdgeColor, line width=1.5pt] (c) edge (a);
\end{tikzpicture}
  }
 &

  \scalebox{0.5}{
    \renewcommand*{\VertexInterMinSize}{9pt}
    \renewcommand*{\VertexSmallMinSize}{9pt}
    \begin{tikzpicture}\fill[opacity=0.45pt,MySimplexColor](0.5,0.866)--(1.5,0.866)--(1,0)--cycle;
\node[](arg) at (1, 0.866/2+0/2+0.12) {\large \textbf 2\plus};
\node[VertexStyle](a) at (0.5, 0.866) {};
\node[VertexStyle](b) at (1.5, 0.866) {};
\node[VertexStyle](c) at (1, 0) {};
\renewcommand*{\VertexLightFillColor}{white}
\node[VertexStyle](y) at (1, 1.732) {};
\renewcommand*{\VertexLineColor}{white}
\node[VertexStyle](x) at (0, 0) {};
\node[VertexStyle](z) at (2, 0) {};
\path[-, MyEdgeColor, line width=1.5pt] (a) edge (y);
\path[-, MyEdgeColor, line width=1.5pt] (y) edge (b);
\path[-, MyEdgeColor, line width=1.5pt] (a) edge (b);
\path[-, MyEdgeColor, line width=1.5pt] (b) edge (c);
\path[-, MyEdgeColor, line width=1.5pt] (c) edge (a);
\end{tikzpicture}
  }
 &

  \scalebox{0.5}{
    \renewcommand*{\VertexInterMinSize}{9pt}
    \renewcommand*{\VertexSmallMinSize}{9pt}
    \begin{tikzpicture}\fill[opacity=0.15pt,MySimplexColor](1,1.732)--(0.5,0.866)--(1.5,0.866)--cycle;
\node[](arg) at (1, 1.732/2+0.866/2-0.12) {\large \textbf 1};
\fill[opacity=0.15pt,MySimplexColor](0.5,0.866)--(1.5,0.866)--(1,0)--cycle;
\node[](arg) at (1, 0.866/2+0/2+0.12) {\large \textbf 1};
\node[VertexStyle](a) at (0.5, 0.866) {};
\node[VertexStyle](b) at (1.5, 0.866) {};
\node[VertexStyle](c) at (1, 0) {};
\renewcommand*{\VertexLightFillColor}{white}
\node[VertexStyle](y) at (1, 1.732) {};
\renewcommand*{\VertexLineColor}{white}
\node[VertexStyle](x) at (0, 0) {};
\node[VertexStyle](z) at (2, 0) {};
\path[-, MyEdgeColor, line width=1.5pt] (a) edge (y);
\path[-, MyEdgeColor, line width=1.5pt] (y) edge (b);
\path[-, MyEdgeColor, line width=1.5pt] (a) edge (b);
\path[-, MyEdgeColor, line width=1.5pt] (b) edge (c);
\path[-, MyEdgeColor, line width=1.5pt] (c) edge (a);
\end{tikzpicture}
  }
 &

  \scalebox{0.5}{
    \renewcommand*{\VertexInterMinSize}{9pt}
    \renewcommand*{\VertexSmallMinSize}{9pt}
    \begin{tikzpicture}\fill[opacity=0.15pt,MySimplexColor](1,1.732)--(0.5,0.866)--(1.5,0.866)--cycle;
\node[](arg) at (1, 1.732/2+0.866/2-0.12) {\large \textbf 1};
\fill[opacity=0.45pt,MySimplexColor](0.5,0.866)--(1.5,0.866)--(1,0)--cycle;
\node[](arg) at (1, 0.866/2+0/2+0.12) {\large \textbf 2\plus};
\node[VertexStyle](a) at (0.5, 0.866) {};
\node[VertexStyle](b) at (1.5, 0.866) {};
\node[VertexStyle](c) at (1, 0) {};
\renewcommand*{\VertexLightFillColor}{white}
\node[VertexStyle](y) at (1, 1.732) {};
\renewcommand*{\VertexLineColor}{white}
\node[VertexStyle](x) at (0, 0) {};
\node[VertexStyle](z) at (2, 0) {};
\path[-, MyEdgeColor, line width=1.5pt] (a) edge (y);
\path[-, MyEdgeColor, line width=1.5pt] (y) edge (b);
\path[-, MyEdgeColor, line width=1.5pt] (a) edge (b);
\path[-, MyEdgeColor, line width=1.5pt] (b) edge (c);
\path[-, MyEdgeColor, line width=1.5pt] (c) edge (a);
\end{tikzpicture}
  }
 &

  \scalebox{0.5}{
    \renewcommand*{\VertexInterMinSize}{9pt}
    \renewcommand*{\VertexSmallMinSize}{9pt}
    \begin{tikzpicture}\fill[opacity=0.45pt,MySimplexColor](1,1.732)--(0.5,0.866)--(1.5,0.866)--cycle;
\node[](arg) at (1, 1.732/2+0.866/2-0.12) {\large \textbf 2\plus};
\fill[opacity=0.45pt,MySimplexColor](0.5,0.866)--(1.5,0.866)--(1,0)--cycle;
\node[](arg) at (1, 0.866/2+0/2+0.12) {\large \textbf 2\plus};
\node[VertexStyle](a) at (0.5, 0.866) {};
\node[VertexStyle](b) at (1.5, 0.866) {};
\node[VertexStyle](c) at (1, 0) {};
\renewcommand*{\VertexLightFillColor}{white}
\node[VertexStyle](y) at (1, 1.732) {};
\renewcommand*{\VertexLineColor}{white}
\node[VertexStyle](x) at (0, 0) {};
\node[VertexStyle](z) at (2, 0) {};
\path[-, MyEdgeColor, line width=1.5pt] (a) edge (y);
\path[-, MyEdgeColor, line width=1.5pt] (y) edge (b);
\path[-, MyEdgeColor, line width=1.5pt] (a) edge (b);
\path[-, MyEdgeColor, line width=1.5pt] (b) edge (c);
\path[-, MyEdgeColor, line width=1.5pt] (c) edge (a);
\end{tikzpicture}
  }
 \\

  \scalebox{0.25}{
    \renewcommand*{\VertexInterMinSize}{16pt}
    \renewcommand*{\VertexSmallMinSize}{16pt}
    \begin{tikzpicture}[line join = round, line cap = round]\coordinate [] (AA) at (2,0,-1.414);
\coordinate [] (BB) at (-2,0,-1.414);
\coordinate [] (CC) at (0,2,1.414);
\coordinate [] (DD) at (0,-2,1.414);
\node[VertexStyle](A) at (2,0,-1.414) {};
\node[VertexStyle](B) at (-2,0,-1.414) {};
\node[VertexStyle](C) at (0,2,1.414) {};
\renewcommand*{\VertexLightFillColor}{white};
\node[VertexStyle](D) at (0,-2,1.414) {};
\path[-, MyEdgeColor, line width=1.25pt] (A) edge (B);
\path[-, MyEdgeColor, line width=1.25pt] (B) edge (C);
\path[-, MyEdgeColor, line width=1.25pt] (C) edge (A);
\path[-, MyEdgeColor, line width=1.25pt] (C) edge (A);
\path[-, MyEdgeColor, line width=1.25pt] (B) edge (D);
\path[dashed, MyEdgeColor, line width=0.5*1.25pt] (C) edge (D);
\end{tikzpicture}
  }
 &

  \scalebox{0.25}{
    \renewcommand*{\VertexInterMinSize}{16pt}
    \renewcommand*{\VertexSmallMinSize}{16pt}
    \begin{tikzpicture}[line join = round, line cap = round]\coordinate [] (AA) at (2,0,-1.414);
\coordinate [] (BB) at (-2,0,-1.414);
\coordinate [] (CC) at (0,2,1.414);
\coordinate [] (DD) at (0,-2,1.414);
\fill[opacity=0.15,FaceColor] (AA)--(BB)--(CC)--cycle;
\node[sloped,xslant=0.6,yslant=0] at (0.4,1.2,0.8) {\LARGE \textbf 1};
\node[VertexStyle](A) at (2,0,-1.414) {};
\node[VertexStyle](B) at (-2,0,-1.414) {};
\node[VertexStyle](C) at (0,2,1.414) {};
\renewcommand*{\VertexLightFillColor}{white};
\node[VertexStyle](D) at (0,-2,1.414) {};
\path[-, MyEdgeColor, line width=1.25pt] (A) edge (B);
\path[-, MyEdgeColor, line width=1.25pt] (B) edge (C);
\path[-, MyEdgeColor, line width=1.25pt] (C) edge (A);
\path[-, MyEdgeColor, line width=1.25pt] (C) edge (A);
\path[-, MyEdgeColor, line width=1.25pt] (B) edge (D);
\path[dashed, MyEdgeColor, line width=0.5*1.25pt] (C) edge (D);
\end{tikzpicture}
  }
 &

  \scalebox{0.25}{
    \renewcommand*{\VertexInterMinSize}{16pt}
    \renewcommand*{\VertexSmallMinSize}{16pt}
    \begin{tikzpicture}[line join = round, line cap = round]\coordinate [] (AA) at (2,0,-1.414);
\coordinate [] (BB) at (-2,0,-1.414);
\coordinate [] (CC) at (0,2,1.414);
\coordinate [] (DD) at (0,-2,1.414);
\fill[opacity=0.45,FaceColor] (AA)--(BB)--(CC)--cycle;
\node[sloped,xslant=0.6,yslant=0] at (0.4,1.2,0.8) {\LARGE \textbf 2+};
\node[VertexStyle](A) at (2,0,-1.414) {};
\node[VertexStyle](B) at (-2,0,-1.414) {};
\node[VertexStyle](C) at (0,2,1.414) {};
\renewcommand*{\VertexLightFillColor}{white};
\node[VertexStyle](D) at (0,-2,1.414) {};
\path[-, MyEdgeColor, line width=1.25pt] (A) edge (B);
\path[-, MyEdgeColor, line width=1.25pt] (B) edge (C);
\path[-, MyEdgeColor, line width=1.25pt] (C) edge (A);
\path[-, MyEdgeColor, line width=1.25pt] (C) edge (A);
\path[-, MyEdgeColor, line width=1.25pt] (B) edge (D);
\path[dashed, MyEdgeColor, line width=0.5*1.25pt] (C) edge (D);
\end{tikzpicture}
  }
 &

  \scalebox{0.25}{
    \renewcommand*{\VertexInterMinSize}{16pt}
    \renewcommand*{\VertexSmallMinSize}{16pt}
    \begin{tikzpicture}[line join = round, line cap = round]\coordinate [] (AA) at (2,0,-1.414);
\coordinate [] (BB) at (-2,0,-1.414);
\coordinate [] (CC) at (0,2,1.414);
\coordinate [] (DD) at (0,-2,1.414);
\fill[opacity=0.15,FaceColor] (BB)--(CC)--(DD)--cycle;
\fill[opacity=0.15,FaceColor] (AA)--(BB)--(CC)--cycle;
\node[sloped,xslant=0,yslant=0.8] at (0,1,2.214) {\LARGE \textbf 1};
\node[sloped,xslant=0.6,yslant=0] at (0.4,1.2,0.8) {\LARGE \textbf 1};
\node[VertexStyle](A) at (2,0,-1.414) {};
\node[VertexStyle](B) at (-2,0,-1.414) {};
\node[VertexStyle](C) at (0,2,1.414) {};
\renewcommand*{\VertexLightFillColor}{white};
\node[VertexStyle](D) at (0,-2,1.414) {};
\path[-, MyEdgeColor, line width=1.25pt] (A) edge (B);
\path[-, MyEdgeColor, line width=1.25pt] (B) edge (C);
\path[-, MyEdgeColor, line width=1.25pt] (C) edge (A);
\path[-, MyEdgeColor, line width=1.25pt] (C) edge (A);
\path[-, MyEdgeColor, line width=1.25pt] (B) edge (D);
\path[dashed, MyEdgeColor, line width=0.5*1.25pt] (C) edge (D);
\end{tikzpicture}
  }
 &

  \scalebox{0.25}{
    \renewcommand*{\VertexInterMinSize}{16pt}
    \renewcommand*{\VertexSmallMinSize}{16pt}
    \begin{tikzpicture}[line join = round, line cap = round]\coordinate [] (AA) at (2,0,-1.414);
\coordinate [] (BB) at (-2,0,-1.414);
\coordinate [] (CC) at (0,2,1.414);
\coordinate [] (DD) at (0,-2,1.414);
\fill[opacity=0.15,FaceColor] (BB)--(CC)--(DD)--cycle;
\fill[opacity=0.45,FaceColor] (AA)--(BB)--(CC)--cycle;
\node[sloped,xslant=0,yslant=0.8] at (0,1,2.214) {\LARGE \textbf 1};
\node[sloped,xslant=0.6,yslant=0] at (0.4,1.2,0.8) {\LARGE \textbf 2+};
\node[VertexStyle](A) at (2,0,-1.414) {};
\node[VertexStyle](B) at (-2,0,-1.414) {};
\node[VertexStyle](C) at (0,2,1.414) {};
\renewcommand*{\VertexLightFillColor}{white};
\node[VertexStyle](D) at (0,-2,1.414) {};
\path[-, MyEdgeColor, line width=1.25pt] (A) edge (B);
\path[-, MyEdgeColor, line width=1.25pt] (B) edge (C);
\path[-, MyEdgeColor, line width=1.25pt] (C) edge (A);
\path[-, MyEdgeColor, line width=1.25pt] (C) edge (A);
\path[-, MyEdgeColor, line width=1.25pt] (B) edge (D);
\path[dashed, MyEdgeColor, line width=0.5*1.25pt] (C) edge (D);
\end{tikzpicture}
  }
 &

  \scalebox{0.25}{
    \renewcommand*{\VertexInterMinSize}{16pt}
    \renewcommand*{\VertexSmallMinSize}{16pt}
    \begin{tikzpicture}[line join = round, line cap = round]\coordinate [] (AA) at (2,0,-1.414);
\coordinate [] (BB) at (-2,0,-1.414);
\coordinate [] (CC) at (0,2,1.414);
\coordinate [] (DD) at (0,-2,1.414);
\fill[opacity=0.45,FaceColor] (BB)--(CC)--(DD)--cycle;
\fill[opacity=0.45,FaceColor] (AA)--(BB)--(CC)--cycle;
\node[sloped,xslant=0,yslant=0.8] at (0,1,2.214) {\LARGE \textbf 2+};
\node[sloped,xslant=0.6,yslant=0] at (0.4,1.2,0.8) {\LARGE \textbf 2+};
\node[VertexStyle](A) at (2,0,-1.414) {};
\node[VertexStyle](B) at (-2,0,-1.414) {};
\node[VertexStyle](C) at (0,2,1.414) {};
\renewcommand*{\VertexLightFillColor}{white};
\node[VertexStyle](D) at (0,-2,1.414) {};
\path[-, MyEdgeColor, line width=1.25pt] (A) edge (B);
\path[-, MyEdgeColor, line width=1.25pt] (B) edge (C);
\path[-, MyEdgeColor, line width=1.25pt] (C) edge (A);
\path[-, MyEdgeColor, line width=1.25pt] (C) edge (A);
\path[-, MyEdgeColor, line width=1.25pt] (B) edge (D);
\path[dashed, MyEdgeColor, line width=0.5*1.25pt] (C) edge (D);
\end{tikzpicture}
  }
 \\
7; $\fiveedgecount_{0,0}$ & 8; $\fiveedgecount_{0,1}$ & 9; $\fiveedgecount_{0,2}$ & 10; $\fiveedgecount_{1,1}$ & 11; $\fiveedgecount_{1,2}$ & 12; $\fiveedgecount_{2,2}$ \\ \midrule

  \scalebox{0.5}{
    \renewcommand*{\VertexInterMinSize}{9pt}
    \renewcommand*{\VertexSmallMinSize}{9pt}
    \begin{tikzpicture}\node[VertexStyle](a) at (0.5, 0.866) {};
\node[VertexStyle](b) at (1.5, 0.866) {};
\node[VertexStyle](c) at (1, 0) {};
\renewcommand*{\VertexLightFillColor}{white}
\node[VertexStyle](x) at (0, 0) {};
\node[VertexStyle](y) at (1, 1.732) {};
\node[VertexStyle](z) at (2, 0) {};
\path[-, MyEdgeColor, line width=1.5pt] (x) edge (a);
\path[-, MyEdgeColor, line width=1.5pt] (a) edge (y);
\path[-, MyEdgeColor, line width=1.5pt] (y) edge (b);
\path[-, MyEdgeColor, line width=1.5pt] (b) edge (z);
\path[-, MyEdgeColor, line width=1.5pt] (z) edge (c);
\path[-, MyEdgeColor, line width=1.5pt] (c) edge (x);
\path[-, MyEdgeColor, line width=1.5pt] (a) edge (b);
\path[-, MyEdgeColor, line width=1.5pt] (b) edge (c);
\path[-, MyEdgeColor, line width=1.5pt] (c) edge (a);\end{tikzpicture}
  }
 &

  \scalebox{0.5}{
    \renewcommand*{\VertexInterMinSize}{9pt}
    \renewcommand*{\VertexSmallMinSize}{9pt}
    \begin{tikzpicture}\fill[opacity=0.15pt,MySimplexColor](2,0)--(1,0)--(1.5,0.866)--cycle;
\node[](arg) at (1.5, 0.866/2+0/2-0.12) {\large \textbf 1};
\node[VertexStyle](a) at (0.5, 0.866) {};
\node[VertexStyle](b) at (1.5, 0.866) {};
\node[VertexStyle](c) at (1, 0) {};
\renewcommand*{\VertexLightFillColor}{white}
\node[VertexStyle](x) at (0, 0) {};
\node[VertexStyle](y) at (1, 1.732) {};
\node[VertexStyle](z) at (2, 0) {};
\path[-, MyEdgeColor, line width=1.5pt] (x) edge (a);
\path[-, MyEdgeColor, line width=1.5pt] (a) edge (y);
\path[-, MyEdgeColor, line width=1.5pt] (y) edge (b);
\path[-, MyEdgeColor, line width=1.5pt] (b) edge (z);
\path[-, MyEdgeColor, line width=1.5pt] (z) edge (c);
\path[-, MyEdgeColor, line width=1.5pt] (c) edge (x);
\path[-, MyEdgeColor, line width=1.5pt] (a) edge (b);
\path[-, MyEdgeColor, line width=1.5pt] (b) edge (c);
\path[-, MyEdgeColor, line width=1.5pt] (c) edge (a);\end{tikzpicture}
  }
 &

  \scalebox{0.5}{
    \renewcommand*{\VertexInterMinSize}{9pt}
    \renewcommand*{\VertexSmallMinSize}{9pt}
    \begin{tikzpicture}\fill[opacity=0.45pt,MySimplexColor](2,0)--(1,0)--(1.5,0.866)--cycle;
\node[](arg) at (1.5, 0.866/2+0/2-0.12) {\large \textbf 2\plus};
\node[VertexStyle](a) at (0.5, 0.866) {};
\node[VertexStyle](b) at (1.5, 0.866) {};
\node[VertexStyle](c) at (1, 0) {};
\renewcommand*{\VertexLightFillColor}{white}
\node[VertexStyle](x) at (0, 0) {};
\node[VertexStyle](y) at (1, 1.732) {};
\node[VertexStyle](z) at (2, 0) {};
\path[-, MyEdgeColor, line width=1.5pt] (x) edge (a);
\path[-, MyEdgeColor, line width=1.5pt] (a) edge (y);
\path[-, MyEdgeColor, line width=1.5pt] (y) edge (b);
\path[-, MyEdgeColor, line width=1.5pt] (b) edge (z);
\path[-, MyEdgeColor, line width=1.5pt] (z) edge (c);
\path[-, MyEdgeColor, line width=1.5pt] (c) edge (x);
\path[-, MyEdgeColor, line width=1.5pt] (a) edge (b);
\path[-, MyEdgeColor, line width=1.5pt] (b) edge (c);
\path[-, MyEdgeColor, line width=1.5pt] (c) edge (a);\end{tikzpicture}
  }
 &

  \scalebox{0.5}{
    \renewcommand*{\VertexInterMinSize}{9pt}
    \renewcommand*{\VertexSmallMinSize}{9pt}
    \begin{tikzpicture}\fill[opacity=0.15pt,MySimplexColor](1,1.732)--(0.5,0.866)--(1.5,0.866)--cycle;
\node[](arg) at (1, 1.732/2+0.866/2-0.12) {\large \textbf 1};
\fill[opacity=0.15pt,MySimplexColor](2,0)--(1,0)--(1.5,0.866)--cycle;
\node[](arg) at (1.5, 0.866/2+0/2-0.12) {\large \textbf 1};
\node[VertexStyle](a) at (0.5, 0.866) {};
\node[VertexStyle](b) at (1.5, 0.866) {};
\node[VertexStyle](c) at (1, 0) {};
\renewcommand*{\VertexLightFillColor}{white}
\node[VertexStyle](x) at (0, 0) {};
\node[VertexStyle](y) at (1, 1.732) {};
\node[VertexStyle](z) at (2, 0) {};
\path[-, MyEdgeColor, line width=1.5pt] (x) edge (a);
\path[-, MyEdgeColor, line width=1.5pt] (a) edge (y);
\path[-, MyEdgeColor, line width=1.5pt] (y) edge (b);
\path[-, MyEdgeColor, line width=1.5pt] (b) edge (z);
\path[-, MyEdgeColor, line width=1.5pt] (z) edge (c);
\path[-, MyEdgeColor, line width=1.5pt] (c) edge (x);
\path[-, MyEdgeColor, line width=1.5pt] (a) edge (b);
\path[-, MyEdgeColor, line width=1.5pt] (b) edge (c);
\path[-, MyEdgeColor, line width=1.5pt] (c) edge (a);\end{tikzpicture}
  }
 &

  \scalebox{0.5}{
    \renewcommand*{\VertexInterMinSize}{9pt}
    \renewcommand*{\VertexSmallMinSize}{9pt}
    \begin{tikzpicture}\fill[opacity=0.15pt,MySimplexColor](1,1.732)--(0.5,0.866)--(1.5,0.866)--cycle;
\node[](arg) at (1, 1.732/2+0.866/2-0.12) {\large \textbf 1};
\fill[opacity=0.45pt,MySimplexColor](2,0)--(1,0)--(1.5,0.866)--cycle;
\node[](arg) at (1.5, 0.866/2+0/2-0.12) {\large \textbf 2\plus};
\node[VertexStyle](a) at (0.5, 0.866) {};
\node[VertexStyle](b) at (1.5, 0.866) {};
\node[VertexStyle](c) at (1, 0) {};
\renewcommand*{\VertexLightFillColor}{white}
\node[VertexStyle](x) at (0, 0) {};
\node[VertexStyle](y) at (1, 1.732) {};
\node[VertexStyle](z) at (2, 0) {};
\path[-, MyEdgeColor, line width=1.5pt] (x) edge (a);
\path[-, MyEdgeColor, line width=1.5pt] (a) edge (y);
\path[-, MyEdgeColor, line width=1.5pt] (y) edge (b);
\path[-, MyEdgeColor, line width=1.5pt] (b) edge (z);
\path[-, MyEdgeColor, line width=1.5pt] (z) edge (c);
\path[-, MyEdgeColor, line width=1.5pt] (c) edge (x);
\path[-, MyEdgeColor, line width=1.5pt] (a) edge (b);
\path[-, MyEdgeColor, line width=1.5pt] (b) edge (c);
\path[-, MyEdgeColor, line width=1.5pt] (c) edge (a);\end{tikzpicture}
  }
 &

  \scalebox{0.5}{
    \renewcommand*{\VertexInterMinSize}{9pt}
    \renewcommand*{\VertexSmallMinSize}{9pt}
    \begin{tikzpicture}\fill[opacity=0.45pt,MySimplexColor](1,1.732)--(0.5,0.866)--(1.5,0.866)--cycle;
\node[](arg) at (1, 1.732/2+0.866/2-0.12) {\large \textbf 2\plus};
\fill[opacity=0.45pt,MySimplexColor](2,0)--(1,0)--(1.5,0.866)--cycle;
\node[](arg) at (1.5, 0.866/2+0/2-0.12) {\large \textbf 2\plus};
\node[VertexStyle](a) at (0.5, 0.866) {};
\node[VertexStyle](b) at (1.5, 0.866) {};
\node[VertexStyle](c) at (1, 0) {};
\renewcommand*{\VertexLightFillColor}{white}
\node[VertexStyle](x) at (0, 0) {};
\node[VertexStyle](y) at (1, 1.732) {};
\node[VertexStyle](z) at (2, 0) {};
\path[-, MyEdgeColor, line width=1.5pt] (x) edge (a);
\path[-, MyEdgeColor, line width=1.5pt] (a) edge (y);
\path[-, MyEdgeColor, line width=1.5pt] (y) edge (b);
\path[-, MyEdgeColor, line width=1.5pt] (b) edge (z);
\path[-, MyEdgeColor, line width=1.5pt] (z) edge (c);
\path[-, MyEdgeColor, line width=1.5pt] (c) edge (x);
\path[-, MyEdgeColor, line width=1.5pt] (a) edge (b);
\path[-, MyEdgeColor, line width=1.5pt] (b) edge (c);
\path[-, MyEdgeColor, line width=1.5pt] (c) edge (a);\end{tikzpicture}
  }
 &

  \scalebox{0.5}{
    \renewcommand*{\VertexInterMinSize}{9pt}
    \renewcommand*{\VertexSmallMinSize}{9pt}
    \begin{tikzpicture}\fill[opacity=0.15pt,MySimplexColor](0,0)--(0.5,0.866)--(1,0)--cycle;
\node[](arg) at (0.5, 0.866/2+0/2-0.12) {\large \textbf 1};
\fill[opacity=0.15pt,MySimplexColor](1,1.732)--(0.5,0.866)--(1.5,0.866)--cycle;
\node[](arg) at (1, 1.732/2+0.866/2-0.12) {\large \textbf 1};
\fill[opacity=0.15pt,MySimplexColor](2,0)--(1,0)--(1.5,0.866)--cycle;
\node[](arg) at (1.5, 0.866/2+0/2-0.12) {\large \textbf 1};
\node[VertexStyle](a) at (0.5, 0.866) {};
\node[VertexStyle](b) at (1.5, 0.866) {};
\node[VertexStyle](c) at (1, 0) {};
\renewcommand*{\VertexLightFillColor}{white}
\node[VertexStyle](x) at (0, 0) {};
\node[VertexStyle](y) at (1, 1.732) {};
\node[VertexStyle](z) at (2, 0) {};
\path[-, MyEdgeColor, line width=1.5pt] (x) edge (a);
\path[-, MyEdgeColor, line width=1.5pt] (a) edge (y);
\path[-, MyEdgeColor, line width=1.5pt] (y) edge (b);
\path[-, MyEdgeColor, line width=1.5pt] (b) edge (z);
\path[-, MyEdgeColor, line width=1.5pt] (z) edge (c);
\path[-, MyEdgeColor, line width=1.5pt] (c) edge (x);
\path[-, MyEdgeColor, line width=1.5pt] (a) edge (b);
\path[-, MyEdgeColor, line width=1.5pt] (b) edge (c);
\path[-, MyEdgeColor, line width=1.5pt] (c) edge (a);\end{tikzpicture}
  }
 &

  \scalebox{0.5}{
    \renewcommand*{\VertexInterMinSize}{9pt}
    \renewcommand*{\VertexSmallMinSize}{9pt}
    \begin{tikzpicture}\fill[opacity=0.15pt,MySimplexColor](0,0)--(0.5,0.866)--(1,0)--cycle;
\node[](arg) at (0.5, 0.866/2+0/2-0.12) {\large \textbf 1};
\fill[opacity=0.15pt,MySimplexColor](1,1.732)--(0.5,0.866)--(1.5,0.866)--cycle;
\node[](arg) at (1, 1.732/2+0.866/2-0.12) {\large \textbf 1};
\fill[opacity=0.45pt,MySimplexColor](2,0)--(1,0)--(1.5,0.866)--cycle;
\node[](arg) at (1.5, 0.866/2+0/2-0.12) {\large \textbf 2\plus};
\node[VertexStyle](a) at (0.5, 0.866) {};
\node[VertexStyle](b) at (1.5, 0.866) {};
\node[VertexStyle](c) at (1, 0) {};
\renewcommand*{\VertexLightFillColor}{white}
\node[VertexStyle](x) at (0, 0) {};
\node[VertexStyle](y) at (1, 1.732) {};
\node[VertexStyle](z) at (2, 0) {};
\path[-, MyEdgeColor, line width=1.5pt] (x) edge (a);
\path[-, MyEdgeColor, line width=1.5pt] (a) edge (y);
\path[-, MyEdgeColor, line width=1.5pt] (y) edge (b);
\path[-, MyEdgeColor, line width=1.5pt] (b) edge (z);
\path[-, MyEdgeColor, line width=1.5pt] (z) edge (c);
\path[-, MyEdgeColor, line width=1.5pt] (c) edge (x);
\path[-, MyEdgeColor, line width=1.5pt] (a) edge (b);
\path[-, MyEdgeColor, line width=1.5pt] (b) edge (c);
\path[-, MyEdgeColor, line width=1.5pt] (c) edge (a);\end{tikzpicture}
  }
 &

  \scalebox{0.5}{
    \renewcommand*{\VertexInterMinSize}{9pt}
    \renewcommand*{\VertexSmallMinSize}{9pt}
    \begin{tikzpicture}\fill[opacity=0.15pt,MySimplexColor](0,0)--(0.5,0.866)--(1,0)--cycle;
\node[](arg) at (0.5, 0.866/2+0/2-0.12) {\large \textbf 1};
\fill[opacity=0.45pt,MySimplexColor](1,1.732)--(0.5,0.866)--(1.5,0.866)--cycle;
\node[](arg) at (1, 1.732/2+0.866/2-0.12) {\large \textbf 2\plus};
\fill[opacity=0.45pt,MySimplexColor](2,0)--(1,0)--(1.5,0.866)--cycle;
\node[](arg) at (1.5, 0.866/2+0/2-0.12) {\large \textbf 2\plus};
\node[VertexStyle](a) at (0.5, 0.866) {};
\node[VertexStyle](b) at (1.5, 0.866) {};
\node[VertexStyle](c) at (1, 0) {};
\renewcommand*{\VertexLightFillColor}{white}
\node[VertexStyle](x) at (0, 0) {};
\node[VertexStyle](y) at (1, 1.732) {};
\node[VertexStyle](z) at (2, 0) {};
\path[-, MyEdgeColor, line width=1.5pt] (x) edge (a);
\path[-, MyEdgeColor, line width=1.5pt] (a) edge (y);
\path[-, MyEdgeColor, line width=1.5pt] (y) edge (b);
\path[-, MyEdgeColor, line width=1.5pt] (b) edge (z);
\path[-, MyEdgeColor, line width=1.5pt] (z) edge (c);
\path[-, MyEdgeColor, line width=1.5pt] (c) edge (x);
\path[-, MyEdgeColor, line width=1.5pt] (a) edge (b);
\path[-, MyEdgeColor, line width=1.5pt] (b) edge (c);
\path[-, MyEdgeColor, line width=1.5pt] (c) edge (a);\end{tikzpicture}
  }
 &

  \scalebox{0.5}{
    \renewcommand*{\VertexInterMinSize}{9pt}
    \renewcommand*{\VertexSmallMinSize}{9pt}
    \begin{tikzpicture}\fill[opacity=0.45pt,MySimplexColor](0,0)--(0.5,0.866)--(1,0)--cycle;
\node[](arg) at (0.5, 0.866/2+0/2-0.12) {\large \textbf 2\plus};
\fill[opacity=0.45pt,MySimplexColor](1,1.732)--(0.5,0.866)--(1.5,0.866)--cycle;
\node[](arg) at (1, 1.732/2+0.866/2-0.12) {\large \textbf 2\plus};
\fill[opacity=0.45pt,MySimplexColor](2,0)--(1,0)--(1.5,0.866)--cycle;
\node[](arg) at (1.5, 0.866/2+0/2-0.12) {\large \textbf 2\plus};
\node[VertexStyle](a) at (0.5, 0.866) {};
\node[VertexStyle](b) at (1.5, 0.866) {};
\node[VertexStyle](c) at (1, 0) {};
\renewcommand*{\VertexLightFillColor}{white}
\node[VertexStyle](x) at (0, 0) {};
\node[VertexStyle](y) at (1, 1.732) {};
\node[VertexStyle](z) at (2, 0) {};
\path[-, MyEdgeColor, line width=1.5pt] (x) edge (a);
\path[-, MyEdgeColor, line width=1.5pt] (a) edge (y);
\path[-, MyEdgeColor, line width=1.5pt] (y) edge (b);
\path[-, MyEdgeColor, line width=1.5pt] (b) edge (z);
\path[-, MyEdgeColor, line width=1.5pt] (z) edge (c);
\path[-, MyEdgeColor, line width=1.5pt] (c) edge (x);
\path[-, MyEdgeColor, line width=1.5pt] (a) edge (b);
\path[-, MyEdgeColor, line width=1.5pt] (b) edge (c);
\path[-, MyEdgeColor, line width=1.5pt] (c) edge (a);\end{tikzpicture}
  }
 & \\

  \scalebox{0.25}{
    \renewcommand*{\VertexInterMinSize}{16pt}
    \renewcommand*{\VertexSmallMinSize}{16pt}
    \begin{tikzpicture}[line join = round, line cap = round]\coordinate [] (AA) at (2,0,-1.414);
\coordinate [] (BB) at (-2,0,-1.414);
\coordinate [] (CC) at (0,2,1.414);
\coordinate [] (DD) at (0,-2,1.414);
\node[VertexStyle](A) at (2,0,-1.414) {};
\node[VertexStyle](B) at (-2,0,-1.414) {};
\node[VertexStyle](C) at (0,2,1.414) {};
\renewcommand*{\VertexLightFillColor}{white};
\node[VertexStyle](D) at (0,-2,1.414) {};
\path[-, MyEdgeColor, line width=1.25pt] (A) edge (B);
\path[-, MyEdgeColor, line width=1.25pt] (B) edge (C);
\path[-, MyEdgeColor, line width=1.25pt] (C) edge (A);
\path[-, MyEdgeColor, line width=1.25pt] (C) edge (A);
\path[-, MyEdgeColor, line width=1.25pt] (A) edge (D);
\path[-, MyEdgeColor, line width=1.25pt] (B) edge (D);
\path[dashed, MyEdgeColor, line width=0.5*1.25pt] (C) edge (D);\end{tikzpicture}
  }
 &

  \scalebox{0.25}{
    \renewcommand*{\VertexInterMinSize}{16pt}
    \renewcommand*{\VertexSmallMinSize}{16pt}
    \begin{tikzpicture}[line join = round, line cap = round]\coordinate [] (AA) at (2,0,-1.414);
\coordinate [] (BB) at (-2,0,-1.414);
\coordinate [] (CC) at (0,2,1.414);
\coordinate [] (DD) at (0,-2,1.414);
\fill[opacity=0.15,FaceColor] (AA)--(BB)--(CC)--cycle;
\node[sloped,xslant=0.6,yslant=0] at (0.4,1.2,0.8) {\LARGE \textbf 1};
\node[VertexStyle](A) at (2,0,-1.414) {};
\node[VertexStyle](B) at (-2,0,-1.414) {};
\node[VertexStyle](C) at (0,2,1.414) {};
\renewcommand*{\VertexLightFillColor}{white};
\node[VertexStyle](D) at (0,-2,1.414) {};
\path[-, MyEdgeColor, line width=1.25pt] (A) edge (B);
\path[-, MyEdgeColor, line width=1.25pt] (B) edge (C);
\path[-, MyEdgeColor, line width=1.25pt] (C) edge (A);
\path[-, MyEdgeColor, line width=1.25pt] (C) edge (A);
\path[-, MyEdgeColor, line width=1.25pt] (A) edge (D);
\path[-, MyEdgeColor, line width=1.25pt] (B) edge (D);
\path[dashed, MyEdgeColor, line width=0.5*1.25pt] (C) edge (D);
\end{tikzpicture}
  }
 &

  \scalebox{0.25}{
    \renewcommand*{\VertexInterMinSize}{16pt}
    \renewcommand*{\VertexSmallMinSize}{16pt}
    \begin{tikzpicture}[line join = round, line cap = round]\coordinate [] (AA) at (2,0,-1.414);
\coordinate [] (BB) at (-2,0,-1.414);
\coordinate [] (CC) at (0,2,1.414);
\coordinate [] (DD) at (0,-2,1.414);
\fill[opacity=0.45,FaceColor] (AA)--(BB)--(CC)--cycle;
\node[sloped,xslant=0.6,yslant=0] at (0.4,1.2,0.8) {\LARGE \textbf 2+};
\node[VertexStyle](A) at (2,0,-1.414) {};
\node[VertexStyle](B) at (-2,0,-1.414) {};
\node[VertexStyle](C) at (0,2,1.414) {};
\renewcommand*{\VertexLightFillColor}{white};
\node[VertexStyle](D) at (0,-2,1.414) {};
\path[-, MyEdgeColor, line width=1.25pt] (A) edge (B);
\path[-, MyEdgeColor, line width=1.25pt] (B) edge (C);
\path[-, MyEdgeColor, line width=1.25pt] (C) edge (A);
\path[-, MyEdgeColor, line width=1.25pt] (C) edge (A);
\path[-, MyEdgeColor, line width=1.25pt] (A) edge (D);
\path[-, MyEdgeColor, line width=1.25pt] (B) edge (D);
\path[dashed, MyEdgeColor, line width=0.5*1.25pt] (C) edge (D);
\end{tikzpicture}
  }
 &

  \scalebox{0.25}{
    \renewcommand*{\VertexInterMinSize}{16pt}
    \renewcommand*{\VertexSmallMinSize}{16pt}
    \begin{tikzpicture}[line join = round, line cap = round]\coordinate [] (AA) at (2,0,-1.414);
\coordinate [] (BB) at (-2,0,-1.414);
\coordinate [] (CC) at (0,2,1.414);
\coordinate [] (DD) at (0,-2,1.414);
\fill[opacity=0.15,FaceColor] (BB)--(CC)--(DD)--cycle;
\fill[opacity=0.15,FaceColor] (AA)--(BB)--(CC)--cycle;
\node[sloped,xslant=0,yslant=0.8] at (0,1,2.214) {\LARGE \textbf 1};
\node[sloped,xslant=0.6,yslant=0] at (0.4,1.2,0.8) {\LARGE \textbf 1};
\node[VertexStyle](A) at (2,0,-1.414) {};
\node[VertexStyle](B) at (-2,0,-1.414) {};
\node[VertexStyle](C) at (0,2,1.414) {};
\renewcommand*{\VertexLightFillColor}{white};
\node[VertexStyle](D) at (0,-2,1.414) {};
\path[-, MyEdgeColor, line width=1.25pt] (A) edge (B);
\path[-, MyEdgeColor, line width=1.25pt] (B) edge (C);
\path[-, MyEdgeColor, line width=1.25pt] (C) edge (A);
\path[-, MyEdgeColor, line width=1.25pt] (C) edge (A);
\path[-, MyEdgeColor, line width=1.25pt] (A) edge (D);
\path[-, MyEdgeColor, line width=1.25pt] (B) edge (D);
\path[dashed, MyEdgeColor, line width=0.5*1.25pt] (C) edge (D);
\end{tikzpicture}
  }
 &

  \scalebox{0.25}{
    \renewcommand*{\VertexInterMinSize}{16pt}
    \renewcommand*{\VertexSmallMinSize}{16pt}
    \begin{tikzpicture}[line join = round, line cap = round]\coordinate [] (AA) at (2,0,-1.414);
\coordinate [] (BB) at (-2,0,-1.414);
\coordinate [] (CC) at (0,2,1.414);
\coordinate [] (DD) at (0,-2,1.414);
\fill[opacity=0.15,FaceColor] (BB)--(CC)--(DD)--cycle;
\fill[opacity=0.45,FaceColor] (AA)--(BB)--(CC)--cycle;
\node[sloped,xslant=0,yslant=0.8] at (0,1,2.214) {\LARGE \textbf 1};
\node[sloped,xslant=0.6,yslant=0] at (0.4,1.2,0.8) {\LARGE \textbf 2+};
\node[VertexStyle](A) at (2,0,-1.414) {};
\node[VertexStyle](B) at (-2,0,-1.414) {};
\node[VertexStyle](C) at (0,2,1.414) {};
\renewcommand*{\VertexLightFillColor}{white};
\node[VertexStyle](D) at (0,-2,1.414) {};
\path[-, MyEdgeColor, line width=1.25pt] (A) edge (B);
\path[-, MyEdgeColor, line width=1.25pt] (B) edge (C);
\path[-, MyEdgeColor, line width=1.25pt] (C) edge (A);
\path[-, MyEdgeColor, line width=1.25pt] (C) edge (A);
\path[-, MyEdgeColor, line width=1.25pt] (A) edge (D);
\path[-, MyEdgeColor, line width=1.25pt] (B) edge (D);
\path[dashed, MyEdgeColor, line width=0.5*1.25pt] (C) edge (D);
\end{tikzpicture}
  }
 &

  \scalebox{0.25}{
    \renewcommand*{\VertexInterMinSize}{16pt}
    \renewcommand*{\VertexSmallMinSize}{16pt}
    \begin{tikzpicture}[line join = round, line cap = round]\coordinate [] (AA) at (2,0,-1.414);
\coordinate [] (BB) at (-2,0,-1.414);
\coordinate [] (CC) at (0,2,1.414);
\coordinate [] (DD) at (0,-2,1.414);
\fill[opacity=0.45,FaceColor] (BB)--(CC)--(DD)--cycle;
\fill[opacity=0.45,FaceColor] (AA)--(BB)--(CC)--cycle;
\node[sloped,xslant=0,yslant=0.8] at (0,1,2.214) {\LARGE \textbf 2+};
\node[sloped,xslant=0.6,yslant=0] at (0.4,1.2,0.8) {\LARGE \textbf 2+};
\node[VertexStyle](A) at (2,0,-1.414) {};
\node[VertexStyle](B) at (-2,0,-1.414) {};
\node[VertexStyle](C) at (0,2,1.414) {};
\renewcommand*{\VertexLightFillColor}{white};
\node[VertexStyle](D) at (0,-2,1.414) {};
\path[-, MyEdgeColor, line width=1.25pt] (A) edge (B);
\path[-, MyEdgeColor, line width=1.25pt] (B) edge (C);
\path[-, MyEdgeColor, line width=1.25pt] (C) edge (A);
\path[-, MyEdgeColor, line width=1.25pt] (C) edge (A);
\path[-, MyEdgeColor, line width=1.25pt] (A) edge (D);
\path[-, MyEdgeColor, line width=1.25pt] (B) edge (D);
\path[dashed, MyEdgeColor, line width=0.5*1.25pt] (C) edge (D);
\end{tikzpicture}
  }
 &

  \scalebox{0.25}{
    \renewcommand*{\VertexInterMinSize}{16pt}
    \renewcommand*{\VertexSmallMinSize}{16pt}
    \begin{tikzpicture}[line join = round, line cap = round]\coordinate [] (AA) at (2,0,-1.414);
\coordinate [] (BB) at (-2,0,-1.414);
\coordinate [] (CC) at (0,2,1.414);
\coordinate [] (DD) at (0,-2,1.414);
\fill[opacity=0.15,FaceColor] (AA)--(CC)--(DD)--cycle;
\fill[opacity=0.15,FaceColor] (BB)--(CC)--(DD)--cycle;
\fill[opacity=0.15,FaceColor] (AA)--(BB)--(CC)--cycle;
\node[sloped,xslant=0.5,yslant=-0.3] at (0.8,0.5,1.414) {\LARGE \textbf 1};
\node[sloped,xslant=0,yslant=0.8] at (0,1,2.214) {\LARGE \textbf 1};
\node[sloped,xslant=0.6,yslant=0] at (0.4,1.2,0.8) {\LARGE \textbf 1};
\node[VertexStyle](A) at (2,0,-1.414) {};
\node[VertexStyle](B) at (-2,0,-1.414) {};
\node[VertexStyle](C) at (0,2,1.414) {};
\renewcommand*{\VertexLightFillColor}{white};
\node[VertexStyle](D) at (0,-2,1.414) {};
\path[-, MyEdgeColor, line width=1.25pt] (A) edge (B);
\path[-, MyEdgeColor, line width=1.25pt] (B) edge (C);
\path[-, MyEdgeColor, line width=1.25pt] (C) edge (A);
\path[-, MyEdgeColor, line width=1.25pt] (C) edge (A);
\path[-, MyEdgeColor, line width=1.25pt] (A) edge (D);
\path[-, MyEdgeColor, line width=1.25pt] (B) edge (D);
\path[dashed, MyEdgeColor, line width=0.5*1.25pt] (C) edge (D);
\end{tikzpicture}
  }
 &

  \scalebox{0.25}{
    \renewcommand*{\VertexInterMinSize}{16pt}
    \renewcommand*{\VertexSmallMinSize}{16pt}
    \begin{tikzpicture}[line join = round, line cap = round]\coordinate [] (AA) at (2,0,-1.414);
\coordinate [] (BB) at (-2,0,-1.414);
\coordinate [] (CC) at (0,2,1.414);
\coordinate [] (DD) at (0,-2,1.414);
\fill[opacity=0.15,FaceColor] (AA)--(CC)--(DD)--cycle;
\fill[opacity=0.15,FaceColor] (BB)--(CC)--(DD)--cycle;
\fill[opacity=0.45,FaceColor] (AA)--(BB)--(CC)--cycle;
\node[sloped,xslant=0.5,yslant=-0.3] at (0.8,0.5,1.414) {\LARGE \textbf 1};
\node[sloped,xslant=0,yslant=0.8] at (0,1,2.214) {\LARGE \textbf 1};
\node[sloped,xslant=0.6,yslant=0] at (0.4,1.2,0.8) {\LARGE \textbf 2+};
\node[VertexStyle](A) at (2,0,-1.414) {};
\node[VertexStyle](B) at (-2,0,-1.414) {};
\node[VertexStyle](C) at (0,2,1.414) {};
\renewcommand*{\VertexLightFillColor}{white};
\node[VertexStyle](D) at (0,-2,1.414) {};
\path[-, MyEdgeColor, line width=1.25pt] (A) edge (B);
\path[-, MyEdgeColor, line width=1.25pt] (B) edge (C);
\path[-, MyEdgeColor, line width=1.25pt] (C) edge (A);
\path[-, MyEdgeColor, line width=1.25pt] (C) edge (A);
\path[-, MyEdgeColor, line width=1.25pt] (A) edge (D);
\path[-, MyEdgeColor, line width=1.25pt] (B) edge (D);
\path[dashed, MyEdgeColor, line width=0.5*1.25pt] (C) edge (D);
\end{tikzpicture}
  }
 &

  \scalebox{0.25}{
    \renewcommand*{\VertexInterMinSize}{16pt}
    \renewcommand*{\VertexSmallMinSize}{16pt}
    \begin{tikzpicture}[line join = round, line cap = round]\coordinate [] (AA) at (2,0,-1.414);
\coordinate [] (BB) at (-2,0,-1.414);
\coordinate [] (CC) at (0,2,1.414);
\coordinate [] (DD) at (0,-2,1.414);
\fill[opacity=0.15,FaceColor] (AA)--(CC)--(DD)--cycle;
\fill[opacity=0.45,FaceColor] (BB)--(CC)--(DD)--cycle;
\fill[opacity=0.45,FaceColor] (AA)--(BB)--(CC)--cycle;
\node[sloped,xslant=0.5,yslant=-0.3] at (0.8,0.5,1.414) {\LARGE \textbf 1};
\node[sloped,xslant=0,yslant=0.8] at (0,1,2.214) {\LARGE \textbf 2+};
\node[sloped,xslant=0.6,yslant=0] at (0.4,1.2,0.8) {\LARGE \textbf 2+};
\node[VertexStyle](A) at (2,0,-1.414) {};
\node[VertexStyle](B) at (-2,0,-1.414) {};
\node[VertexStyle](C) at (0,2,1.414) {};
\renewcommand*{\VertexLightFillColor}{white};
\node[VertexStyle](D) at (0,-2,1.414) {};
\path[-, MyEdgeColor, line width=1.25pt] (A) edge (B);
\path[-, MyEdgeColor, line width=1.25pt] (B) edge (C);
\path[-, MyEdgeColor, line width=1.25pt] (C) edge (A);
\path[-, MyEdgeColor, line width=1.25pt] (C) edge (A);
\path[-, MyEdgeColor, line width=1.25pt] (A) edge (D);
\path[-, MyEdgeColor, line width=1.25pt] (B) edge (D);
\path[dashed, MyEdgeColor, line width=0.5*1.25pt] (C) edge (D);
\end{tikzpicture}
  }
 &

  \scalebox{0.25}{
    \renewcommand*{\VertexInterMinSize}{16pt}
    \renewcommand*{\VertexSmallMinSize}{16pt}
    \begin{tikzpicture}[line join = round, line cap = round]\coordinate [] (AA) at (2,0,-1.414);
\coordinate [] (BB) at (-2,0,-1.414);
\coordinate [] (CC) at (0,2,1.414);
\coordinate [] (DD) at (0,-2,1.414);
\fill[opacity=0.45,FaceColor] (AA)--(CC)--(DD)--cycle;
\fill[opacity=0.45,FaceColor] (BB)--(CC)--(DD)--cycle;
\fill[opacity=0.45,FaceColor] (AA)--(BB)--(CC)--cycle;
\node[sloped,xslant=0.5,yslant=-0.3] at (0.8,0.5,1.414) {\LARGE \textbf 2+};
\node[sloped,xslant=0,yslant=0.8] at (0,1,2.214) {\LARGE \textbf 2+};
\node[sloped,xslant=0.6,yslant=0] at (0.4,1.2,0.8) {\LARGE \textbf 2+};
\node[VertexStyle](A) at (2,0,-1.414) {};
\node[VertexStyle](B) at (-2,0,-1.414) {};
\node[VertexStyle](C) at (0,2,1.414) {};
\renewcommand*{\VertexLightFillColor}{white};
\node[VertexStyle](D) at (0,-2,1.414) {};
\path[-, MyEdgeColor, line width=1.25pt] (A) edge (B);
\path[-, MyEdgeColor, line width=1.25pt] (B) edge (C);
\path[-, MyEdgeColor, line width=1.25pt] (C) edge (A);
\path[-, MyEdgeColor, line width=1.25pt] (C) edge (A);
\path[-, MyEdgeColor, line width=1.25pt] (A) edge (D);
\path[-, MyEdgeColor, line width=1.25pt] (B) edge (D);
\path[dashed, MyEdgeColor, line width=0.5*1.25pt] (C) edge (D);
\end{tikzpicture}
  }
 & \\
13; $\opentetracount_{0,0,0,0}$ & 14; $\opentetracount_{0,0,0,1}$ & 15; $\opentetracount_{0,0,0,2}$ & 16; $\opentetracount_{0,0,1,1}$ & 17; $\opentetracount_{0,0,1,2}$ &
18; $\opentetracount_{0,0,2,2}$ & 19; $\opentetracount_{0,1,1,1}$ & 20; $\opentetracount_{0,1,1,2}$ & 21; $\opentetracount_{0,1,2,2}$ & 22; $\opentetracount_{0,2,2,2}$ \\ \midrule

  \scalebox{0.5}{
    \renewcommand*{\VertexInterMinSize}{9pt}
    \renewcommand*{\VertexSmallMinSize}{9pt}
    \begin{tikzpicture}\fill[opacity=0.15pt,MySimplexColor](0.5,0.866)--(1.5,0.866)--(1,0)--cycle;
\node[](arg) at (1, 0.866/2+0/2+0.12) {\large \textbf 1};
\fill[opacity=0.15pt,MySimplexColor](0,0)--(0.5,0.866)--(1,0)--cycle;
\node[](arg) at (0.5, 0.866/2+0/2-0.12) {\large \textbf 1};
\fill[opacity=0.15pt,MySimplexColor](1,1.732)--(0.5,0.866)--(1.5,0.866)--cycle;
\node[](arg) at (1, 1.732/2+0.866/2-0.12) {\large \textbf 1};
\fill[opacity=0.15pt,MySimplexColor](2,0)--(1,0)--(1.5,0.866)--cycle;
\node[](arg) at (1.5, 0.866/2+0/2-0.12) {\large \textbf 1};
\node[VertexStyle](a) at (0.5, 0.866) {};
\node[VertexStyle](b) at (1.5, 0.866) {};
\node[VertexStyle](c) at (1, 0) {};
\renewcommand*{\VertexLightFillColor}{white}
\node[VertexStyle](x) at (0, 0) {};
\node[VertexStyle](y) at (1, 1.732) {};
\node[VertexStyle](z) at (2, 0) {};
\path[-, MyEdgeColor, line width=1.5pt] (x) edge (a);
\path[-, MyEdgeColor, line width=1.5pt] (a) edge (y);
\path[-, MyEdgeColor, line width=1.5pt] (y) edge (b);
\path[-, MyEdgeColor, line width=1.5pt] (b) edge (z);
\path[-, MyEdgeColor, line width=1.5pt] (z) edge (c);
\path[-, MyEdgeColor, line width=1.5pt] (c) edge (x);
\path[-, MyEdgeColor, line width=1.5pt] (a) edge (b);
\path[-, MyEdgeColor, line width=1.5pt] (b) edge (c);
\path[-, MyEdgeColor, line width=1.5pt] (c) edge (a);\end{tikzpicture}
  }
 &

  \scalebox{0.5}{
    \renewcommand*{\VertexInterMinSize}{9pt}
    \renewcommand*{\VertexSmallMinSize}{9pt}
    \begin{tikzpicture}\fill[opacity=0.15pt,MySimplexColor](0.5,0.866)--(1.5,0.866)--(1,0)--cycle;
\node[](arg) at (1, 0.866/2+0/2+0.12) {\large \textbf 1};
\fill[opacity=0.15pt,MySimplexColor](0,0)--(0.5,0.866)--(1,0)--cycle;
\node[](arg) at (0.5, 0.866/2+0/2-0.12) {\large \textbf 1};
\fill[opacity=0.15pt,MySimplexColor](1,1.732)--(0.5,0.866)--(1.5,0.866)--cycle;
\node[](arg) at (1, 1.732/2+0.866/2-0.12) {\large \textbf 1};
\fill[opacity=0.45pt,MySimplexColor](2,0)--(1,0)--(1.5,0.866)--cycle;
\node[](arg) at (1.5, 0.866/2+0/2-0.12) {\large \textbf 2\plus};
\node[VertexStyle](a) at (0.5, 0.866) {};
\node[VertexStyle](b) at (1.5, 0.866) {};
\node[VertexStyle](c) at (1, 0) {};
\renewcommand*{\VertexLightFillColor}{white}
\node[VertexStyle](x) at (0, 0) {};
\node[VertexStyle](y) at (1, 1.732) {};
\node[VertexStyle](z) at (2, 0) {};
\path[-, MyEdgeColor, line width=1.5pt] (x) edge (a);
\path[-, MyEdgeColor, line width=1.5pt] (a) edge (y);
\path[-, MyEdgeColor, line width=1.5pt] (y) edge (b);
\path[-, MyEdgeColor, line width=1.5pt] (b) edge (z);
\path[-, MyEdgeColor, line width=1.5pt] (z) edge (c);
\path[-, MyEdgeColor, line width=1.5pt] (c) edge (x);
\path[-, MyEdgeColor, line width=1.5pt] (a) edge (b);
\path[-, MyEdgeColor, line width=1.5pt] (b) edge (c);
\path[-, MyEdgeColor, line width=1.5pt] (c) edge (a);\end{tikzpicture}
  }
 &

  \scalebox{0.5}{
    \renewcommand*{\VertexInterMinSize}{9pt}
    \renewcommand*{\VertexSmallMinSize}{9pt}
    \begin{tikzpicture}\fill[opacity=0.15pt,MySimplexColor](0.5,0.866)--(1.5,0.866)--(1,0)--cycle;
\node[](arg) at (1, 0.866/2+0/2+0.12) {\large \textbf 1};
\fill[opacity=0.15pt,MySimplexColor](0,0)--(0.5,0.866)--(1,0)--cycle;
\node[](arg) at (0.5, 0.866/2+0/2-0.12) {\large \textbf 1};
\fill[opacity=0.45pt,MySimplexColor](1,1.732)--(0.5,0.866)--(1.5,0.866)--cycle;
\node[](arg) at (1, 1.732/2+0.866/2-0.12) {\large \textbf 2\plus};
\fill[opacity=0.45pt,MySimplexColor](2,0)--(1,0)--(1.5,0.866)--cycle;
\node[](arg) at (1.5, 0.866/2+0/2-0.12) {\large \textbf 2\plus};
\node[VertexStyle](a) at (0.5, 0.866) {};
\node[VertexStyle](b) at (1.5, 0.866) {};
\node[VertexStyle](c) at (1, 0) {};
\renewcommand*{\VertexLightFillColor}{white}
\node[VertexStyle](x) at (0, 0) {};
\node[VertexStyle](y) at (1, 1.732) {};
\node[VertexStyle](z) at (2, 0) {};
\path[-, MyEdgeColor, line width=1.5pt] (x) edge (a);
\path[-, MyEdgeColor, line width=1.5pt] (a) edge (y);
\path[-, MyEdgeColor, line width=1.5pt] (y) edge (b);
\path[-, MyEdgeColor, line width=1.5pt] (b) edge (z);
\path[-, MyEdgeColor, line width=1.5pt] (z) edge (c);
\path[-, MyEdgeColor, line width=1.5pt] (c) edge (x);
\path[-, MyEdgeColor, line width=1.5pt] (a) edge (b);
\path[-, MyEdgeColor, line width=1.5pt] (b) edge (c);
\path[-, MyEdgeColor, line width=1.5pt] (c) edge (a);\end{tikzpicture}
  }
 &

  \scalebox{0.5}{
    \renewcommand*{\VertexInterMinSize}{9pt}
    \renewcommand*{\VertexSmallMinSize}{9pt}
    \begin{tikzpicture}\fill[opacity=0.15pt,MySimplexColor](0.5,0.866)--(1.5,0.866)--(1,0)--cycle;
\node[](arg) at (1, 0.866/2+0/2+0.12) {\large \textbf 1};
\fill[opacity=0.45pt,MySimplexColor](0,0)--(0.5,0.866)--(1,0)--cycle;
\node[](arg) at (0.5, 0.866/2+0/2-0.12) {\large \textbf 2\plus};
\fill[opacity=0.45pt,MySimplexColor](1,1.732)--(0.5,0.866)--(1.5,0.866)--cycle;
\node[](arg) at (1, 1.732/2+0.866/2-0.12) {\large \textbf 2\plus};
\fill[opacity=0.45pt,MySimplexColor](2,0)--(1,0)--(1.5,0.866)--cycle;
\node[](arg) at (1.5, 0.866/2+0/2-0.12) {\large \textbf 2\plus};
\node[VertexStyle](a) at (0.5, 0.866) {};
\node[VertexStyle](b) at (1.5, 0.866) {};
\node[VertexStyle](c) at (1, 0) {};
\renewcommand*{\VertexLightFillColor}{white}
\node[VertexStyle](x) at (0, 0) {};
\node[VertexStyle](y) at (1, 1.732) {};
\node[VertexStyle](z) at (2, 0) {};
\path[-, MyEdgeColor, line width=1.5pt] (x) edge (a);
\path[-, MyEdgeColor, line width=1.5pt] (a) edge (y);
\path[-, MyEdgeColor, line width=1.5pt] (y) edge (b);
\path[-, MyEdgeColor, line width=1.5pt] (b) edge (z);
\path[-, MyEdgeColor, line width=1.5pt] (z) edge (c);
\path[-, MyEdgeColor, line width=1.5pt] (c) edge (x);
\path[-, MyEdgeColor, line width=1.5pt] (a) edge (b);
\path[-, MyEdgeColor, line width=1.5pt] (b) edge (c);
\path[-, MyEdgeColor, line width=1.5pt] (c) edge (a);\end{tikzpicture}
  }
 &

  \scalebox{0.5}{
    \renewcommand*{\VertexInterMinSize}{9pt}
    \renewcommand*{\VertexSmallMinSize}{9pt}
    \begin{tikzpicture}\fill[opacity=0.45pt,MySimplexColor](0.5,0.866)--(1.5,0.866)--(1,0)--cycle;
\node[](arg) at (1, 0.866/2+0/2+0.12) {\large \textbf 2\plus};
\fill[opacity=0.45pt,MySimplexColor](0,0)--(0.5,0.866)--(1,0)--cycle;
\node[](arg) at (0.5, 0.866/2+0/2-0.12) {\large \textbf 2\plus};
\fill[opacity=0.45pt,MySimplexColor](1,1.732)--(0.5,0.866)--(1.5,0.866)--cycle;
\node[](arg) at (1, 1.732/2+0.866/2-0.12) {\large \textbf 2\plus};
\fill[opacity=0.45pt,MySimplexColor](2,0)--(1,0)--(1.5,0.866)--cycle;
\node[](arg) at (1.5, 0.866/2+0/2-0.12) {\large \textbf 2\plus};
\node[VertexStyle](a) at (0.5, 0.866) {};
\node[VertexStyle](b) at (1.5, 0.866) {};
\node[VertexStyle](c) at (1, 0) {};
\renewcommand*{\VertexLightFillColor}{white}
\node[VertexStyle](x) at (0, 0) {};
\node[VertexStyle](y) at (1, 1.732) {};
\node[VertexStyle](z) at (2, 0) {};
\path[-, MyEdgeColor, line width=1.5pt] (x) edge (a);
\path[-, MyEdgeColor, line width=1.5pt] (a) edge (y);
\path[-, MyEdgeColor, line width=1.5pt] (y) edge (b);
\path[-, MyEdgeColor, line width=1.5pt] (b) edge (z);
\path[-, MyEdgeColor, line width=1.5pt] (z) edge (c);
\path[-, MyEdgeColor, line width=1.5pt] (c) edge (x);
\path[-, MyEdgeColor, line width=1.5pt] (a) edge (b);
\path[-, MyEdgeColor, line width=1.5pt] (b) edge (c);
\path[-, MyEdgeColor, line width=1.5pt] (c) edge (a);\end{tikzpicture}
  }
 \\

  \scalebox{0.25}{
    \renewcommand*{\VertexInterMinSize}{16pt}
    \renewcommand*{\VertexSmallMinSize}{16pt}
    \begin{tikzpicture}[line join = round, line cap = round]\coordinate [] (AA) at (2,0,-1.414);
\coordinate [] (BB) at (-2,0,-1.414);
\coordinate [] (CC) at (0,2,1.414);
\coordinate [] (DD) at (0,-2,1.414);
\fill[opacity=0.15,FaceColor] (AA)--(BB)--(DD)--cycle;
\fill[opacity=0.15,FaceColor] (AA)--(CC)--(DD)--cycle;
\fill[opacity=0.15,FaceColor] (BB)--(CC)--(DD)--cycle;
\fill[opacity=0.15,FaceColor] (AA)--(BB)--(CC)--cycle;
\node[sloped,xslant=0.2,yslant=0.2] at (0.3,-0.25,0.914) {\LARGE \textbf 1};
\node[sloped,xslant=0.5,yslant=-0.3] at (0.8,0.5,1.414) {\LARGE \textbf 1};
\node[sloped,xslant=0,yslant=0.8] at (0,1,2.214) {\LARGE \textbf 1};
\node[sloped,xslant=0.6,yslant=0] at (0.4,1.2,0.8) {\LARGE \textbf 1};
\node[VertexStyle](A) at (2,0,-1.414) {};
\node[VertexStyle](B) at (-2,0,-1.414) {};
\node[VertexStyle](C) at (0,2,1.414) {};
\renewcommand*{\VertexLightFillColor}{white};
\node[VertexStyle](D) at (0,-2,1.414) {};
\path[-, MyEdgeColor, line width=1.25pt] (A) edge (B);
\path[-, MyEdgeColor, line width=1.25pt] (B) edge (C);
\path[-, MyEdgeColor, line width=1.25pt] (C) edge (A);
\path[-, MyEdgeColor, line width=1.25pt] (C) edge (A);
\path[-, MyEdgeColor, line width=1.25pt] (A) edge (D);
\path[-, MyEdgeColor, line width=1.25pt] (B) edge (D);
\path[dashed, MyEdgeColor, line width=0.5*1.25pt] (C) edge (D);
\end{tikzpicture}
  }
 &

  \scalebox{0.25}{
    \renewcommand*{\VertexInterMinSize}{16pt}
    \renewcommand*{\VertexSmallMinSize}{16pt}
    \begin{tikzpicture}[line join = round, line cap = round]\coordinate [] (AA) at (2,0,-1.414);
\coordinate [] (BB) at (-2,0,-1.414);
\coordinate [] (CC) at (0,2,1.414);
\coordinate [] (DD) at (0,-2,1.414);
\fill[opacity=0.15,FaceColor] (AA)--(BB)--(DD)--cycle;
\fill[opacity=0.15,FaceColor] (AA)--(CC)--(DD)--cycle;
\fill[opacity=0.15,FaceColor] (BB)--(CC)--(DD)--cycle;
\fill[opacity=0.45,FaceColor] (AA)--(BB)--(CC)--cycle;
\node[sloped,xslant=0.2,yslant=0.2] at (0.3,-0.25,0.914) {\LARGE \textbf 1};
\node[sloped,xslant=0.5,yslant=-0.3] at (0.8,0.5,1.414) {\LARGE \textbf 1};
\node[sloped,xslant=0,yslant=0.8] at (0,1,2.214) {\LARGE \textbf 1};
\node[sloped,xslant=0.6,yslant=0] at (0.4,1.2,0.8) {\LARGE \textbf 2+};
\node[VertexStyle](A) at (2,0,-1.414) {};
\node[VertexStyle](B) at (-2,0,-1.414) {};
\node[VertexStyle](C) at (0,2,1.414) {};
\renewcommand*{\VertexLightFillColor}{white};
\node[VertexStyle](D) at (0,-2,1.414) {};
\path[-, MyEdgeColor, line width=1.25pt] (A) edge (B);
\path[-, MyEdgeColor, line width=1.25pt] (B) edge (C);
\path[-, MyEdgeColor, line width=1.25pt] (C) edge (A);
\path[-, MyEdgeColor, line width=1.25pt] (C) edge (A);
\path[-, MyEdgeColor, line width=1.25pt] (A) edge (D);
\path[-, MyEdgeColor, line width=1.25pt] (B) edge (D);
\path[dashed, MyEdgeColor, line width=0.5*1.25pt] (C) edge (D);
\end{tikzpicture}
  }
 &

  \scalebox{0.25}{
    \renewcommand*{\VertexInterMinSize}{16pt}
    \renewcommand*{\VertexSmallMinSize}{16pt}
    \begin{tikzpicture}[line join = round, line cap = round]\coordinate [] (AA) at (2,0,-1.414);
\coordinate [] (BB) at (-2,0,-1.414);
\coordinate [] (CC) at (0,2,1.414);
\coordinate [] (DD) at (0,-2,1.414);
\fill[opacity=0.15,FaceColor] (AA)--(BB)--(DD)--cycle;
\fill[opacity=0.15,FaceColor] (AA)--(CC)--(DD)--cycle;
\fill[opacity=0.45,FaceColor] (BB)--(CC)--(DD)--cycle;
\fill[opacity=0.45,FaceColor] (AA)--(BB)--(CC)--cycle;
\node[sloped,xslant=0.2,yslant=0.2] at (0.3,-0.25,0.914) {\LARGE \textbf 1};
\node[sloped,xslant=0.5,yslant=-0.3] at (0.8,0.5,1.414) {\LARGE \textbf 1};
\node[sloped,xslant=0,yslant=0.8] at (0,1,2.214) {\LARGE \textbf 2+};
\node[sloped,xslant=0.6,yslant=0] at (0.4,1.2,0.8) {\LARGE \textbf 2+};
\node[VertexStyle](A) at (2,0,-1.414) {};
\node[VertexStyle](B) at (-2,0,-1.414) {};
\node[VertexStyle](C) at (0,2,1.414) {};
\renewcommand*{\VertexLightFillColor}{white};
\node[VertexStyle](D) at (0,-2,1.414) {};
\path[-, MyEdgeColor, line width=1.25pt] (A) edge (B);
\path[-, MyEdgeColor, line width=1.25pt] (B) edge (C);
\path[-, MyEdgeColor, line width=1.25pt] (C) edge (A);
\path[-, MyEdgeColor, line width=1.25pt] (C) edge (A);
\path[-, MyEdgeColor, line width=1.25pt] (A) edge (D);
\path[-, MyEdgeColor, line width=1.25pt] (B) edge (D);
\path[dashed, MyEdgeColor, line width=0.5*1.25pt] (C) edge (D);
\end{tikzpicture}
  }
 &

  \scalebox{0.25}{
    \renewcommand*{\VertexInterMinSize}{16pt}
    \renewcommand*{\VertexSmallMinSize}{16pt}
    \begin{tikzpicture}[line join = round, line cap = round]\coordinate [] (AA) at (2,0,-1.414);
\coordinate [] (BB) at (-2,0,-1.414);
\coordinate [] (CC) at (0,2,1.414);
\coordinate [] (DD) at (0,-2,1.414);
\fill[opacity=0.15,FaceColor] (AA)--(BB)--(DD)--cycle;
\fill[opacity=0.45,FaceColor] (AA)--(CC)--(DD)--cycle;
\fill[opacity=0.45,FaceColor] (BB)--(CC)--(DD)--cycle;
\fill[opacity=0.45,FaceColor] (AA)--(BB)--(CC)--cycle;
\node[sloped,xslant=0.2,yslant=0.2] at (0.3,-0.25,0.914) {\LARGE \textbf 1};
\node[sloped,xslant=0.5,yslant=-0.3] at (0.8,0.5,1.414) {\LARGE \textbf 2+};
\node[sloped,xslant=0,yslant=0.8] at (0,1,2.214) {\LARGE \textbf 2+};
\node[sloped,xslant=0.6,yslant=0] at (0.4,1.2,0.8) {\LARGE \textbf 2+};
\node[VertexStyle](A) at (2,0,-1.414) {};
\node[VertexStyle](B) at (-2,0,-1.414) {};
\node[VertexStyle](C) at (0,2,1.414) {};
\renewcommand*{\VertexLightFillColor}{white};
\node[VertexStyle](D) at (0,-2,1.414) {};
\path[-, MyEdgeColor, line width=1.25pt] (A) edge (B);
\path[-, MyEdgeColor, line width=1.25pt] (B) edge (C);
\path[-, MyEdgeColor, line width=1.25pt] (C) edge (A);
\path[-, MyEdgeColor, line width=1.25pt] (C) edge (A);
\path[-, MyEdgeColor, line width=1.25pt] (A) edge (D);
\path[-, MyEdgeColor, line width=1.25pt] (B) edge (D);
\path[dashed, MyEdgeColor, line width=0.5*1.25pt] (C) edge (D);
\end{tikzpicture}
  }
 &

  \scalebox{0.25}{
    \renewcommand*{\VertexInterMinSize}{16pt}
    \renewcommand*{\VertexSmallMinSize}{16pt}
    \begin{tikzpicture}[line join = round, line cap = round]\coordinate [] (AA) at (2,0,-1.414);
\coordinate [] (BB) at (-2,0,-1.414);
\coordinate [] (CC) at (0,2,1.414);
\coordinate [] (DD) at (0,-2,1.414);
\fill[opacity=0.45,FaceColor] (AA)--(BB)--(DD)--cycle;
\fill[opacity=0.45,FaceColor] (AA)--(CC)--(DD)--cycle;
\fill[opacity=0.45,FaceColor] (BB)--(CC)--(DD)--cycle;
\fill[opacity=0.45,FaceColor] (AA)--(BB)--(CC)--cycle;
\node[sloped,xslant=0.2,yslant=0.2] at (0.3,-0.25,0.914) {\LARGE \textbf 2\plus};
\node[sloped,xslant=0.5,yslant=-0.3] at (0.8,0.5,1.414) {\LARGE \textbf 2\plus};
\node[sloped,xslant=0,yslant=0.8] at (0,1,2.214) {\LARGE \textbf 2\plus};
\node[sloped,xslant=0.6,yslant=0] at (0.4,1.2,0.8) {\LARGE \textbf 2\plus};
\node[VertexStyle](A) at (2,0,-1.414) {};
\node[VertexStyle](B) at (-2,0,-1.414) {};
\node[VertexStyle](C) at (0,2,1.414) {};
\renewcommand*{\VertexLightFillColor}{white};
\node[VertexStyle](D) at (0,-2,1.414) {};
\path[-, MyEdgeColor, line width=1.25pt] (A) edge (B);
\path[-, MyEdgeColor, line width=1.25pt] (B) edge (C);
\path[-, MyEdgeColor, line width=1.25pt] (C) edge (A);
\path[-, MyEdgeColor, line width=1.25pt] (C) edge (A);
\path[-, MyEdgeColor, line width=1.25pt] (A) edge (D);
\path[-, MyEdgeColor, line width=1.25pt] (B) edge (D);
\path[dashed, MyEdgeColor, line width=0.5*1.25pt] (C) edge (D);
\end{tikzpicture}
  }
 \\
23; $\opentetracount_{1,1,1,1}$ & 24; $\opentetracount_{1,1,1,2}$ & 25; $\opentetracount_{1,1,2,2}$ & 26; $\opentetracount_{1,2,2,2}$ & 27; $\opentetracount_{2,2,2,2}$ \\
\bottomrule
\end{tabular}
\label{tab:configuration_labels}
\end{table*}

%%%%%

An open configuration on three or four nodes is a set of nodes that have not
jointly appeared in a simplex in the training set comprising the first 80\% of
the timestamped simplices (the training set). An instance of a subgraph
configuration ``closes'' if the nodes subsequently all appear in one of the
final 20\% of timestamped simplices (the test set). For all newly formed
simplices in the test set, we can check their prior configuration $c$ in the
training set, which provides the number of times each configuration closes.
Dividing the number of closures of a configuration $c$ by the total number of
instances it was open in the training set gives the probability of a simplicial
closure event. Most of the datasets we study are large enough that naively
computing the simplicial closure event probabilities is infeasible. We need to
develop efficient algorithms for computing the closure probabilities.

The key idea of our approach is that we do not need to \emph{enumerate} all of
the configurations in the training set and check if they close.
Instead, we only need the \emph{total count} of open configurations in the
training data.
We then count how many close by examining the test data directly.
The idea of avoiding enumeration when simply counting suffices has been used in other
fast graph configuration counting
algorithms~\cite{Paranjape-2017-motifs,Pinar-2017-escape}.

\subsection{Counting for 3-node configurations}
We first show how to count the number of each 3-node subgraph configuration (the
top row of \cref{tab:configuration_labels}).
Recall that a weak tie corresponds to an edge in the projected graph with
a weight of $1$, whereas a strong tie corresponds to an edge with a weight of at
least $2$. Subscripts of $1$ and $2$ denote weak and strong ties in our
notation. (Note that we use ``2+'' for strong ties in the illustrations in \cref{tab:configuration_labels};
however, it will be convenient to use the integer 2 in our description of the algorithms.)

Let $\tricount_{i,j,k}$, $1 \le i \le j \le k \le 2$, be the number of (open or
closed) triangles whose edges have the tie strengths given by the subscripts.
For instance, $\tricount_{1,1,1}$ is the number of triangles whose edges are all
weak ties.
Similarly, let $\opentricount_{i,j,k}$ be the number of triangles with given tie
strengths that are open (see the right-most configurations in the first row
of \cref{tab:configuration_labels}).
We can count the number of all triangles $\tricount_{i,j,k}$ using a number of
efficient triangle enumeration algorithms for sparse
graphs~\cite{Latapy-2008-triangles}.
For each of these triangles we then determine whether it is closed by examining
the entries of a simplex-to-node adjacency matrix (this can be efficiently read
from our set-based data).
The difference between the total number of triangles and the number of closed
triangles gives us the open triangle counts $\opentricount_{i,j,k}$.

Next, consider the number of $2$-edge, $3$-node induced ``wedge'' subgraphs.
Let the symbols $\wedgecount_{1,1}$, $\wedgecount_{1,2}$, and
$\wedgecount_{2,2}$ denote these configurations, where the tie strengths of
the two edges are given by the subscripts (see the first row of
\cref{tab:configuration_labels}).
Furthermore, let $d_1(u)$ and $d_2(u)$ be the number of weak and strong ties
containing node $u$ as an endpoint.
Then $\wedgecount_{i,j}$ is given by the number of (non-induced) $2$-edge,
$3$-node subgraphs with tie strengths $i$ and $j$ minus the ones that appear in
triangles:
\begin{align}
  \wedgecount_{1,1} &=\textstyle \sum_{u} \binom{d_1(u)}{2} - 3\tricount_{1,1,1} - \tricount_{1,1,2} \\
  \wedgecount_{2,2} &=\textstyle \sum_{u} \binom{d_2(u)}{2} - 3\tricount_{2,2,2} - \tricount_{1,2,2} \\
  \wedgecount_{1,2} &=\textstyle \sum_{u} d_1(u)d_2(u)      - 2\tricount_{1,1,2} - 2\tricount_{1,2,2}
\end{align}

Now let $\edgeisocount_{1}$ and $\edgeisocount_{2}$ be the counts of the
$1$-edge, $3$-node induced subgraphs, where again the tie strength of the edge
is given by the subscript (see the first row of
\cref{tab:configuration_labels}).  Denote the total number of weak and strong
ties by $m_{s}=\frac{1}{2}\sum_{u}d_{s}(u)$, $s = 1, 2$, and the total number of
nodes by $n$.  Then the total number of (non-induced) $1$-edge, $3$-node
subgraphs with tie strength $s$ is then $m_s(n-2)$. Induced $1$-edge, $3$-node
subgraph are given by the non-induced counts minus the $2$- and $3$-edge induced
counts discussed above:
\begin{align}
  \edgeisocount_{1} &= m_1(n-2) - 2\wedgecount_{1,1} - \wedgecount_{1,2} - 3\tricount_{1,1,1} - 2\tricount_{1,1,2} - \tricount_{1,2,2} \\
  \edgeisocount_{2} &= m_2(n-2) - 2\wedgecount_{2,2} - \wedgecount_{1,2} - 3\tricount_{2,2,2} - 2\tricount_{1,2,2} - \tricount_{1,1,2}
\end{align}

Finally, let $\emptycount$ be the number of empty $3$-node induced subgraphs of
the projected graph (the top left of \cref{tab:configuration_labels}).  The
number of subsets of 3 nodes minus all other induced $3$-node subgraphs gives
the value of $\emptycount$:
\begin{align}
  \emptycount =
  \binom{n}{3}
  - \sum_{s=1}^{2}\edgeisocount_{s}
  - \sum_{1 \le i, j \le 2}\wedgecount_{i,j}
  - \sum_{1 \le i \le j \le k \le 2}\tricount_{i,j,k}.
\end{align}

\subsection{Counting for 4-node configurations}
Now we describe how we compute the simplicial closure event probabilities conditioned
on the 27 subgraph configurations on four nodes in
\cref{fig:closure_probs4} (these are the 4-node configurations in the
second through fifth rows of \cref{tab:configuration_labels}).
Recall that the simplicial tie strength of a triangle is (i) \emph{open} if the three
nodes form an open triangle; (ii) \emph{weak} if the three nodes have jointly appeared
in exactly one simplex; or (iii) \emph{strong} if the three nodes have jointly appeared
in at least two simplices.
We use subscripts $0$, $1$, and $2$ to denote these bins.

There are 15 total 4-node, 6-edge tetrahedral subgraph configurations.
Each configuration corresponds to a non-decreasing 4-tuple of the simplicial tie strengths of the
four triangles in the configuration. We denote the sum of open and closed
tetrahedral counts by $\tetracount_{i,j,k,l}$, where $i$, $j$, $k$, and $l$
denote the simplicial tie strengths, and the open tetrahedral counts by
$\opentetracount_{i,j,k,l}$ ($0 \le i \le j \le k \le l \le 2$; the 15 configurations in the bottom two row blocks
of \cref{tab:configuration_labels}).
We may count both $\tetracount_{i,j,k,l}$ and
$\opentetracount_{i,j,k,l}$ by enumerating 4-cliques using, e.g., the Chiba and
Nishizeki algorithm~\cite{Chiba-1985-arboricity} and checking if each 4-clique
is closed or open by examining the simplex-node adjacency matrix.

Next, we consider counts of the six 4-node, 5-edge subgraph configurations
$\fiveedgecount_{i,j}$, where each configuration is given by a non-decreasing pair of
simplicial tie strengths for the two triangles in the configuration
(the third row block of \cref{tab:configuration_labels}).
Each instance of this configuration consists of two
triangles sharing one edge.
We first use a fast triangle enumeration algorithm
to compute matrices $Y^{(s)}$, $s \in \{0,1,2\}$, where $Y^{(s)}_{uv}$ is the
number of triangles with simplicial tie strength $s$ containing nodes $u$ and $v$.
The counts of the non-induced configuration, which we denote by
$\fiveedgecount'_{i,j}$, are then given by:
\begin{align}
  \textstyle  \fiveedgecount'_{s, s} &=\textstyle \sum_{(u,v)} {Y^{(s)}_{uv} \choose 2}, \quad s = 0, 1, 2 \\
  \textstyle  \fiveedgecount'_{i, j} &=\textstyle \sum_{(u,v)} Y^{(i)}_{uv}Y^{(j)}_{uv}, \quad 0 \le i < j \le 2.
\end{align}
The summations are over the edges $(u, v)$ in the projected graph. Each
non-induced instance of these subgraph configurations may correspond to a 6-edge
tetrahedral configuration, and we need to adjust for these cases. Each (open or
closed) 6-edge tetrahedron count $\tetracount_{i,j,k,l}$ contributes to the
non-induced counts $\fiveedgecount'_{i,j}$, $\fiveedgecount'_{i,k}$,
$\fiveedgecount'_{i,l}$, $\fiveedgecount'_{j,k}$, $\fiveedgecount'_{j,l}$, and
$\fiveedgecount'_{k,l}$. To get the count $\fiveedgecount_{i, j}$, we subtract
the portion of $\fiveedgecount'_{i,j}$ coming from the tetrahedra. Denote the
set of valid 4-tuples of indices for the counts $\tetracount_{i,j,k,l}$ by
$\validinds$. Formally, $\validinds = \{(i, j, k, l) \;\vert\; 0 \le i \le j \le k \le l \le 2\}$.
Then $\fiveedgecount_{i,j}$ is given by
\begin{align}
  \fiveedgecount_{i, j} = \fiveedgecount'_{i, j}
  & - \sum_{k, l \;:\; (i,j,k,l) \in \validinds} \tetracount_{i,j,k,l}
  - \sum_{k, l \;:\; (i,k,j,l) \in \validinds} \tetracount_{i,k,j,l}  \\
  &\quad
  - \sum_{k, l \;:\; (i,k,l,j) \in \validinds} \tetracount_{i,k,l,j}
  - \sum_{k, l \;:\; (k,i,j,l) \in \validinds} \tetracount_{k,i,j,l} \nonumber \\
  &\quad 
  - \sum_{k, l \;:\; (k,i,l,j) \in \validinds} \tetracount_{k,i,l,j}
  - \sum_{k, l \;:\; (k,l,i,j) \in \validinds} \tetracount_{k,l,i,j}. \nonumber
\end{align}

Next, we show how to count 4-node, 4-edge subgraph configurations that contain one
triangle. There are three such configurations,
corresponding to the three possible simplicial ties in the triangle,
and we denote the counts by $\triedgecount_{s}$, $s \in \{0,1,2\}$ (the three right-most configurations in the second row of \cref{tab:configuration_labels}).
We again compute non-induced counts and then subtract the induced counts of
subgraphs with more edges, for which we showed how to compute above. Some
additional notation will be helpful for these counts. Let $\triangleset_s$ be
the set of triangles with simplicial tie strength $s \in \{0,1,2\}$, and let $a_s$
and $b_s$ count how many times triangles with a particular tie strength appear
in 5-edge configuration patterns and 6-edge configuration patterns:
\begin{align}
  a_s &= \sum_{0 \le i \le j \le 2}(\indicator{i = s} + \indicator{j = s})\fiveedgecount_{i,j} \\
  b_s &= \sum_{(i, j, k, l) \in \validinds}(\indicator{i = s} + \indicator{j = s} + \indicator{k = s} + \indicator{l = s})\tetracount_{i, j, k, l} \nonumber.
\end{align}

Consider a fixed triangle $(u, v, w)$ with simplicial tie strength $s$. We
would like to count the number of times this triangle appears in a 4-node,
4-edge subgraph configuration. Each neighbor of each of the three nodes in the
triangle is either (i) the neighbor of just one node in the triangle (ii) the
neighbor of exactly two nodes in the triangle, or (iii) the neighbor of all three
nodes in the triangle. The first case corresponds to the induced subgraph in
which we are interested, the second case to counts $\fiveedgecount_{i, j}$, and
the third case to counts $\tetracount_{i,j,k,l}$. By the inclusion-exclusion
principle,
\begin{align}
  \textstyle \triedgecount_{s} &=\textstyle \sum_{(u,v,w) \in \triangleset_s}(d_u + d_v + d_w - 6) - 2a_s -3b_s,
\end{align}
where $d$ is the degree vector of nodes in the unweighted projected graph.

Finally, we count the 4-node subgraph configuration consisting of a triangle and an
isolated node (the three leftmost configurations in the second row block of
\cref{tab:configuration_labels}).
Again, we count three types of this
configuration ($\triisocount_{s}$, $s \in \{0, 1, 2\}$), one for each of the three
simplicial tie strengths of the triangle. Every triangle appears in $(n-3)$
non-induced subgraphs with an isolated node, so we only need to subtract induced
subgraph counts with more edges. We already counted these above, so the counts
$\triisocount_{s}$ are given by
\begin{align}
  \triisocount_{s} &= \lvert \triangleset_s \rvert (n - 3) - \triedgecount_{s} -a_s -b_s.
\end{align}

\clearpage
%!TEX root = higher-order-link-prediction-postprint.tex

\section{Score functions, higher-order link prediction performance, and example predictions}\label{sec:all_score_funcs}
We derive algorithms for higher-order link prediction, which fall into four
broad categories for determining the score $\scoreijk$ of a triple of nodes:
\begin{compactenum}
\item $\scoreijk$ depends only on the weights of the edges $(i,j)$, $(i,k)$, and $(j,k)$ in the projected graph
\item $\scoreijk$ is based on the local neighborhood features in the projected graph such as the common neighbors of nodes $i$, $j$, and $k$;
\item $\scoreijk$ comes from a random-walk-based similarity score
\item $\scoreijk$ is a learned logistic regression model in a feature-based
  supervised learning setting.
\end{compactenum}
Several of these score functions are generalizations of traditional approaches
for dyadic link prediction~\cite{LibenNowell-2007-link-prediction} to account
for higher-order structure.

Here we introduce some notation for this section.
We denote the set of simplices that node $u$ appears in by $\snbrhood{u}$;
formally, $\snbrhood{u} = \{S_i \given u \in S_i\}$. The (weighted) projected
graph of a dataset is the graph on node set $V$, where the weight of edge
$(u,v)$ is the number of simplices containing both $u$ and $v$. In other words,
the $\lvert V \rvert \times \lvert V \rvert$ weighted adjacency matrix $W$ of
the projected graph is defined by
\begin{equation}\label{eqn:proj_graph}
  W_{uv} = \begin{cases}
    \cardinality{\snbrhood{u} \cap \snbrhood{v}} & u \neq v \\
    0 & u = v
  \end{cases}
\end{equation}
Sometimes, we will only need to consider unweighted version of the projected
graph, which is encoded by the adjacency matrix $A$ with entries
$A_{uv} = \min(W_{uv}, 1)$.
Finally, we denote the neighbors of a node $u$ in the projected graph by
$\nbrhood{u} = \{v \in V \given W_{uv} > 0\}$.

\subsection{Weights in the projected graph}
We use three score functions based on the weights of the pair-wise edges in the
subgraph induced by nodes $i$, $j$, and $k$. The motivation for these methods is
that weight-based tie strength positively correlates with probabilities of
simplicial closure events in an aggregate sense. Therefore,
larger weights amongst the edges between nodes $i$, $j$, and $k$ should yield
larger scores. To this end, we use the following as score functions:
\begin{align}
  \text{\emph{the harmonic mean:}}&
  \enspace  \scoreijk = 3 / (W_{ij}^{-1} + W_{ik}^{-1} + W_{jk}^{-1}) \label{eqn:harm_mean} \\
  \text{\emph{the geometric mean:}}&
  \enspace  \scoreijk = \left(W_{ij}W_{ik}W_{jk}\right)^{1/3} \label{eqn:geom_mean} \\
  \text{\emph{the arithmetic mean:}}&
  \enspace  \scoreijk = (W_{ij} + W_{ik} + W_{jk}) / 3. \label{eqn:arith_mean}
\end{align}
As discussed in the \cref{sec:prediction}, these functions
are all special cases of the generalized mean function.

\subsection{Local neighborhood features}
The next set of score functions use local neighborhood features such as common
neighbors of a triple of nodes. The reasoning here is that common neighborhood
structure amongst a triple of nodes are positive indicators of association
of the nodes; in fact, these score functions are generalizations of traditional methods used
in dyadic link prediction~\cite{LibenNowell-2007-link-prediction}.  The
common neighbors score function for a triple of nodes $i$, $j$, and
$k$ is the number of nodes that have appeared in at least one simplex with each of
the three nodes in the candidate set:
\begin{align}
  \text{\emph{3-way common neighbors:}}& 
  \enspace \scoreijk = \cardinality{\nbrhood{i} \cap \nbrhood{j} \cap \nbrhood{k}}, \label{eqn:common3}
\end{align}
where again $\nbrhood{x}$ is the set of neighbors of node $x$ in the projected graph.

The Jaccard coefficient score normalizes the number of common
neighbors by the total number of neighbors of the three candidate nodes:
\begin{align}
    \text{\emph{3-way Jaccard coefficient:}}& 
    \enspace \scoreijk = 
    \frac{ \cardinality{\nbrhood{i} \cap \nbrhood{j} \cap \nbrhood{k}}  }{
           \cardinality{\nbrhood{i} \cup \nbrhood{j} \cup \nbrhood{k}} }.\label{eqn:jaccard3}
\end{align}
This score function has been used as a general multi-way
similarity measurement for binary vectors~\cite{Heiser-1997-triadic}, but has
not been employed for a link prediction task until now.

Adamic and Adar proposed log-scaled normalization for features of common
neighbors between two nodes~\cite{Adamic-2003-friends}.
Here we adapt this to a score that performs the same
normalization over the common neighbors of three nodes:
\begin{align}\label{eqn:AA3}
  \text{\emph{3-way Adamic-Adar:}}&
  \enspace \scoreijk = \sum_{l \in \nbrhood{i} \cap \nbrhood{j} \cap \nbrhood{k}} \frac{1}{\log \cardinality{\nbrhood{l}}}.
\end{align}

Prior studies on the evolution of coauthorship have suggested preferential
attachment (PA)---in terms of degree in the coauthorship network---as a mechanism for
dyadic link formation~\cite{Newman-2001-clustering,Barabasi-2002-evolution}. We
use two scores based on a preferential attachment model of link formation:
first is 
\begin{align}
  \text{\emph{projected graph degree based PA:}}&
  \enspace \scoreijk = \cardinality{\nbrhood{i}} \cdot \cardinality{\nbrhood{j}} \cdot \cardinality{\nbrhood{k}}\label{eqn:PA_proj_deg} \\
  \text{\emph{simplicial degree based PA:}}&
  \enspace \scoreijk = \cardinality{\snbrhood{i}} \cdot \cardinality{\snbrhood{j}} \cdot \cardinality{\snbrhood{k}}.\label{eqn:PA_simp_deg}
\end{align}

\subsection{Paths and random walks}
The next set of scores functions are path-based metrics that ascribe higher scores when
there are more paths in the projected graph between a candidate triple of nodes.
Recall that $A$ and $W$ are the unweighted and weighted adjacency matrices for
the projected graph of a dataset.

The Katz score between two nodes is the sum of geometrically damped length-$l$
paths between two nodes~\cite{Katz-1953-status}.
Katz scores have been used as a criterion for predicting dyadic
links~\cite{LibenNowell-2007-link-prediction,Wang-2007-local}. Formally, the
Katz score between two nodes $i$ and $j$ in the unweighted projected graph is
$\sum_{l=1}^{\infty}\beta^lA^l_{ij}$, where $\beta$ is the damping parameter and
$A^l_{ij}$ counts the number of length-$l$ paths between $i$ and $j$. All
pairwise Katz scores can be computed in matrix form as:
\begin{equation}
  K^{(u)} = (I - \beta A)^{-1} - I.
\end{equation}
In order to guarantee that the weighted sum of length-$l$ path lengths
converges, we require that $\beta < 1 / \sigma_1(A)$, the principal singular value
of $A$ (this guarantees that $I - \beta A$ is nonsingular). 
We chose $\beta = \frac{1}{4\sigma_1(A)}$ in our experiments.

We can also use paths in the original (weighted) projected graph, where $W^l_{ij}$ is the number
of length-$l$ paths between $i$ and $j$ if we interpret the integer weights in
$W$ to be parallel edges. This leads to the weighted pairwise Katz scores
\begin{equation}
  K^{(w)} = (I - \beta W)^{-1} - I.
\end{equation}
Again, $\beta$ must be less than $1 / \sigma_1(W)$ to guarantee that $(I - \beta W)$ is nonsingular, 
and we choose $\beta = \frac{1}{4\sigma_1(W)}$ in our experiments.

Given the pairwise Katz scores, we define score functions for triples of nodes as follows:
\begin{align}
  \text{\emph{unweighted 3-way Katz:}}&
  \enspace \scoreijk = K^{(u)}_{ij} + K^{(u)}_{ik} + K^{(u)}_{jk} \label{eqn:Katz} \\
  \text{\emph{weighted 3-way Katz:}}&
  \enspace \scoreijk = K^{(w)}_{ij} + K^{(w)}_{ik} + K^{(w)}_{jk}. \label{eqn:WKatz}
\end{align}

For many of our datasets, storing the $K$ matrices in a dense format requires too much
memory. In these cases, we use the Krylov subspace method
MINRES~\cite{Paige-1975-MINRES} to solve the linear systems
\begin{equation}
  (I - \beta A)k_j = e_j, \quad j = 1, \ldots, \lvert V \rvert,
\end{equation}
where $e_j$ is the $j$th standard basis vector. After computing $k_j$, we store
only the entries of the $j$th column of $K$ corresponding to the sparsity
pattern of the $j$th column of $A$. These are the only entries of $K$ needed
for computing the scores in \cref{eqn:Katz}.

The personalized PageRank (PPR) score is another path-based score used in dyadic link
prediction~\cite{LibenNowell-2007-link-prediction,Bahmani-2010-fast}.
PPR is based on the random walk underlying the classical PageRank ranking
system for web pages~\cite{Page-1999-pagerank}. More specifically, consider a
Markov chain, where at each step, with probability $0 < \alpha < 1$, the chain
transitions according to a random walk in a graph, and with
probability $1 - \alpha$ transitions to node $i$. The PPR
score of node $j$ with respect to node $i$ is then the stationary probability of
the state $j$ for the Markov chain. The PPR scores are given by the matrix
\begin{equation}
  F^{(u)} = (1 - \alpha)(I - \alpha AD^{-1})^{-1},
\end{equation}
where $F^{(u)}_{ji}$ is the PPR score of $j$ with respect to node $i$.
Here $D$ is the diagonal degree matrix, $D_{jj} = \sum_{i}A_{ij}$.
We can again provide an analog for the weighted case:
\begin{equation}
  F^{(w)} = (1 - \alpha)(I - \alpha WD_W^{-1})^{-1},
\end{equation}
where $[D_W]_{jj} = \sum_{i}W_{ij}$ is the weighted diagonal degree matrix.

As we did with the Katz scores, we construct
three-way scores from the pairwise PPR scores:
\begin{align}
  &\text{\emph{unweighted 3-way PPR:}} \nonumber \\
  &\scoreijk = F^{(u)}_{ij} + F^{(u)}_{ji} + F^{(u)}_{ik} + F^{(u)}_{ki} + F^{(u)}_{jk} + F^{(u)}_{kj} \label{eqn:PPR} \\
  &\text{\emph{weighted 3-way PPR:}} \nonumber \\
  &\scoreijk =  F^{(w)}_{ij} + F^{(w)}_{ji} + F^{(w)}_{ik} + F^{(w)}_{ki} + F^{(w)}_{jk} + F^{(w)}_{kj}. \label{eqn:WPPR}
\end{align}
(Unlike the Katz score matrices $K$, the PPR matrices are not symmetric, so we
account for both directions of the edges.)

We also use a recent generalization of PPR scores for abstract simplicial
complexes, based on tools from algebraic topology~\cite{Schaub-2018-random}.
Here, we describe the computations necessary for these scores, assuming a basic knowledge of algebraic topology.

We consider the abstract simplicial complex defined by the union of the set of
closed triangles $T$, the set of edges $E$, and the set of vertices $V$. We
orient the edges and triangles so that $(i, j)$ for $i < j$ corresponds to an
edge $\{i, j\}$ and $(i, j, k)$ for $i < j < k$ corresponds to a closed triangle
$\{i, j, k\}$.  Following the ideas of Schaub et al.,
we define the normalized combinatorial Hodge Laplacian as
\begin{equation}
  \normedgelap = (G D^{-1} G^T + C^TC) M^{-1},
\end{equation}
where the ``gradient operator'' $G$ is a $\lvert E \rvert \times \lvert V \rvert$ matrix defined by
\begin{equation}
  G_{(i, j), x} =
  \begin{cases}
    1 & x = j \\
    -1 & x = i \\
    0 & \text{otherwise},
  \end{cases}
\end{equation}
the ``curl operator'' $C$ is a $\lvert T \rvert \times \lvert E \rvert$ matrix defined by
\begin{equation}
  C_{(i, j, k), (x, y)} =
  \begin{cases}
    1  & (x, y) = (i, j) \text{ or } (x, y) = (j, k) \\
    -1 & (x, y) = (i, k)\\
    0 & \text{otherwise},
  \end{cases}
\end{equation}
$D$ is a diagonal matrix defined by
\begin{equation}
  \textstyle D_{xx} = \sum_{(i, j)} \lvert G_{(i, j), x} \rvert,
\end{equation}
and $M$ is a diagonal matrix defined by
\begin{equation}
  \textstyle M_{(x,y), (x,y)} = 2 + \sum_{(i, j, k)} \lvert C_{(i, j, k), (x, y)} \rvert.
\end{equation}

The matrix $P = \frac{1}{2}(I - \normedgelap)$ defines a Markov-like operator.
The simplicial PageRank scores (defined on each pair of edges) can thus be defined analogously to the standard PageRank:
\begin{equation}
  \sppr = (I - \alpha P)^{-1}(1 - \alpha).
\end{equation}
Here, the matrix $\sppr$ defines pairwise scores between \emph{edges}, and we construct
a score function on triples of nodes by taking the sum of pairwise scores:
\begin{align}
 &\text{\emph{3-way simplicial PPR:}} \nonumber \\
 &\enspace \scoreijk =
  \lvert\sppr_{(i, j), (j, k)} \rvert + \lvert \sppr_{(j, k), (i, j)} \rvert +\lvert\sppr_{(i, j), (i, k)} \rvert \nonumber \\
  &+ \lvert\sppr_{(i, k), (i, j)} \rvert +\lvert \sppr_{(j, k), (i, k)} \rvert + \lvert \sppr_{(j, k), (i, k)} \rvert.\label{eqn:sppr}
\end{align}

%%%%%%
% tab:prediction_performance_full
%!TEX root = higher-order-link-prediction-postprint.tex

\begin{table*}[tb]
\setlength{\tabcolsep}{2pt}
\centering
\caption{Open triangle closure prediction performance based on several score functions:
  random (Rand.);
  harmonic, geometric, and arithmetic means of the 3 edge weights;
  3-way common neighbors (Common);
  3-way Jaccard coefficient (Jaccard);
  3-way Adamic-Adar (A-A);
  projected graph degree and simplicial degree preferential attachment (PGD-PA and SD-PA);
  unweighted and weighted Katz similarity (U-Katz and W-Katz);
  unweighted and weighted personalized PageRank (U-PPR and W-PPR);
  simplicial personalized PageRank (S-PPR; missing entries are cases where
  computations did not finish within 2 weeks); and
  a feature-based supervised model using logistic regression (Log.\ reg.).
  Performance is AUC-PR relative to the random baseline.
  The random baseline is listed in absolute terms and equals the fraction of
  open triangles that close.
  The harmonic and geometric means of edge weights perform well across
  many datasets, further highlighting the role of tie strength in
  predicting simplicial closure events. This signal from local structure
  contrasts from traditional pairwise link prediction, where longer paths
  are needed for effective prediction~\cite{LibenNowell-2007-link-prediction}.
  The supervised method also performs well, suggesting that combinations of features
  capture the rich variety of structure observed across datasets.
}
\scalebox{0.83}{
\begin{tabular}{l c @{\quad} c c c c c c c c c c c c c c c c c c}
\toprule
Dataset                & Rand.    & Harm.\ mean & Geom.\ mean & Arith.\ mean & Common & Jaccard & A-A   & PGD-PA & SD-PA & U-Katz  & W-Katz & U-PPR  & W-PPR & S-PPR & Log.\ reg. \\
\midrule
coauth-DBLP            & 1.68e-03 & 1.49  & 1.59  & 1.50  & 1.33  & 1.84  & 1.60  & 0.74  & 0.74 & 0.97  & 1.51  & 1.62 & 1.83 & 1.21 & 3.37   \\
coauth-MAG-History     & 7.16e-04 & 1.69  & 2.72  & 3.20  & 5.11  & 2.24  & 5.82  & 1.50  & 2.49 & 6.30  & 3.40  & 1.66 & 1.88 & 1.35 & 6.75   \\
coauth-MAG-Geology     & 3.35e-03 & 2.01  & 1.97  & 1.69  & 2.43  & 1.84  & 2.71  & 1.31  & 0.97 & 1.99  & 1.74  & 1.06 & 1.26 & 0.94 & 4.74   \\
music-rap-genius       & 6.82e-04 & 5.44  & 6.92  & 1.98  & 1.85  & 1.62  & 2.10  & 1.82  & 2.15 & 1.93  & 2.00  & 1.78 & 2.09 & 1.39 & 2.67   \\
tags-stack-overflow    & 1.84e-04 & 13.08 & 10.42 & 3.97  & 6.45  & 9.43  & 6.63  & 3.37  & 2.74 & 2.95  & 3.60  & 1.08 & 1.85 &  --   & 3.37   \\
tags-math-sx           & 1.08e-03 & 9.08  & 8.67  & 2.88  & 6.19  & 9.37  & 6.34  & 3.48  & 2.81 & 4.53  & 2.71  & 1.19 & 1.55 & 1.86 & 13.99  \\
tags-ask-ubuntu        & 1.08e-03 & 12.29 & 12.64 & 4.24  & 7.15  & 4.96  & 7.51  & 7.48  & 5.63 & 7.10  & 4.15  & 1.75 & 2.54 & 1.19 & 7.48   \\
threads-stack-overflow & 1.14e-05 & 23.85 & 31.12 & 12.97 & 2.73  & 3.85  & 3.19  & 5.20  & 3.89 & 1.06  & 11.54 & 1.66 & 4.06 & -- & 1.53   \\
threads-math-sx        & 5.63e-05 & 20.86 & 16.01 & 5.03  & 25.08 & 28.13 & 23.32 & 10.46 & 7.46 & 11.04 & 4.86  & 0.90 & 1.18 & 0.61 & 47.18  \\
threads-ask-ubuntu     & 1.31e-04 & 78.12 & 80.94 & 29.00 & 21.04 & 2.80  & 30.82 & 7.09  & 6.62 & 16.63 & 32.31 & 0.94 & 1.51 & 1.78 & 9.82   \\
NDC-substances         & 1.17e-03 & 4.90  & 5.27  & 2.90  & 5.92  & 3.36  & 5.97  & 4.76  & 4.46 & 5.35  & 2.93  & 1.39 & 1.83 & 1.86 & 8.17   \\
NDC-classes            & 6.72e-03 & 4.43  & 3.38  & 1.82  & 1.27  & 1.19  & 0.99  & 0.94  & 2.14 & 0.92  & 1.34  & 0.78 & 0.91 & 2.45 & 0.62   \\
DAWN                   & 8.47e-03 & 4.43  & 3.86  & 2.13  & 4.73  & 3.76  & 4.77  & 3.76  & 1.45 & 4.61  & 2.04  & 1.57 & 1.37 & 1.55 & 2.86   \\
congress-committees    & 6.99e-04 & 3.59  & 3.28  & 2.48  & 4.83  & 2.49  & 5.04  & 1.06  & 1.31 & 3.21  & 2.59  & 1.50 & 3.89 & 2.13 & 7.67   \\
congress-bills         & 1.71e-04 & 0.93  & 0.90  & 0.88  & 0.65  & 1.23  & 0.66  & 0.60  & 0.55 & 0.60  & 0.78  & 3.16 & 1.07 & 6.01 & 107.19 \\
email-Enron            & 1.40e-02 & 1.78  & 1.62  & 1.33  & 0.85  & 0.83  & 0.87  & 1.27  & 0.83 & 0.99  & 1.28  & 3.69 & 3.16 & 2.02 & 0.72   \\ 
email-Eu               & 5.34e-03 & 1.98  & 2.15  & 1.78  & 1.28  & 2.69  & 1.37  & 0.88  & 1.55 & 1.01  & 1.79  & 1.59 & 1.75 & 1.26 & 3.47   \\
contact-high-school    & 2.47e-03 & 3.86  & 4.16  & 2.54  & 1.92  & 3.61  & 2.00  & 0.96  & 1.13 & 1.72  & 2.53  & 1.39 & 2.41 & 0.78 & 2.86   \\
contact-primary-school & 2.59e-03 & 5.63  & 6.40  & 3.96  & 2.98  & 2.95  & 3.21  & 0.92  & 0.94 & 1.63  & 4.02  & 1.41 & 4.31 & 0.93 & 6.91   \\
\bottomrule
\end{tabular}
}
\label{tab:prediction_performance_full}
\end{table*}

%%%%%%

\subsection{Supervised learning}
Finally, we used a supervised machine learning approach that learns the
appropriate score function given features of the open triangle.  To this end, we
further divide the training data into a sub-training set (simplices appearing in
the first 60\% of the entire dataset) and a validation set (simplices appearing
between the 60th and 80th percentile of the time spanned by the entire dataset).
We trained an $\ell_2$-regularized logistic regression model using
the scikit learn library\footnote{\url{http://scikit-learn.org/}}~\cite{Pedregosa-2011-scikit-learn}
for predicting closure on the validation set using features of open structures in the
sub-training set. The features for each open triangle $(i, j, k)$ were
\begin{compactenum}
\item\label{item:feats1} the number of simplices containing pairs of nodes $i$ and $j$, $i$ and $k$, and $j$ and $k$;
\item\label{item:feats2} the degree of nodes $i$, $j$, and $k$ in the projected graph: 
$\lvert N(i) \rvert$, $\lvert N(j) \rvert$, and $\lvert N(k) \rvert$;
\item\label{item:feats3} the number of simplices containing nodes $i$, $j$, and $k$:
$\lvert R(i) \rvert$, $\lvert R(j) \rvert$, and $\lvert R(k) \rvert$;
\item\label{item:feats4} the number of common neighbors in the projected graph of nodes $i$ and $j$,
  $i$ and $k$, and $j$ and $k$:
$\lvert N(i) \cap N(j) \rvert$, $\lvert N(i) \cap N(k) \rvert$, and $\lvert N(j) \cap N(k) \rvert$;  
\item\label{item:feats5} the number of common neighbors of all three nodes $i$, $j$, and $k$
in the projected graph:
$\lvert N(i) \cap N(j) \cap N(k) \rvert$
\item the log of the features in \cref{item:feats1,item:feats2,item:feats3} and the
  log of the sum of 1 and the feature value for the features in \cref{item:feats4,item:feats5}.
\end{compactenum}
After learning the model, we predicted on the test set using the same features
computed on the entire training set (first 80\% of the dataset).

\subsection{Prediction performance}

Using the ranking induced by the score functions described above, we evaluated
the prediction performance on each dataset by the area under the
precision-recall curve (AUC-PR) metric (\cref{tab:prediction_performance_full}).
We use random scores---more specifically, a random ranking---as a baseline,
and report scores relative to this baseline.

As seen in \cref{sec:prediction}, our proposed algorithms can achieve much higher
performance than randomly guessing which open triangles go through a simplicial
closure event. We also still see good performance of the harmonic and geometric
means, as well as the supervised problem, with respect to this expanded set of
score functions.

We may further decompose the pairwise scores of simplicial PageRank scores in
\cref{eqn:sppr} into the gradient, harmonic, and curl components given by the
Hodge decomposition~\cite{Schaub-2018-random}. Computationally, we solve the
least squares problems
\begin{align}
  \min_{X} \| GX - S \|_F, \qquad \min_{Y} \| C^TY - S \|_F \label{eqn:lsqr_grad_curl}
\end{align}
using the iterative method LSQR~\cite{Paige-1982-LSQR} (with tolerances
$10^{-3}$) on each column. Given the minimizers $X^*$ and $Y^*$ of
\cref{eqn:lsqr_grad_curl}, the components of the Hodge decomposition are
\begin{align}
  \spprgrad = GX^*, \quad \spprcurl &= C^TY^*, \quad \spprharm = \sppr - \spprgrad - \spprcurl.
\end{align}
Each of $\spprgrad$, $\spprcurl$, and $\spprharm$ defines pairwise scores
between edges, and we construct score functions on triples of nodes in the same
way as in \cref{eqn:sppr}.

We report the performance results in \cref{tab:Hodge_perf} for the datasets that
were small enough on which computing the Hodge decomposition was computationally
feasible. We observe that the components from the Hodge decomposition can
provide substantially better results than the ``combined'' simplicial PageRank
score reported \cref{tab:prediction_performance_full}. However, no component
consistently out-performs the others.

%%%%%%
% tab:Hodge_perf
%!TEX root = higher-order-link-prediction-postprint.tex

\begin{table}[tb]
\setlength{\tabcolsep}{2pt}
\centering
\caption{%
  Open triangle closure prediction performance based on score functions from the
  Hodge decomposition of the simplicial personalized PageRank vector.
}
\begin{tabular}{l c @{\quad} c c c c c c c c c c c c c c c c c}
\toprule
Dataset                & Rand.    & combined  & gradient & harmonic & curl \\
\midrule
coauth-MAG-History     & 7.16e-04 & 1.35 & 1.25 & 1.13 & 1.27 \\
music-rap-genius       & 6.82e-04 & 1.39 & 1.44 & 1.40 & 1.47 \\
tags-math-sx           & 1.08e-03 & 1.86 & 0.73 & 0.66 & 0.74 \\
tags-ask-ubuntu        & 1.08e-03 & 1.19 & 0.61 & 0.59 & 0.71 \\
threads-ask-ubuntu     & 1.31e-04 & 0.61 & 0.58 & 0.61 & 4.59 \\
NDC-substances         & 1.17e-03 & 1.86 & 0.63 & 0.72 & 0.60 \\
NDC-classes            & 6.72e-03 & 2.45 & 1.37 & 0.83 & 1.74 \\
DAWN                   & 8.47e-03 & 1.55 & 0.59 & 0.60 & 0.65 \\
congress-committees    & 6.99e-04 & 2.13 & 1.22 & 1.13 & 1.63 \\
email-Enron            & 1.40e-02 & 2.02 & 2.90 & 1.98 & 2.46 \\ 
email-Eu               & 5.34e-03 & 1.26 & 1.28 & 0.82 & 1.63 \\
contact-high-school    & 2.47e-03 & 0.78 & 0.99 & 1.68 & 2.38 \\
contact-primary-school & 2.59e-03 & 0.93 & 1.45 & 1.84 & 3.26 \\
\bottomrule
\end{tabular}
\label{tab:Hodge_perf}
\end{table}

%%%%%%

\subsection{Example predictions}
Lastly, we provide a concrete example of predictions.
\Cref{tab:top_predictions} shows the top 25 predictions of the Adamic-Adar score
function on the DAWN dataset. In this dataset, fewer than one in a hundred open
triangles in the training set experience a simplicial closure event in the test
set, but 4 of the top 25 predictions from this score function go through a
simplicial closure event. Three of the correct predictions relate to novel
combinations with proton pump inhibitors.

%%%%%
% tab:top_predictions
%!TEX root = higher-order-link-prediction-postprint.tex

\begin{table}[tb]
\setlength{\tabcolsep}{2pt}
\centering
\caption{Top 25 predictions from the 3-way Adamic-Adar algorithm for open
  triangles to go through a simplicial closure event in the DAWN dataset. An
  ``\mycheck'' marks open triangles that actually go through a simplicial closure event
  final 20\% of the time spanned by the dataset. Four of the top 25 predictions
  do indeed have a simplicial closure event.}
  \begin{tabular}{l @{\quad} c l}
    \toprule
  1 & &  methyldopa; gentamicin; proton pump inhibitors \\
  2 & \mycheck & norepinephrine; chlormezanone; proton pump inhibitors \\
  3 & &  ranitidine; gentamicin; proton pump inhibitors \\
  4 & &  dihydroergotamine; methyldopa; asa/butalbital/caffeine/codeine \\
  5 & &  ranitidine; gentamicin; levodopa \\
  6 & &  praziquantel; diazepam; alfentanil \\
  7 & &  asa/caffeine/dihydrocodeine; praziquantel; proton pump inhibitors  \\
  8 & &  chloral hydrate; tobramycin; sumatriptan \\
  9 & &  oxybutynin; gentamicin; tobramycin \\
  10 & & asa/caffeine/dihydrocodeine; norepinephrine; sumatriptan \\
  11 & & ampicillin; chlormezanone; proton pump inhibitors \\
  12 & & bepridil; diazepam; alfentanil \\
  13 & &  colestipol; oxybutynin; proton pump inhibitors \\
  14 & \mycheck & nadolol; benazepril; proton pump inhibitors \\
  15 & &  thalidomide; amiloride; maprotiline \\
  16 & \mycheck &  nadolol; lamivudine-zidovudine; proton pump inhibitors \\  
  17 & &  chloral hydrate; verapamil; methyldopa \\
  18 & &  chlorzoxazone; benazepril; proton pump inhibitors \\
  19 & &  heparin; asa/caffeine/dihydrocodeine; proton pump inhibitors \\
  20 & &  oxcarbazepine; norepinephrine; proton pump inhibitors \\
  21 & &  dihydroergotamine; tobramycin; alfentanil \\
  22 & &  maprotiline; norepinephrine; proton pump inhibitors \\
  23 & &  oxybutynin; methyldopa; dihydroergotamine \\
  24 & &  heparin; dihydroergotamine; proton pump inhibitors \\
  25 & \mycheck & ampicillin; methyldopa; diazepam \\
\bottomrule
\end{tabular}
\label{tab:top_predictions}
\end{table}

%%%%%

\end{document}